\documentclass[a4paper,fleqn]{cas-dc}
\usepackage{xcolor}
\usepackage{units}
\usepackage{upgreek}
\usepackage{revsymb}

\usepackage{lineno}

\usepackage[normalem]{ulem}

\def\tsc#1{\csdef{#1}{\textsc{\lowercase{#1}}\xspace}}
\tsc{WGM}
\tsc{QE}
\tsc{EP}
\tsc{PMS}
\tsc{BEC}
\tsc{DE}

\usepackage{overpic}

\usepackage[numbers]{natbib}
\let\oldsim\sim 
\renewcommand{\sim}{{\oldsim}}
\newcommand{\PTARMIGAN}{\textsc{Ptarmigan}\xspace}
\newcommand{\GEANT}{\textsc{Geant4}\xspace}
\newcommand{\ALL}{Allpix$^2$\xspace}

\usepackage{subcaption}
\usepackage[title]{appendix}

\begin{document}
\let\WriteBookmarks\relax
\def\floatpagepagefraction{1}
\def\textpagefraction{.001}
\shorttitle{Prospects for the production and detection of Breit-Wheeler tunneling positrons in Experiment 320 at the FACET\textsf{--}II accelerator.}
\shortauthors{O. Borysov et~al.}

\title [mode = title]{Prospects for the production and detection of Breit-Wheeler tunneling positrons in Experiment 320 at the FACET\textsf{--}II accelerator.}                      

\author[1]{Oleksandr {Borysov}}

\author[1]{Alon {Levi}}

\author[2]{Sebastian {Meuren}}[type=editor,orcid=0000-0002-2744-7756]
\ead{sebastian.meuren@polytechnique.edu}
\cormark[1]

\author[1]{Nathaly {Nofech-Mozes}}

\author[4]{Ivan {Rajkovic}}

\author[5]{David A. {Reis}}

\author[1,3,6]{Arka {Santra}}

\author[4]{Doug W. {Storey}}

\author[1]{Noam {Tal Hod}}[type=editor,orcid=0000-0001-5241-0544]
\ead{noam.hod@weizmann.ac.il}
\cormark[1]

\author[1]{Roman {Urmanov}}[type=editor,orcid=0009-0001-5686-9050]
\ead{roman.urmanov@weizmann.ac.il}
\cormark[1]

\affiliation[1]{organization={Department of Particle Physics and Astrophysics, Weizmann Institute of Science},
addressline={234 Herzl Street}, 
city={Rehovot},
postcode={7610001}, 
country={Israel}}

\affiliation[2]{organization={LULI, CNRS, CEA, Sorbonne Université, Ecole Polytechnique,},
addressline={Institut Polytechnique}, 
city={Palaiseau},
postcode={91128}, 
country={France}}

\affiliation[3]{organization={Saha Institute of Nuclear Physics},
addressline={1/AF Bidhannagar}, 
city={Kolkata},
postcode={700064}, 
country={India}}

\affiliation[4]{organization={SLAC National Accelerator Laboratory},
addressline={2575 Sand Hill Rd}, 
city={Menlo Park},
postcode={94025}, 
state={CA}, 
country={USA}}

\affiliation[5]{organization={Stanford PULSE Institute, SLAC National Accelerator Laboratory},
addressline={2575 Sand Hill Rd}, 
city={Menlo Park},
postcode={94025}, 
state={CA}, 
country={USA}}

\affiliation[6]{organization={Homi Bhabha National Institute},
addressline={Training School Complex, Anushakti Nagar}, 
city={Mumbai},
postcode={400094}, 
state={Maharashtra}, 
country={India}}

\cortext[cor1]{Corresponding authors}

\begin{abstract}
The SLAC Experiment 320 collides 10~TW-class laser pulses with the high-quality, 10~GeV electron beam from the FACET\textsf{--}II RF LINAC.
This setup is expected to produce a sizable number of $e^+e^-$ pairs via nonlinear Breit-Wheeler mechanism in the strong-field tunneling regime, with an estimated yield of $\sim 0.01\textsf{--}0.1$ pairs per collision.
The positrons are separated from the electrons with a dipole magnet and are measured with a detector that is placed after it.
This small signal rate typically comes along with large backgrounds originating, e.g., from dumping the high-charge primary beam, secondaries induced by the beam halo, as well as photons and low-energy electrons produced in the electron-laser collision itself.
These backgrounds may reach densities of $\mathcal{O}(100)$ charged particles per cm$^2$ (and even more neutral particles) at the surface of the detector, making it a tremendous challenge for an unambiguous detection of single particles.
In this work, we demonstrate how detectors and methods adapted from high-energy physics experiments, can enable the single NBW positron measurement.
The solution presented is based on a highly granular, multi-layer, radiation-hard pixel detector paired with powerful particle-tracking algorithms.
Using a detailed simulation of the existing experimental setup, we show how the false-positive rate due to background processes can be reduced by more than an order of magnitude relative to the expected signal after full reconstruction.
Furthermore, we show that the high spatial resolution ($<10~\mu{\rm m}$) achievable with tracking using four detector layers allows for positron momentum measurements with a resolution of <2\% within acceptance ($\sim 1\textsf{--}4$~GeV), thus enabling spectral characterization of the nonlinear Breit-Wheeler process.
Based on our extensive simulation, with a conservatively large background assumption, we show that it is possible to measure single Breit-Wheeler positrons in the coming data taking campaign of E320.
That would be the first statistically significant observation and characterization of this elusive process in the (deep) tunneling regime.
\end{abstract}

\begin{keywords}
E320, NBW, Tracking, ALPIDE, ACTS, SF-QED
\end{keywords}

\maketitle

\section{Introduction}
Understanding quantum electrodynamics (QED) in the presence of strong electromagnetic (EM) background fields is a thriving, worldwide experimental effort.
Thanks to the rapid evolvement of laser power in the last few decades, with energies reaching $\sim 100$~J and above, strong EM fields with laboratory-frame amplitudes reaching $\sim\mathcal{O}(10^{-5}\textsf{--}10^{-4})$ of the Schwinger QED critical field strength of $\sim 1.3\times 10^{18}$~V/m~\cite{PhysRev.82.664,Sauter} are now achievable~\cite{danson_petawatt_2019}.
While the QED critical field is currently orders of magnitude above any terrestrially producible field strength, the gap can be significantly narrowed and may even be closed in collisions of these lasers with high-energy probe particles (electrons or photons).
The field seen by particles of mass $m$ and energy $E$ is boosted by $\sim\gamma = E/m$\footnote{Here and in the following we will use natural units with $c = \hbar =\epsilon_0 =1$.}.
For example, the boost factor for a $10~{\rm GeV}$ probe electron, is $\gamma \approx 2\times 10^4$.
Multi-GeV electron beams are available in a few radio-frequency (RF) linear acceleration (LINAC) facilities (see e.g.~\cite{Yakimenko:IPAC2016-TUOBB02,altarelli2007european}).
In addition, such beams are becoming available also in an increasing number of (multi-) PW laser wakefield acceleration (LWFA) facilities worldwide (see e.g.~\cite{weber2017p3,gales2018extreme,papadopoulos2016apollon,yoon2021realization}), where the electrons are accelerated by a high-power laser~\cite{kim2013zsa,yan_high_2017,PhysRevX.8.031004,PhysRevX.8.011020,PhysRevLett.122.084801,hojbota2019accurate,tanaka2020eli,kim2021multigev,miao2022multigev,aniculaesei2023acceleration,rezaei2023laser,PhysRevLett.133.255001,corels_compton_2024,los2024observation,PhysRevLett.132.195001,winkler2025active}.

The emerging experimental capability to collide high-energy electrons with intense laser pulses opens the door for observing physical phenomena that are typically found only in extremely violent environments in nature.
Similar characteristic fields exist, for example, in astrophysical environments~\cite{Ruffini:2009hg,nikishov62,Kouveliotou:1998ze,Harding_2006,Turolla:2015mwa} and in lepton colliders~\cite{yakimenko2019prospect,Bucksbaum:2020,Akhmadaliev:2001ik}.
Furthermore, analogies with strong-field phenomena in non-relativistic atomic, molecular, and condensed matter systems are being explored~\cite{ivanov,doi:10.1142/9789811279461_0008}.

Two main strong-field QED (SF-QED) processes occur when a GeV-scale electron propagates in an intense EM field of a laser pulse: the first is nonlinear Compton scattering (NCS), where the electron absorbs (multiple) laser photons and emits one high-energy photon.
The second process is nonlinear Breit-Wheeler pair production (NBW), where a real $e^+e^-$ pair is produced~\cite{RevModPhys.94.045001, Fedotov:2022ely}.
In the regime accessible to present experiments, the two-step contribution to NBW is dominant, where first a real high-energy NCS photon is produced and then decays into a real $e^+e^-$ pair~\cite{PhysRevD.107.096004,PhysRevD.101.056017,PhysRevLett.105.080401}.
The one-step (nonlinear 
trident) contribution can be neglected in the same regime.\\

To date, there are several running and planned experiments worldwide seeking to study these SF-QED processes in different setups (see e.g.~\cite{corels_compton_2024,PhysRevLett.132.175002,Matheron:2024hwy,LUXETDR}).
While their goals are coherent, the different approaches (e.g., LWFA or RF-LINAC based) are complementary in the prospective precision and the phase-space coverage.
Beyond electron-laser collisions, there are other experiments exploring SF-QED in different environments, e.g., where electrons are collided with oriented single crystals in the NA63 experiment~\cite{PhysRevLett.130.071601} at CERN (see also~\cite{RevModPhys.77.1131}), or using heavy, highly charged ions such as Helium-like Uranium~\cite{HeliumLikeUranium}.
Light-by-light scattering and the Breit–Wheeler process were also measured by the ATLAS~\cite{ATLAS:2020hii} and CMS~\cite{CMS:2024bnt} collaborations in heavy-ion collisions at the LHC.

While the NCS process has been observed multiple times in the past, most recently using an ``all-laser'' demonstration~\cite{corels_compton_2024} with the 4~PW laser at the Center for Relativistic Laser Science (CoReLS) in Korea, we focus on the NBW process in this work.
To date, the only measurement of the NBW process was carried out by the seminal SLAC Experiment 144~\cite{Bamber:1999zt} (E144) in the mid 1990s, where both nonlinear Compton scattering and Breit–Wheeler pair production were measured, albeit not in the strong-field tunneling regime.
Nonetheless, the observed pair-production rate was equally well modeled in the multi-photon (involving $n=5$ laser photons) and the tunneling pictures.
However, the fitted tunneling exponent deviated significantly from the one expected in the strong-field limit.

Thus, a high-priority goal of current and future SF-QED experiments is to measure the NBW process for the first time in the (deep) tunneling regime (see Appendix.~\ref{app:tunneling} for a more detailed discussion on the tunneling regime), where the electron-positron pair is not produced via the collisional absorption of multiple laser photons, but rather via a strong-field tunneling process as discussed below.
These experiments face extreme challenges owing to the small signal production probability and the large background fluxes associated with the collision and the beam electrons as discussed extensively in this work in the context of the running Experiment 320 (E320)~\cite{Chen:22}.

\section{Physics overview}
\label{sec:physics}
The NBW $e^+e^-$ pair production process can happen via two different mechanisms: in the perturbative regime multiple laser photons are absorbed by the NCS photon, providing the necessary four-momentum to produce the $e^+e^-$ pair; this regime has been probed by E144 for the first time.
In the high-intensity regime, however, an electron/positron can exchange so many photons with the laser modes that the collective EM field of the laser becomes dominant over the interaction with individual, quantized laser photons.
In this regime the electric field enables $e^+e^-$ pair production via a strong-field tunneling process, in close analogy to tunnel ionization in atoms: inside the laser field the high-energy NCS photon keeps fluctuating into a virtual $e^+e^-$ pair, which can be separated by the collective electric field of the laser~\cite{PhysRevD.93.085028,PhysRevD.91.013009}.
As in atomic physics, we also expect to see re-collision processes once the pair is real \cite{meuren_high-energy_2015,kuchiev_production_2007}.
The NCS and NBW processes can in principle be cascaded, resulting in an exponentially growing number of produced photons and $e^+e^-$ pairs \cite{pouyez_kinetic_2025,qu_signature_2021,bell_possibility_2008}.

The two processes are characterized by two Lorentz-invariant, dimensionless parameters.
The first is the classical nonlinearity parameter, $a_0=e|\textbf{E}|/(m_e \omega)$.
It is often simply referred to as the laser intensity parameter, and is also denoted by $\xi$ or $\eta$.
Here, $\omega$ and $|\textbf{E}|$ denote the (average) angular frequency and the (sub-cycle) peak electric field of the laser, respectively, and $e$ and $m_e$ are the electron charge and mass, respectively.
To give a physical intuition for $a_0$, we note that it determines the transverse momentum gained by a free electron over the laser cycle, normalized to $m_e c$~\cite{landau_fields_1975}.
We identify that $\gamma_K = 1/a_0$ is the generalized Keldysh parameter~\cite{karnako_current_2015,keldysh_ionization_1965} for pair production; it distinguishes the multi-photon ($a_0 \ll 1$) from the tunneling regime ($a_0 \gg 1$)~\cite{aleksandrov_lcfa_2019,PhysRevD.93.085028}. 
Note that laser intensity is usually given as a cycle-averaged value.
However, for tunneling pair production, the sub-cycle peak electric field strength is the decisive factor and this is the reason for the notation used here.

The second important parameter is the quantum parameter of the high-energy NCS gamma photon, $\chi_\gamma=a_0\eta_\gamma$, where $\eta_\gamma=\kappa\cdot k/m_e^2 = \omega\omega_k\left(1+\cos\theta\right)/m_e^2$ is the photon energy parameter.
Here, $\kappa^\mu$ is the (average) four-momentum of the laser photons ($\kappa^0 = \omega$), $k^\mu$ is the four-momentum of the high-energy NCS photon ($k^0 = \omega_k$), and $\theta$ is the collision angle (with $\theta =0$ corresponding to a head-on collision).
One can also define $\chi_e$ as the quantum parameter of the primary (radiating) electron.
In an intuitive analogy to $a_0$, 
$\chi_e$ represents the transverse momentum gained by a free electron over a Compton wavelength in the rest frame of the electron, normalized to $m_e c$.
This is also equivalent to the ratio between the field experienced by the electron in its rest frame to the Schwinger field.
For manifestly covariant representations of the classical nonlinearity parameter, the energy parameter and the quantum parameter, see~\cite{Bamber:1999zt}.

In an (approximately) plane-wave laser field the NBW rate is expected to scale as $\chi_\gamma e^{-8/(3\chi_\gamma)}$ for $a_0^2\gg 1$ and $\chi_\gamma\ll 1$~\cite{Ritus}.
To a good approximation, this scaling is still valid for $a_0 \gtrsim 1$ and $\chi_\gamma\lesssim 1$~\cite{Fedotov:2022ely,Hartin:2018sha}.
The exponential suppression is typical for a tunneling process.
At $\chi_\gamma\gg 1$ the rate behaves as $\chi_\gamma^{2/3}$.

To probe NBW pair production in a highly nonlinear regime requires a concerted interaction of many laser photons.
This becomes obvious from the energy-momentum conservation law: $p^\mu + n \kappa^\mu = p'^\mu + p_+^\mu + p_-^\mu$, where $p^\mu$ ($p'^\mu$) is the four-momentum of the incoming (outgoing) electron that undergoes the NCS process and stimulates the NBW pair production, $n$ denotes the number of net-absorbed laser photons, and $p_\pm^\mu$ are the four-momenta of the produced $e^+e^-$ pair.
All four-momenta are taken at asymptotic times, i.e., outside the laser field.
We note that in the literature there are also discussions about how to evaluate the energy-momentum conservation in the laser field using dressed momenta~\cite{Bamber:1999zt,Ritus}.
Generally, electron-induced pair production is mediated via a virtual intermediate photon.
However, as mentioned above, the dominant contribution is from the two-step process, where a real NCS photon is created and decays to the NBW pair.
By squaring the conservation relation we find that $n\kappa\cdot p = m_e^2 + p_+\cdot p_- + p_-\cdot p' + p_+\cdot p'$ and since $p'\cdot p_\pm$, $p_+\cdot p_- \geq m_e^2$, we obtain the kinematic threshold $n \geq 4m_e^2/(\kappa\cdot p)$.
For E320, $\kappa\cdot p/m_e^2 \approx 0.11$ (see Sec.~\ref{sec:experiment} for details), implying that an electron needs to absorb at least $n=37$ laser photons in order to produce one $e^+e^-$ pair ($n=5$ for E144).
If the mass dressing is properly taken into account, this lower bound increases even more.
In the perturbative regime ($a_0^2 < 1$, as in E144) the probability for absorbing $n$ laser photons typically scales as $(a_0)^{2n}$~\cite{Ritus}, implying that multi-photon pair production with a large value of $n$ is not viable.

For further intuitive discussion about the different regimes of SF-QED and the relevant parameters we refer the reader to App.~\ref{app:sfqed}.

\section{Experiment overview}
\label{sec:experiment}
The FACET-II National User Facility at the SLAC National Accelerator Laboratory provides high-intensity electron beams for studying their interactions with lasers, plasmas, and solids~\cite{Yakimenko:IPAC2016-TUOBB02,PhysRevLett.134.085001,PhysRevAccelBeams.22.101301}.
Electron bunches of up to 2~nC are presently accelerated to 10~GeV (potentially up to 13~GeV if all klystrons are being used) in the $\sim$1~km S-band LINAC with three stages of bunch compression to deliver bunches with a transverse spot size on the order of $10\times 10~\mu{\rm m}^2$, a longitudinal bunch length of $\sim$10~$\mu{\rm m}$, and thus peak currents in excess of $\sim100$~kA to the experimental area.
The multipurpose experimental area is specifically designed to facilitate a variety of experiments, with a strong emphasis on plasma wakefield acceleration (PWFA) and probing SF-QED, and is outfitted with a suite of diagnostic devices to measure the incoming and outgoing beam properties.
This includes a magnetic imaging spectrometer to measure the properties of electrons after the IP~\cite{PhysRevAccelBeams.27.051302}.
The machine sends electron bunches to the experimental area at a typical rate of 10~Hz (up to 30~Hz is possible).

The highest laser intensities expected for the 2025\textsf{--}2026 run time of E320, with an on-target total laser energy of 0.12\textsf{--}0.3~J, a pulse duration of $\lesssim 50$~fs (intensity FWHM), a wavelength of 800~nm and a $\sim 2~\mu{\rm m}$ waist at focus (FWHM), are as large as $a_0=3\textsf{--}7$ with the existing 10~TW FACET-II laser system.
The repetition rate of the laser is $10$~Hz (rate reduction down to 1~Hz is possible with a pulse chopper).
Here, we assume that the quoted on-target total laser energy is distributed in an approximate Airy pattern, implying that about 50\% of the reported on-target total energy is contained within the FWHM of the focal spot.

\begin{figure*}[pos=!ht]
\centering
\begin{overpic}[width=0.9\textwidth]{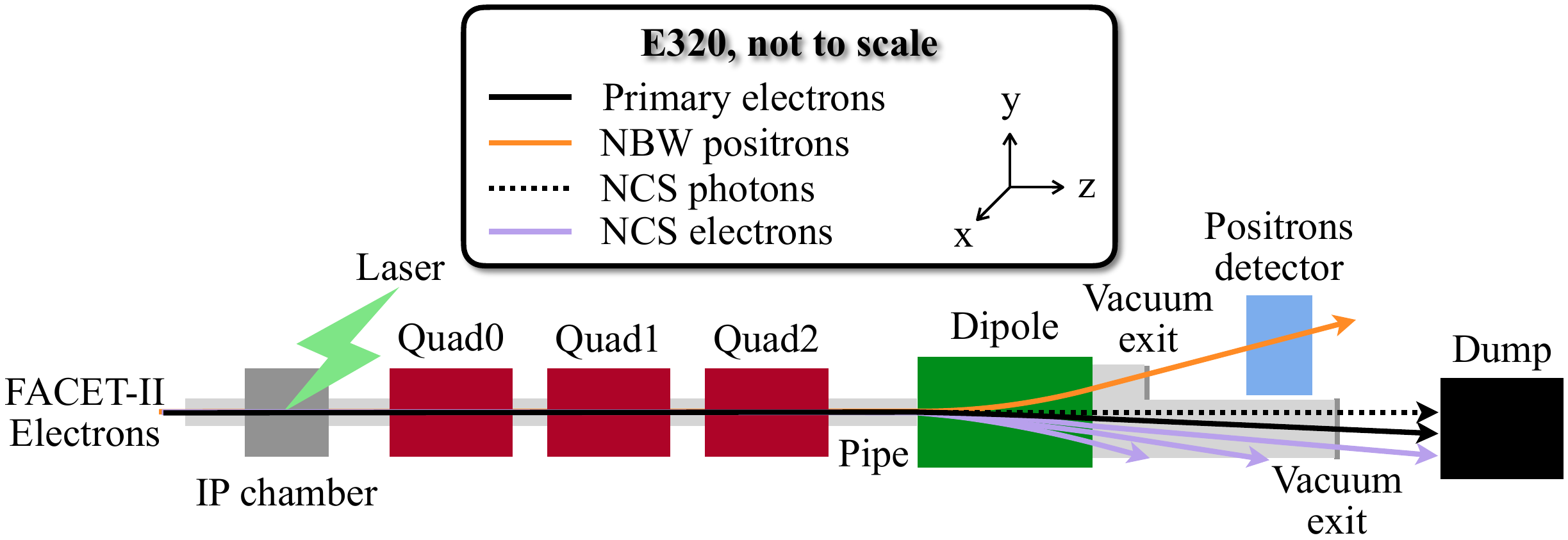}\end{overpic}
\caption{A schematic illustration of the E320 experimental setup from the IP chamber to the dump, focusing on the key elements related to the NBW signal positron detection (the NCS electrons and NCS photons are also detected by E320).}
\label{fig:sketch}
\end{figure*}

Combined with the 10~GeV (spread of $\sim$1\%, required for compression), 1\textsf{--}2~nC, FACET\textsf{--}II~\cite{Yakimenko:IPAC2016-TUOBB02,PhysRevAccelBeams.27.051302} electron beam with a collision angle of $\theta\approx30^\circ$ (the laser is linearly polarized in the collision plane), the corresponding $\chi_\gamma$ parameter is expected to reach values of $\mathcal{O}(0.1\textsf{--}1)$.
Simulations predict $\sim 0.01\textsf{--}0.1$ NBW positrons per collision for $a_0=4\textsf{--}5$ or $\sim 300$ positrons per collision for $a_0=10$ (achievable at FACET-II with a moderately upgraded laser).
A conceptual sketch of the last part of the beamline, showing only the main components of the experiment, is given in Fig.~\ref{fig:sketch}.
A summary of the experiment's key parameters is given in Table~\ref{tab:exp_parameters}.

\begin{table}[!ht]
{\footnotesize
\begin{tabular}{llc}
\toprule
\midrule
\textbf{Parameter}   & \textbf{Value} & \textbf{Units}\\
\toprule
\multicolumn{3}{c}{\textbf{Electron beam}} \\
\midrule
Final beam energy & $\phantom{\sim\,}$10\,(8-13) & [GeV]\\
Final rms energy spread & $\sim\,1$ & [\%]\\
Bunch charge & $\phantom{\sim\,}$1-2 & [nC]\\
Pulse repetition rate & $\phantom{\sim\,}$10\,(1-30) & [Hz]\\
Bunch length & $\lesssim\,20$  & [$\mu$m]\\
Bunch transverse size & $\gtrsim\,30$  & [$\mu$m]\\
$\beta^*$ & $\phantom{\sim\,}$0.5\,(0.1-1) & [m]\\
Emittance (normalized) & $\gtrsim\,35$ & [$\mu$m-rad]\\

\midrule
\multicolumn{3}{c}{\textbf{Laser}}         \\
\midrule
Power & $\sim\,10$ (upgrade: 100) & [TW]\\
Pulse repetition rate  & $\phantom{\sim\,}$10 & [Hz]\\
Wavelength & $\phantom{\sim\,}$0.8 & [$\mu$m] \\
Pulse spot size (FWHM) & 2-3  & [$\mu$m] \\
Pulse duration & $\lesssim\,60$ & [fs]\\
Beam diameter & $\phantom{\sim\,}$40 & [mm]\\
\midrule
Energy sent to tunnel  & $\lesssim\,0.65$ & [J]\\
Probe splitter  & $\phantom{\sim\,}$80 & [\%] \\
Compressor efficiency & $\phantom{\sim\,}$70 & [\%]\\
Transport efficiency  & $\phantom{\sim\,}$90 & [\%]\\
Energy in PB  & $\lesssim\,0.16$ & [J]\\
\midrule
OAP $f_{\#}$ & $\lesssim\,2$ \\
Spot size (FWHM)  & $\lesssim\,2.5$ & [$\mu$m] \\
OAP Strehl & $\gtrsim\,50$ & [\%]\\
Peak intensity  & $\gtrsim\,3.5\times 10^{19}$ & [W/cm$^2$] \\
$a_0$ & $\gtrsim\,4$ (upgrade: $\lesssim 10$) \\
\midrule
\multicolumn{3}{c}{\textbf{Combined}}         \\
\midrule
Effective collision rate & $\phantom{\sim\,}$5 & [Hz]\\
Collision angle  & $\lesssim\,30$ & [degrees]\\
$\chi_e$ & $\gtrsim\,0.5$ \\
$\chi_\gamma$ & $\lesssim\,0.5$ \\
\midrule
\bottomrule
\end{tabular}
}
\caption{Summary of parameters achieved in previous E320 runs; values in parentheses denote the range potentially accessible by the machine. Here,$\beta^*$ denotes the value of the beta-function at the electron-beam waist, the electron bunch length and transverse size are specified at the waist/IP as rms, and the normalized emittance $\epsilon_n = \gamma\epsilon_{x,y}$ is calculated for $\epsilon = 10$~GeV ($\gamma = \epsilon/m_e$). We have estimated the laser energy in the ``picnic basket'' (PB), the vacuum chamber where the E320 IP is located, using the quoted efficiencies and taking a splitting ratio of $80:20$ between main laser and probe arm into account. For the final focusing of the laser, a low surface-roughness ($\lambda/10$) off-axis parabola (OAP) from Space Optics Research Labs (SORL) with a dielectric coating from Alpine Research Optics (ARO) is being used. Exact characteristics will be published elsewhere.}
\label{tab:exp_parameters}
\end{table}

\begin{figure}[pos=!ht]
\centering
\begin{overpic}[width=0.48\textwidth]{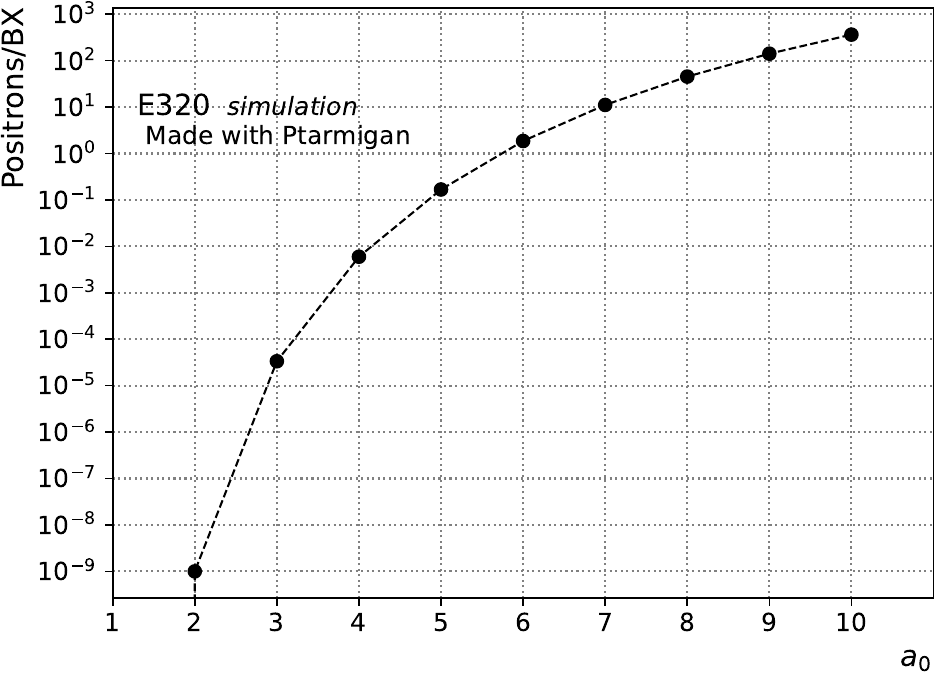}\end{overpic}
\caption{The expected NBW positron production rate versus the laser intensity, $a_0$.
The relevant electron beam parameters used are: Gaussian profile with an average electron energy of $\mathcal{E}=10$~GeV, a transverse beam radius of $r=40~\mu{\rm m}$ (rms) and a longitudinal length of $20~\mu{\rm m}$ (rms).
The laser pulse parameters are: duration of 50~fs (intensity FWHM) and a focal spot size of $2~\mu{\rm m}$ (FWHM).}
\label{fig:rate}
\end{figure}

At a rate of 10~Hz, the expected number of positrons produced within one hour of a stable-conditions run is $\sim 360\textsf{--}3600$ for $a_0=4\textsf{--}5$, respectively.
The expected positron rate per collision (also denoted as bunch-crossing, BX) versus $a_0$, obtained with \PTARMIGAN~\cite{Blackburn:2023mlo}, using the locally monochromatic approximation (LMA)~\cite{LMA2,LMA3,LMA4,LMA5,PhysRevD.106.013010}, is shown in Fig.~\ref{fig:rate}.
As the few NBW electrons cannot be observed due to the presence of the overwhelming number of NCS electrons ($\lesssim 1\%$ of the primary beam, e.g., at $a_0=10$ and for the electron beam parameters quoted in the caption), we are focusing only on the positrons.

The expected observable positron rate is, in fact, lower than the values listed in Fig.~\ref{fig:rate} due to the geometrical acceptance of the different elements along the beamline seen in Fig.~\ref{fig:sketch}.
This acceptance can be controlled by the four magnets (three focusing quadruples and one dispersing dipole) between the interaction point (IP) and the detector.
Given the IP kinematics and the geometrical constraints (e.g., the vacuum exit window size, the location and size of the beampipe, etc.), the achievable acceptance is found to be around 50\%, mostly depending on the dipole settings.
Hence, assuming an ideal detector, one can expect to observe at least $\sim 100\textsf{--}1000$ positrons (for $a_0=4\textsf{--}5$, respectively) within one hour.
Note, however, that in a typical run about 50\% of the shots are used for background-only studies such that this estimate is roughly smaller by a factor of two.

E320 will attempt to measure the number of NBW positrons per BX as a function of $a_0$ and thus $\chi_\gamma$.
The laser intensity (and thus $a_0$) can be varied in different ways.
One naive way to vary it is by simply attenuating the total laser energy. 
However this is not the most effective approach, as it reduces the number of laser photons available to induce pair production.
Instead, it is advantageous to maintain a high total laser energy, while reducing the light intensity by, for example, changing the temporal duration of the laser pulse.
The shot-to-shot diagnostics of the laser parameters as well as the beam vector and the relative electron-laser timing via electro-optical sampling (EOS) are crucial for determining the expected positron rate~\cite{Storey:2023wlu,HUNTSTONE2021165210,PhysRevLett.94.114801}. 

The NBW positrons are accompanied by a much larger number of scattered (low-energy) electrons and photons, produced via NCS with a much higher cross-section at the same IP.
Except for low-energy photons, all particles exiting the IP propagate almost collinear with the primary electron beam along the spectrometer beamline.
Although the spectrometer quadrupoles are chromatic, the primary energy-dependent effect arises from the dipole, which deflects positrons upwards and electrons downwards.

The currently available beam parameters at FACET-II imply that the electron beam is significantly larger than the high-intensity laser focus (see Table~\ref{tab:exp_parameters}).
Therefore, most of the primary electrons ($\gtrsim 99\%$) propagate unimpeded from the IP to the electron beam dump through a 5~mm Aluminum vacuum exit window, located $\sim$4~m upstream the dump.
Contrarily, the lower energy NCS electrons ($0.1\textsf{--}1\%$ of the primary beam, depending on the dipole setting and the laser intensity) are not transmitted all the way to the beam dump due to the stronger deflection imposed by the dipole compared with the primary 10~GeV electrons.
A significant fraction of these NCS electrons hits the beamline elements (the wall of the vacuum chamber, the beampipe, etc.) in close proximity to the positrons detector.

These factors lead to a substantial amount of background due to secondary particles (electrons, positrons, photons, neutrons, pions, etc.) that are produced in the beamline elements and pass through the detector.
The neutrons originate mostly from the dump, with their energy spectrum peaking around 1~keV and truncating approximately at 100~MeV. 
The photons are mostly produced in the dump, but a substantial fraction of them is produced throughout the beamline elements (from the IP to the dump including the detector). 
Their spectrum is peaking below 10~MeV, falls rapidly and vanishes at $\sim 4$~GeV.
The electrons and positrons are similarly produced all along the beamline (including in the detector material) and their energy spectra follow that of the photons, albeit a bit softer.
Other charged background particles come at negligible rates.
Thus, the detector must be capable of resolving a very small number of positrons against an overwhelming background.
The detector is located in air, immediately after a  500~$\mu{\rm m}$ stainless-steel window, through which the NBW positrons exit the vacuum.

Detecting single positrons at a rate of $\mathcal{O}(10^{-3})$ per cm$^2$ per BX in the presence of a large flux of background with $\mathcal{O}(10^2)$ charged particles and $\mathcal{O}(10^4)$ neutral particles per cm$^2$ per BX at the detector surface, is a highly challenging task.
For reference, in E144 the residual background was estimated to be $2 \times 10^{-3}$ positrons per BX~\cite{Bamber:1999zt}.

Therefore, the technology for detecting single positrons at E320 should be selected based on the following criteria:
(i) a sensitive surface area of roughly $1\times 10~{\rm cm}^2$, (ii) insensitivity to neutral particles (photons and neutrons), (iii) ability to measure single-particle energy at $\sim 1\textsf{--}10$~GeV with $\mathcal{O}(1\%)$ resolution, and production vertex with $\mathcal{O}(10\%)$ uncertainties.
Note that for (i), the size is determined by the designated position of the detector relative to the dipole and the expected NBW positron spectrum.
In addition, the technology must be radiation tolerant to withstand the (cumulative) large background fluxes.
Moreover, other FACET-II experiments that use the same IP chamber and beamline generate even larger doses~\cite{PhysRevAccelBeams.27.051302}.
The detector should withstand these as well, although it is not expected to operate during these runs.
The dipole is operating in all runs, so the primary beam can never reach the region close to the detector.
Therefore,  most of the generated flux consists of secondary particles.

Complying with all of the above requirements, the solution proposed here is a highly-granular, multi-layer tracking detector.
The chosen technology is based on a specific thin, state-of-the-art monolithic active pixel sensors, dubbed ``ALPIDEs''~\footnote{ALPIDE stands for ``ALice PIxel DEtector''}~\cite{AGLIERIRINELLA2017583,MAGER2016434,ALPIDEManual,YANG201561}.
The detector design is described below in Sec.~\ref{sec:detector}, and its expected performance is discussed in Sec.~\ref{sec:reconstruction}.

\section{Detector}
\label{sec:detector}
In the following we discuss the design of the E320 tracking detector. The discussion begins with the ALPIDE sensors description and then proceeds to the detector modules, and how they are assembled into the tracking detector. 

\subsection{The ALPIDE sensor}
\label{sec:alpide}
The ALPIDE chips are produced by TowerJazz~\cite{TowerJazz} (based on its 180~nm CMOS imaging process) for the upgrade of the ALICE experiment~\cite{Collaboration_2008,Abelevetal:2014dna}
at the LHC.
The pixels, $27\times 29~\mu{\rm m}^2$, 512 rows and 1024 columns in one $\sim 3\times1.5~{\rm cm}^2$ chip, integrate the sensing volume with the readout circuitry in one all-silicon chip.

The ALPIDEs have been extensively tested at laboratories as well as at a number of test beam facilities in the past few years.
The chips have demonstrated very good performance both before and after irradiation. 
For example, in~\cite{SENYUKOV2013115} (more tests are found e.g.  in~\cite{AGLIERIRINELLA2021164859,DANNHEIM2019187}), the chip was irradiated with the combined dose of 1~MRad and $10^{13}~1\,{\rm MeV}~n_{\rm eq}/{\rm cm}^2$ ($10^{13}~{\rm cm}^{-2}~ 1\,{\rm MeV}$ neutron equivalent) and it was observed to yield a signal to noise ratio ranging between 11 and 23, resulting in particle detection efficiencies above 99.5\% and 98\% before and after irradiation respectively~\footnote{These levels of irradiation are relevant for the ALICE experiment and not for E320, where lower overall radiation ($\mathcal{O}(10^{-2})$ Rad) is expected.}.
In~\cite{MAGER2016434}, the sensor has shown a detection efficiency above 99\%, a fake hit rate much better than $10^{-5}$, a spatial resolution of around $\sim 5~\mu{\rm m}$ and a peaking time of around $2~\mu{\rm s}$.
The new inner tracking system (ITS) of ALICE, which was installed in 2021, is a $\sim 10~{\rm m}^2$ pixel detector - the largest and most complex pixel detector ever built~\cite{ALICE:2023udb,Ravasenga:2023yqd}.

\subsection{Detector modules}
\label{sec:modules}
\begin{figure*}[pos=!ht]
\centering
\begin{overpic}[width=0.45\textwidth]{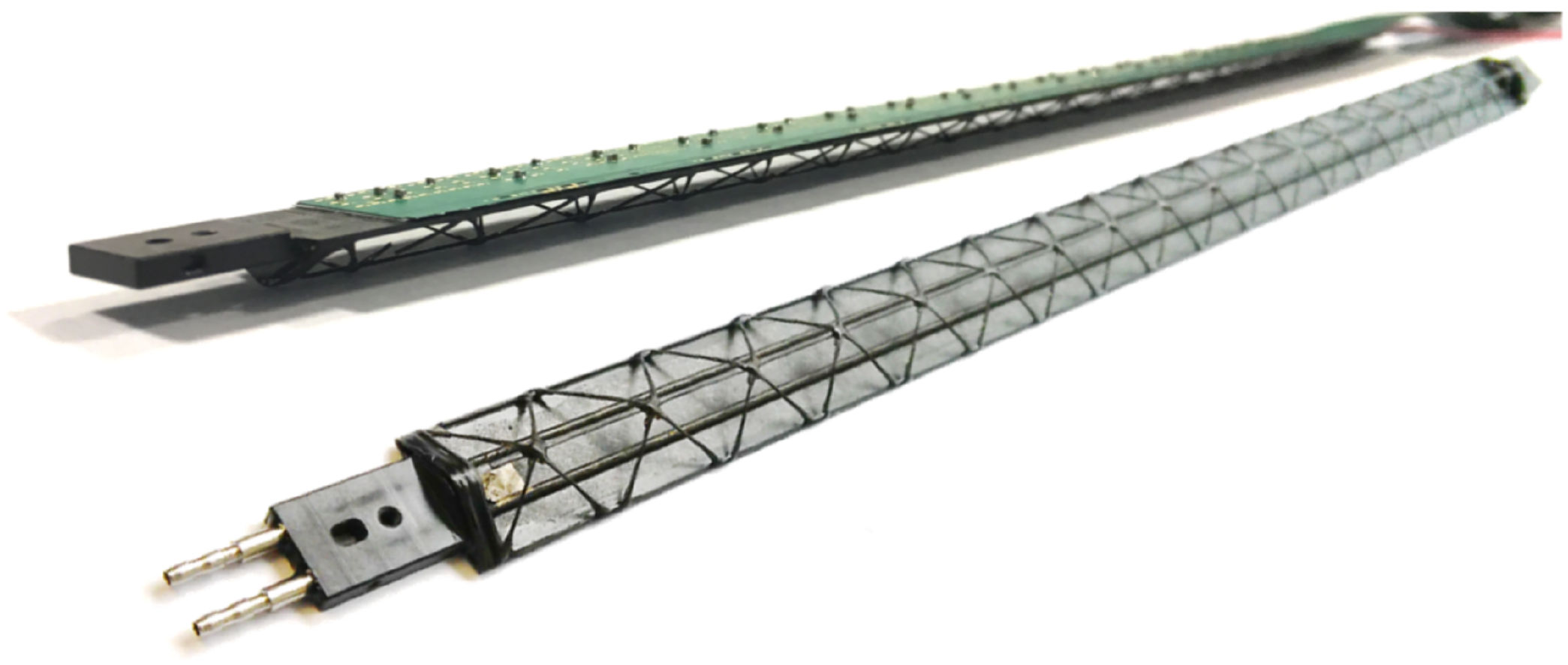}\put(20,65){}\end{overpic}
\begin{overpic}[width=0.45\textwidth]{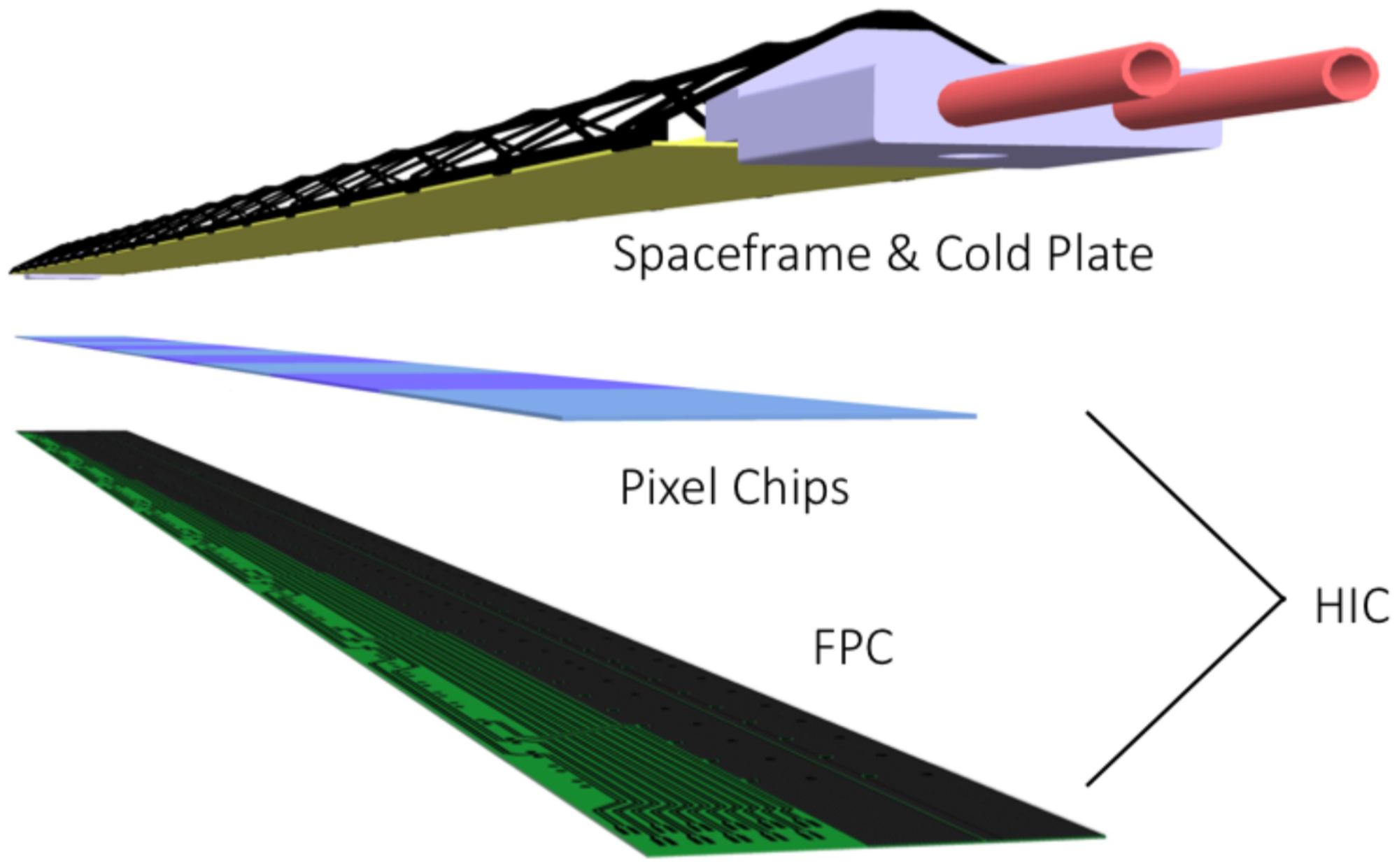}\put(20,65){\textrm{}}\end{overpic}
\caption{{\footnotesize Left: an ALPIDE ``stave''~\cite{Abelevetal:2014dna,AGLIERIRINELLA2017583} including nine ($3 \times 1.5~{\rm cm}^2$) ALPIDE sensors flip-chip mounted on a flexible printed circuit that is glued to a carbon fibre cooling sheet and a mechanical support frame.  Right: a schematic layout of a single stave.}}
\label{fig:stave}
\end{figure*}

The E320 tracker is built using detector assemblies made of the ALPIDE sensors, purchased from CERN.
The basic detector element of the ITS is called a ``stave''.
Each stave is built from nine ALPIDE sensors.
A stave is built from the following elements:
(i) Space Frame: a carbon fibre structure providing the mechanical support and the necessary stiffness;
(ii) Cold Plate: a sheet of high thermal-conductivity carbon fibre with embedded polyimide cooling pipes, which is integrated into the space frame. The cold plate is in thermal contact with the pixel chips to remove the generated heat;
(iii) Hybrid Integrated Circuit (HIC): an assembly consisting of a polyimide flexible printed circuit (FPC) onto which the pixel chips and some passive components are bonded.
The stave picture and structure are shown in Fig.~\ref{fig:stave}.
Each stave is instrumented with one HIC, which consists of nine pixel chips in a row connected to the FPC, covering an effective active area of $\sim 13.8\times 270.8~{\rm mm}^2$ including $100~\mu{\rm m}$ gaps between adjacent chips along the stave.
The sensor’s physical size is $\sim 15\times 30~{\rm mm}^2$, but its sensitive area is smaller due to the chip periphery.
An extension of the FPC, not shown in Fig.~\ref{fig:stave}, connects the stave to the electrical services (power, bias, control and readout) entering the detector only from one side.

The overall structure mean thickness in ALICE’s ITS2 is $X/X_0 = 0.357$\% (per tracking layer), with the material budget broken down to fractions of the total width as: $\sim 15$\% silicon sensor (with absolute thickness of $50~\mu{\rm m}$), $\sim 50$\% electrical substrate (FPC) including the passive components and the glue, $\sim 20$\% cooling circuit and $\sim 15$\% carbon spaceframe~\cite{ALICE-PUBLIC-2018-013}.
This already very small average thickness includes partial overlap between two adjacent staves of the ITS2 layer. Therefore, in the E320 configuration the effective thickness is even smaller, and for $\sim 1\textsf{--}10$~GeV positrons the momentum and position resolution degradation associated with multiple scattering is not significant.

\subsection{Detector assembly}
\label{sec:tracker}
\begin{figure}[pos=!ht]
\centering
\begin{overpic}[width=0.48\textwidth]{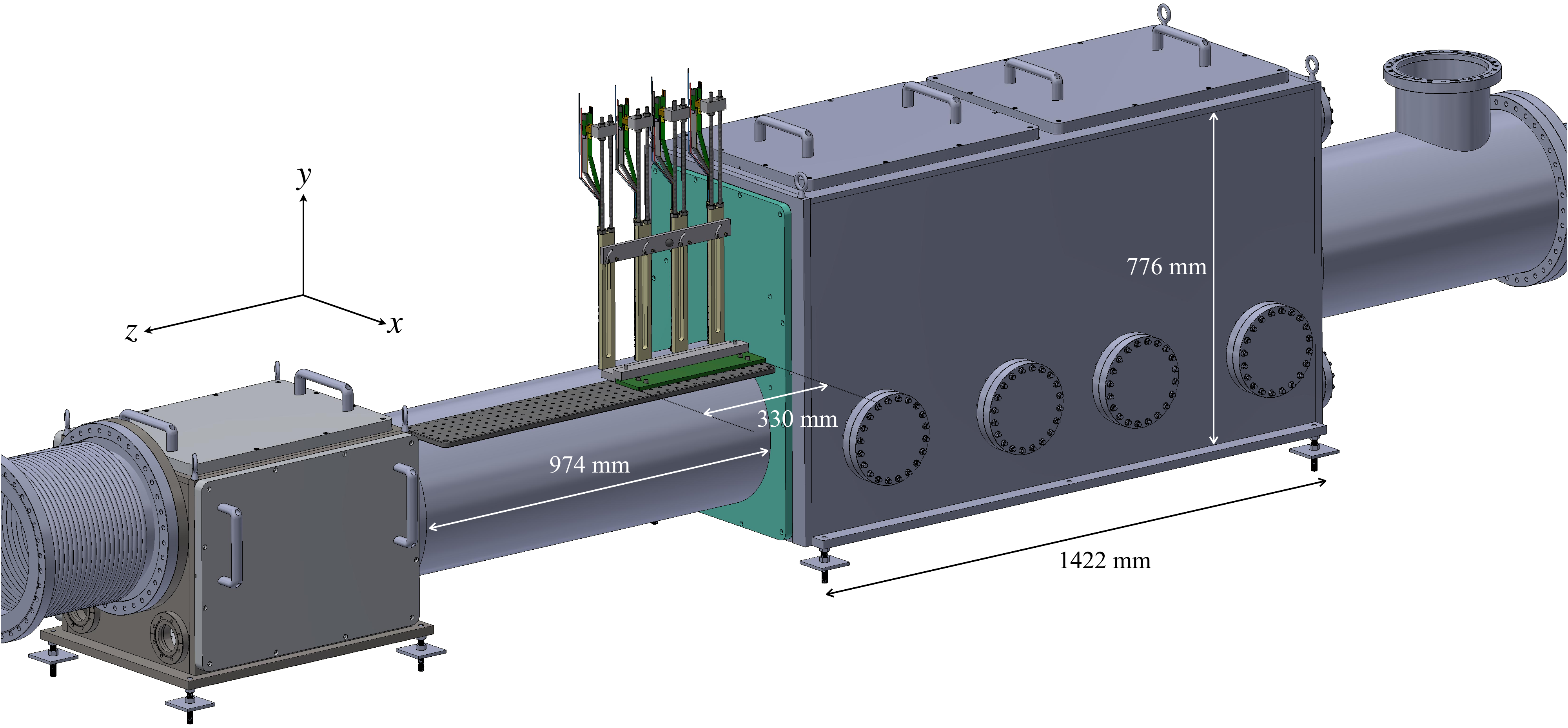}\end{overpic}
\begin{overpic}[width=0.48\textwidth]{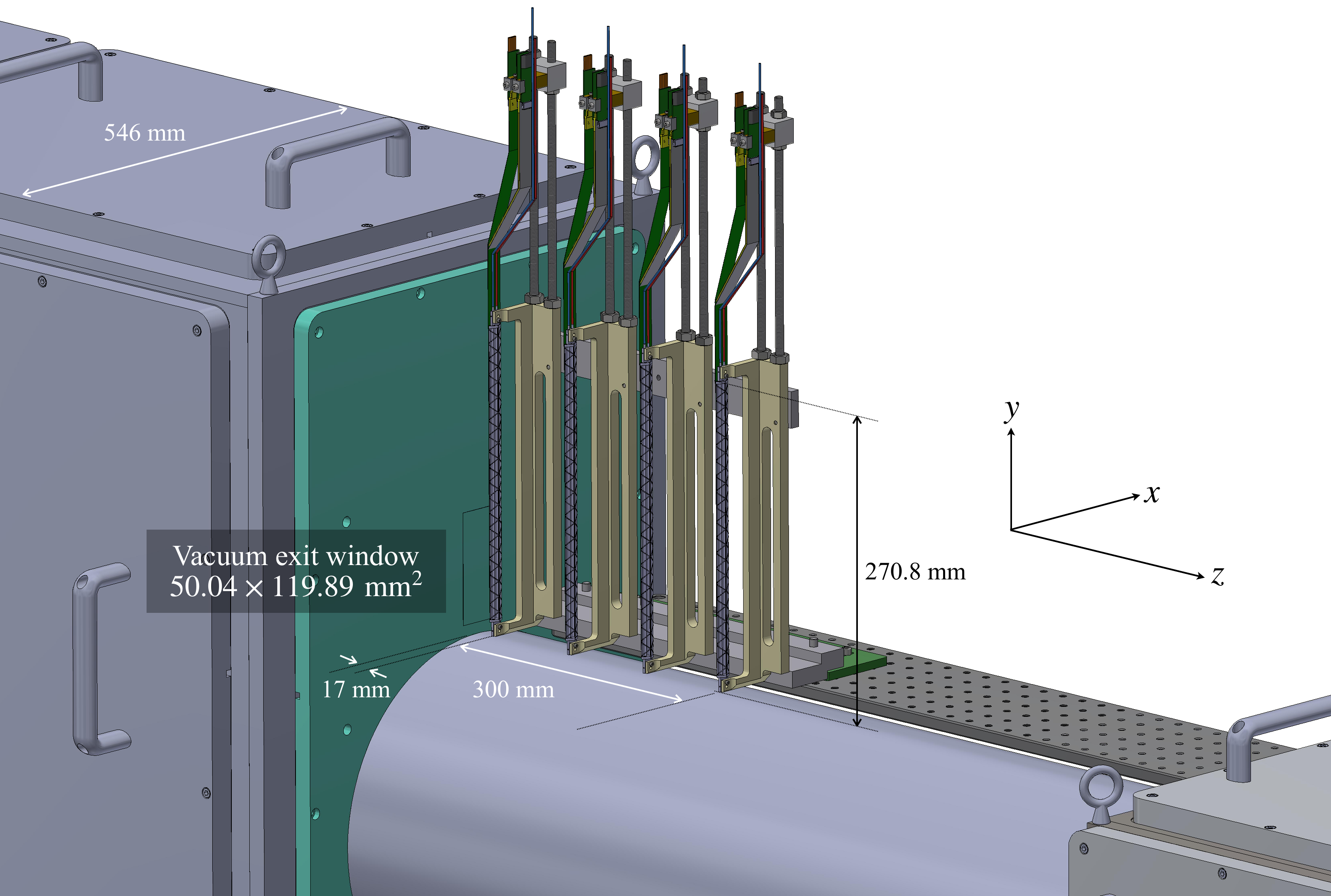}\end{overpic}
\caption{The default detector arrangement, where the staves are placed vertically in the closest approach to the vacuum exit window. Top: a full view around the vacuum exit window as seen from the accelerator's back wall. Bottom: a close-up view of the detector from the opposite side (the accelerator's aisle).}
\label{fig:detector_default}
\end{figure}

The tracker is assembled as an arm of four layers made of the ALPIDE staves, staggered above the beamline, next to the positron exit window.
The distance between two layers is $\sim 10$~cm along the beam direction.
The detector assembly is shown in Fig.~\ref{fig:detector_default} for the default setup where the staves are placed vertically and closest to the vacuum exit window.

\begin{figure}[pos=!ht]
\centering
\begin{overpic}[width=0.48\textwidth]{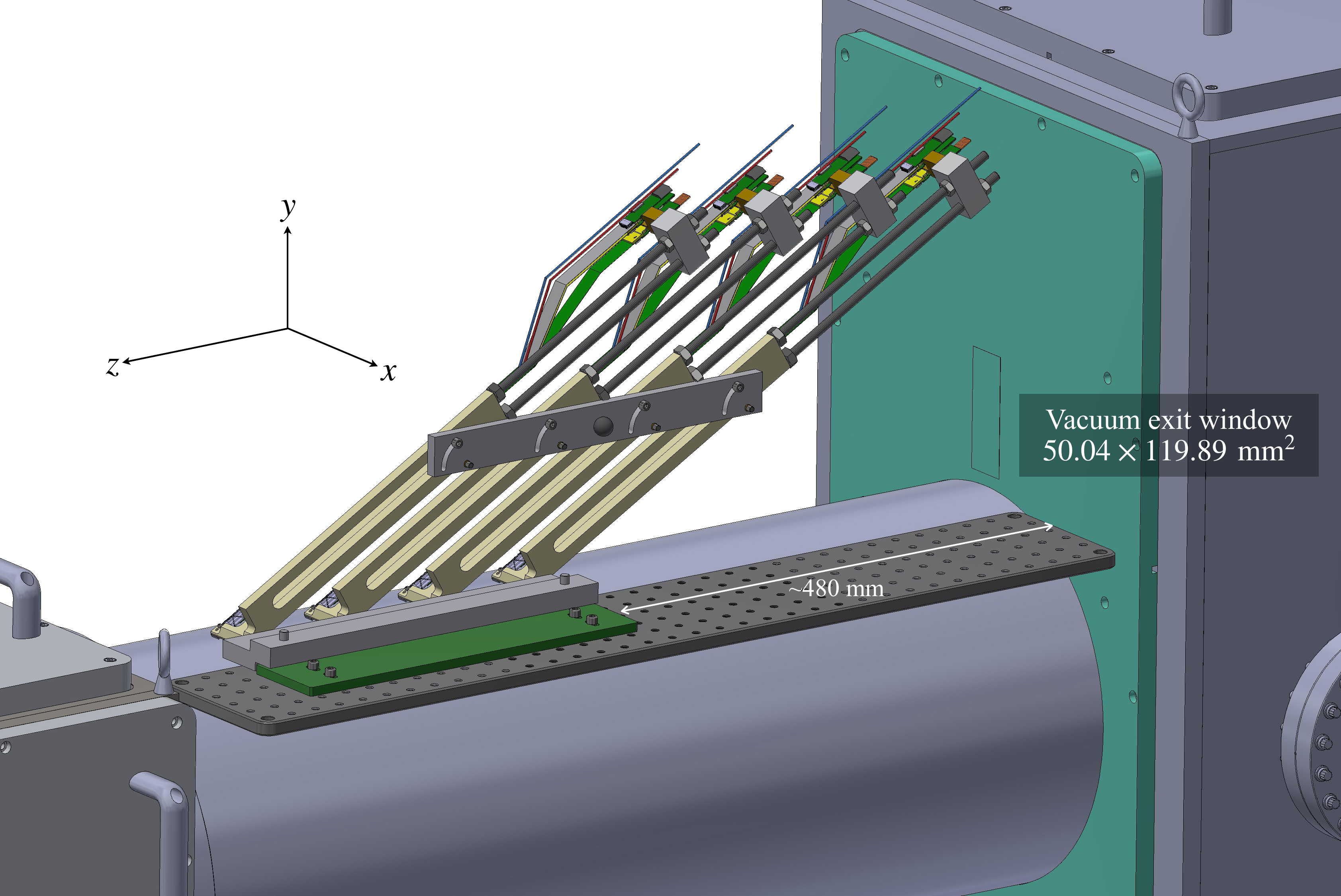}\end{overpic}
\caption{The alternative detector arrangement, where the staves can be tilted and the detector is pushed further downstream the vacuum exit window, to allow larger acceptance. The tilt angle on the picture is chosen arbitrarily for illustration purposes.}
\label{fig:detector_tilted}
\end{figure}

In Fig.~\ref{fig:detector_tilted} the same detector is shown for an alternative setup, where the staves are tilted by $30^\circ$ and the entire assembly is pushed downstream, i.e., away from the exit window.
This latter setup provides a larger geometrical acceptance at the expense of longer travel range in air, as well as slightly larger material budget (in the detector modules) and incident angle seen by the positrons.
The tilt angle and orientation are chosen such that all low-energy positrons can be captured, while traveling a shorter distance in the air (less scattering) compared with the high-energy ones.
The larger acceptance is obtained at the high-$E$ end of the spectrum (i.e., closer to the beampipe) since the staves’ passive ``ears'' (see Fig.~\ref{fig:stave}) interfere less with those particles compared to the case where the staves are positioned vertically.
The design is flexible enough to allow easy switching between these two extreme positions (and effectively any position in between), while always keeping the alignment between the staves intact.
With a total detector volume of $\sim 1.38\times 27.08 \times 30~{\rm cm}^3$, good coverage of the expected signal is achieved: $\sim 80\%$ of positrons that reach the  vacuum exit window are fully contained within the detector volume.

While we leave the detailed alignment discussion for future work, we comment that the alignment between the detector elements can be determined either with cosmic muons (during accelerator shutdowns) or with Bremsstrahlung-produced positrons during special accelerator runs, by placing a thin retractable Aluminum foil (that is present in the IP chamber) in front of the accelerator beam. 
The latter procedure can be also used to determine the overall alignment of the detector with respect to the beamline elements (mainly the IP, quadrupoles and dipole).

\subsection{Power, Readout \& Services}
\label{sec:services}
\begin{figure*}[pos=!ht]
\centering 
\begin{overpic}[width=0.9\textwidth]{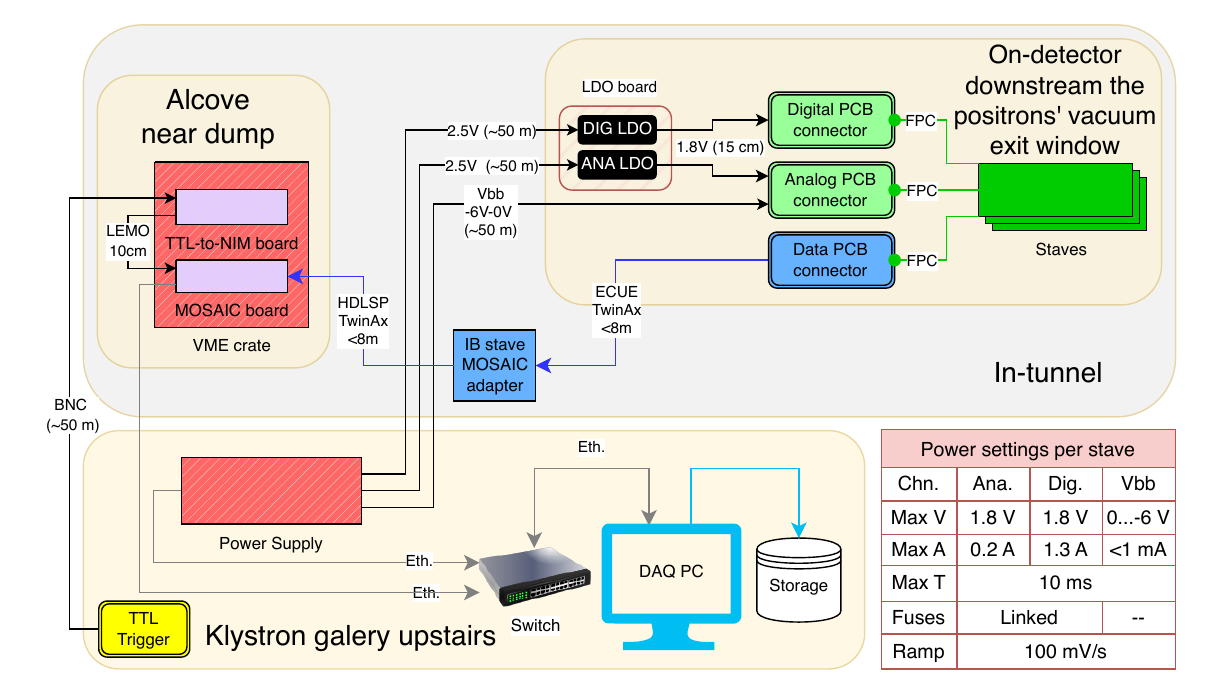}\end{overpic}
\caption{The schematic power supply and readout system for one stave.
The key elements are grouped by location, either in the tunnel or in the Klystron gallery above. 
The different connection types and lengths are specified.}
\label{fig:ps_and_ro}
\end{figure*}

The detector services, namely the power supply, the readout and possibly  water cooling come out from the top of the layers farthest from the beam axis as seen in Fig.~\ref{fig:detector_default} and~\ref{fig:detector_tilted}.
A sketch of the power supply and readout system is shown in Fig.~\ref{fig:ps_and_ro}.
The different elements are illustrated, including their locations and lengths.

Each stave requires three power channels: digital, analog, and bias.
An off-the-shelf power supply with remote sensing support can be used and positioned approximately 50~m from the detector in the Klystron gallery above the tunnel since only a moderate low-voltage ($<6$~V) is needed and the voltage drop due to the current consumption (mostly in the digital channels) is manageable using remote sensing or optimized cable-routing (larger wire gauges and fine-tuned lengths).
A custom low-dropout regulator (LDO) board is providing the 1.8~V power input to the stave and also filters most of the noise picked up on the long line leading to the board.
The LDO chip mounted on the board is found to be radiation-hard for this application~\cite{Abovyan:2022qaz}.
A TTL-formatted trigger signal is obtained from the FACET-II data acquisition (DAQ) system.
It is converted to a NIM-formatted signal and connected to the MOSAIC~\cite{refId0} readout board.
The MOSAIC board and the TTL-to-NIM conversion board are mounted in the same standard VME crate.
Each stave has one readout line that provides the control and data via the MOSAIC board, which is not radiation-hard.
Its location is dictated by the maximal length of the data line (TwinAx cables) at which the staves can still be configured and no readout errors occur during intensive stress testing.
The maximum possible distance is found to be $16$~m.
This means that the VME crate holding the MOSAIC boards must be positioned inside the tunnel and hence heavily shielded.
To minimize radiation exposure, it is placed outside the line-of-sight of the accelerator in a small alcove as far away from the beamline as possible, a few meters away from the detector.
The MOSAIC board is connected with an Ethernet line to a switch that is connected to the DAQ computer; both are located outside the tunnel in the Klystron gallery.
The configuration, control and DAQ of the stave is managed by the EUDAQ~\cite{Liu_2019,EUDAQ2020} framework running on the DAQ computer.
A slightly scaled-down version of this power and readout system is already fully operational at FACET-II with a smaller acceptance prototype detector based on the same ALPIDE technology.

\section{Simulation}
\label{sec:simulation}
This study includes two types of simulation pipelines, full (FullSim) for signal and background and fast (FastSim) for background only.
The FullSim procedures and datasets are discussed in Sec.~\ref{sec:fullsim}.
The FastSim is discussed generically in Sec.~\ref{sec:fastsim} and in more detail in Sec.~\ref{sec:background_fastsim}.
The samples produced via these two pipelines can be blended in different ways to study the physics reconstruction  algorithms' (particle tracking) performance.
For example, one can analyze the signal alone, the background components alone or the combination of all components with different fractions of injected signal.
This is discussed in Sec.~\ref{sec:background_fastsim}.

\subsection{Full simulation}
\label{sec:fullsim}
\begin{figure*}[pos=!ht]
\centering
\begin{overpic}[width=0.9\textwidth]{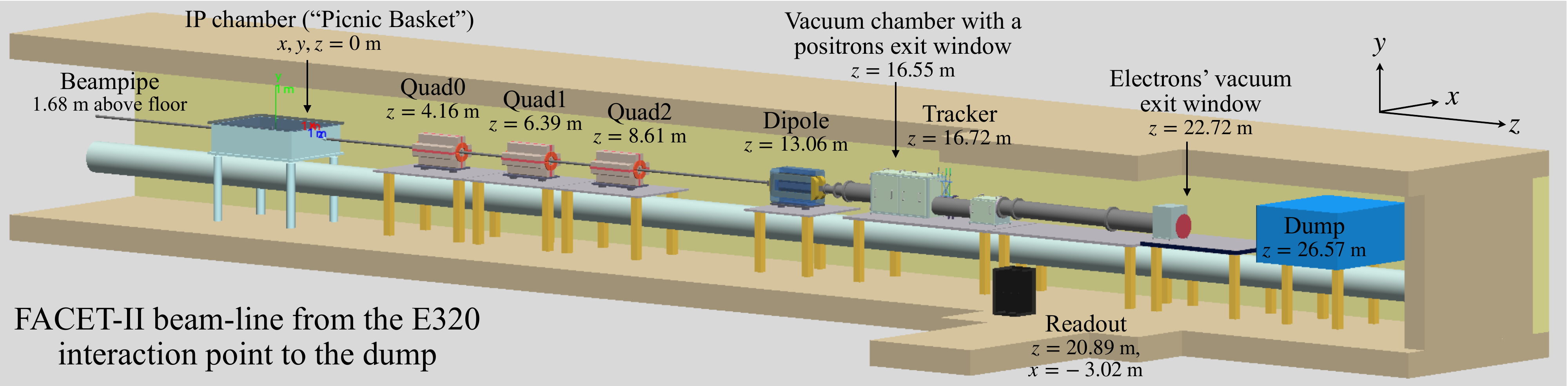}\end{overpic}
\caption{The key components of the E320 experimental setup at the FACET\textsf{--}II tunnel, as implemented in \GEANT.}
\label{fig:exp}
\end{figure*}
The production of NBW signal positrons is simulated with \PTARMIGAN~\cite{Blackburn:2023mlo}.
Besides the NBW positrons, the output of the \PTARMIGAN collision also includes a large collection of photons and electrons resulting from the NCS, as well as the NBW electrons.
For the current E320 parameters, the characteristic number of NCS products can reach $\mathcal{O}(1\%)$ of the primary beam population (larger fractions are possible if the incoming beam is more compressed).
There are typically more NCS photons than NCS electrons since the primary beam electrons which interact with the laser may each radiate a few NCS photons ($\sim 2$ on average for the current E320 parameters), while they pass through the laser pulse.

The signal simulation in this work is done for a choice of $a_0=10$, an intensity that may only be achievable after a prospective upgrade of the laser system, due to two reasons: (i) the NCS photons and electrons yield is higher than at lower $a_0$ values, leading to a higher and hence more conservative background level, and (ii) the signal yield is higher than for lower $a_0$ values as can be seen in Fig.~\ref{fig:rate}, leading to higher positron yields per BX and hence better statistics for the tracking performance studies.
We later scale these results to the parameters that are realistically expected for the upcoming E320 runs.
We stress that the kinematic distributions do not differ in a significant way between the two scenarios, so the positrons used to quantify the signal properties at $a_0=10$ are effectively applicable also for $a_0=5$ (see Appendix~\ref{app:a0510signal}).

Fig.~\ref{fig:Epos} shows the energy, $a_0$, and $\chi_\gamma$ distributions at the interaction point as generated by \PTARMIGAN for an input $a_0=10$.
The distributions shown include particles from twelve events (BXs) simulated, normalized to one BX.
The respective yields per BX are $\sim 8.8\times 10^7$ NCS photons, $\sim 4.7\times 10^7$ NCS electrons, and $\sim 360$ NBW positrons.
One Compton-edge~\cite{Blackburn:2021cuq,Bamber:1999zt} can be seen in the energy distribution around 8.3~GeV (electrons) or 1.7~GeV (photons).
The presence of this edge is induced by those beam electrons that experience only an attenuated laser ($a_0 \lesssim 1$).
Because the LMA used in this \PTARMIGAN simulation averages over the fast oscillation of the laser, one has access only to the local value of the envelope: $\langle a^2 \rangle = a_0^2/2 \cdot g(t,x,y,z)$ for linear polarization (LP), whereas  $\langle a^2 \rangle = a_0^2\cdot g(t,x,y,z)$ for circular polarization (CP).
Hence, the maximum rms value of $a_0$ in the LP case is at $a_0/\sqrt{2} \simeq 7.07$ for $a_0=10$.
The energy distribution of the photons producing NBW $e^+e^-$ pairs ranges in $\sim 2.5\textsf{--}8.5$~GeV and hence the quantum parameter $\chi_\gamma$ distribution ranges roughly between $0.1\textsf{--}0.65$.
The energy spectrum of the positrons (and their parent photons) for the lower intensity benchmark at $a_0=5$ is expected to be only slightly harder (see Appendix~\ref{app:a0510signal}), the $\chi_\gamma$ distribution will shift to values lower by a factor of two, roughly.
\begin{figure*}[pos=!ht]
\centering
\begin{overpic}[width=0.8\textwidth]{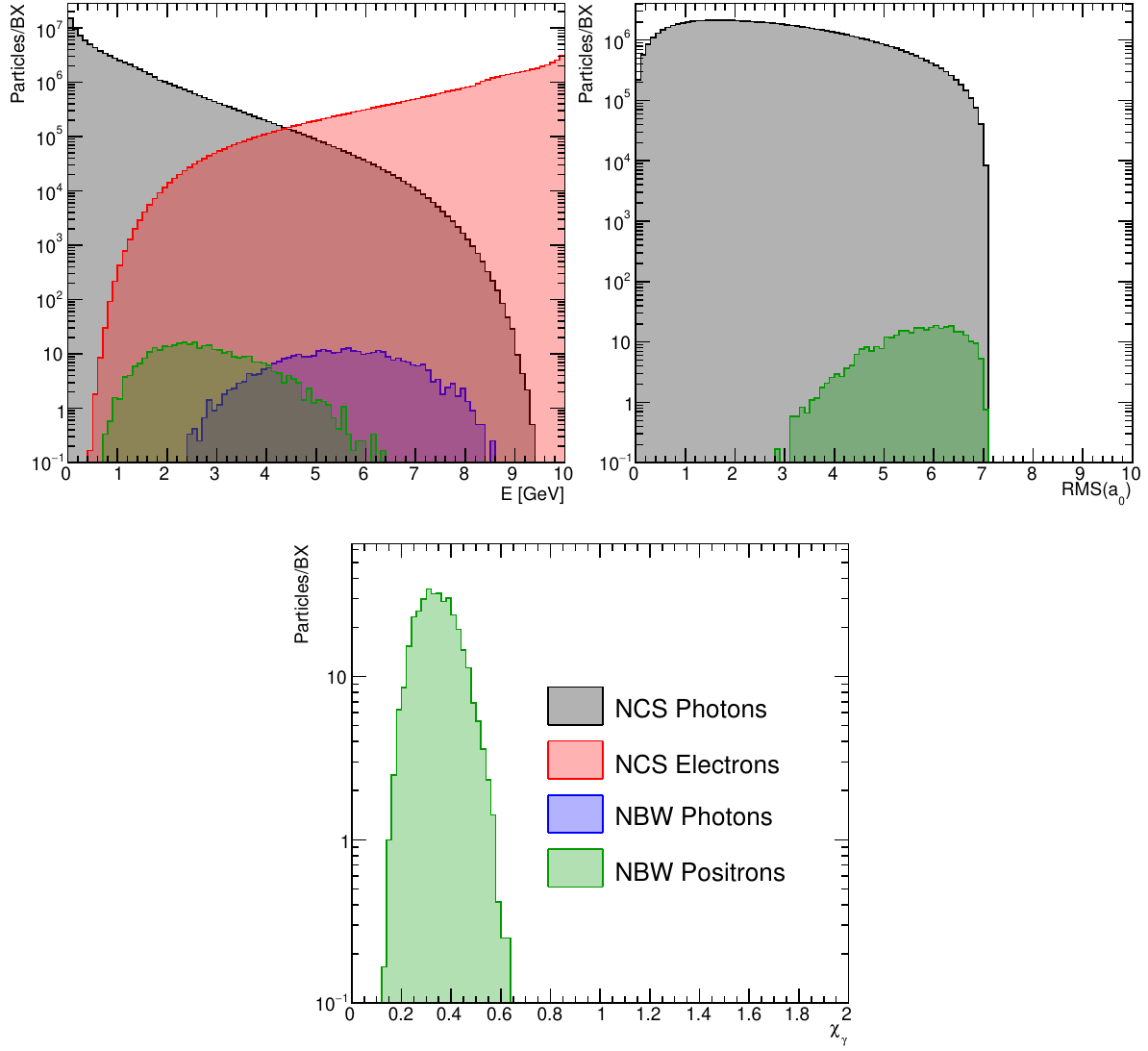}\end{overpic}
\caption{A \PTARMIGAN simulation with the E320 electron-laser collision parameters, assuming an input intensity parameter of $a_0=10$. Top left: The energy spectra of the NCS electrons, NCS photons, NBW positrons and their parent photons at the IP.
Top right: The distribution of the in-situ laser intensity parameter $a_0$ rms value at the IP.
Bottom: The distribution of the in-situ quantum parameter $\chi_\gamma$ value of the positrons' parent photons at the IP.}
\label{fig:Epos}
\end{figure*}

The output from \PTARMIGAN is interfaced with a full \GEANT~\cite{ALLISON2016186,1610988,AGOSTINELLI2003250} simulation for the propagation of all particles in the experimental setup, which includes all key components along the FACET\textsf{--}II beamline, starting from the interaction chamber and ending at the dump, with a detailed description of the materials and magnetic fields.
The beamline, as implemented in the \GEANT model, is shown in Fig.~\ref{fig:exp}.
The magnets, which have an active length of $\sim 1$~m each, are configured in the \GEANT model as follows: the upstream and downstream quadrupoles fields are set to $\vec{B}_{\rm Quad0,2}=(4y,4x,0)$~T/m, while the middle quadrupole is set to $\vec{B}_{\rm Quad1}=(-7y,-7x,0)$~T/m.
The dipole magnet is $\vec{B}_{\rm Dipole}=(0.31,0,0)$~T.
These settings are configurable and allow to focus the center of the NBW positron distribution at any point along the beamline between the dipole exit and dump.
To make sure that the acceptance is maximal, the focusing settings are chosen such that the object plane is set to the IP, while the image plane is set to the first tracking layer.
In the simulation, all magnets are assumed to be ideal, i.e., with perfect fields (perfect uniformity in the dipole and perfect linear gradients in the quadrupoles) that vanish outside the magnet volumes.
While the optimal settings can be easily determined in simulation, in reality they are tuned by using a retractable thin Aluminum foil in the IP chamber ($\sim 30$ from the IP itself) to produce positrons via Bremsstrahlung and examining their focus pattern at the detector face.

The primary beam electrons that do not interact with the laser are discarded in \PTARMIGAN and instead simulated directly by \GEANT using the same characteristics as the primary beam.
The number of beam electrons simulated directly by \GEANT is calculated by subtracting the number of output NCS electrons reported by \PTARMIGAN from the number of initial primary electrons.
These electrons are passed through the experimental setup until they reach the accelerator's dump.
The extremely large amount of energy and charge dissipated in the dump leads to a large back-scattered flux, consisting primarily of photons and neutrons, as well as electrons and other charged particles.

In addition to the back‐scattered flux from dumping the non‐interacting primary beam electrons, the NCS products contribute a significant fraction of the observed background. 
Particularly, the NCS electrons are less energetic than the primary ones and hence they are bent down stronger by the dipole magnet.
They eventually hit the material of different components along the beamline, leading to back-scattered particles that may also reach the detector.

The \GEANT output samples for both signal and background contain the energy depositions per pixel per particle, where the total energy deposited is simply the sum of the individual depositions from all particles.
This includes all possible types of particles that may be produced anywhere in the material and reach the detector, including neutrons, pions, muons, etc.
In this way, the contributions from signal (NBW positrons), signal-induced backgrounds (due to the NCS photons and electrons, as well as secondaries created by the positrons, e.g., at the vacuum exit window) and ``pure'' backgrounds (secondaries created by the primary beam electrons) can be easily blended in the right proportion or studied separately as needed.

Finally, to complete the simulation chain, the \GEANT outputs are processed by the \ALL~\cite{Spannagel:2018usc,Spannagel:2018xhg} digitization simulation software, which given an energy deposition by a particle at a given position in the pixel, converts the deposited energy into electron-hole pairs created in the semiconductor.
These charges are then propagated to the collection nodes via drift or diffusion processes in the epitaxial layer of the pixel.
Given the amount of charge collected in the node, a binary hit/no-hit decision is then made per pixel based on a predefined charge threshold.
The \ALL simulation takes as an input also the characteristic electric field inside the pixel, which in the ALPIDE case is highly nonlinear.
This field is pre-calculated as discussed in~\cite{Santra:2022sjw}.
The \ALL simulation also takes care of charge induced in adjacent pixels, even if there is no actual incident particle that has deposited energy in those pixels.
All ``fired'' pixels (pixels with charge above threshold) are then grouped in ``clusters'' -- sets of fired pixels that have a common edge. 
In the ALPIDE case the average cluster size is $\sim 2.5$ pixels for the charge threshold of 120~$e$ that is used in this study~\cite{Santra:2022sjw,Dannheim:2020yna}. 
The clustering is done using the breadth-first search (BFS) algorithm~\cite{cormen2009introduction,moore1959shortest}.
In case of large pixel multiplicities this is significantly faster than a the  common recursive ``pac-man'' algorithm.
The \ALL simulations have been shown to match data extremely well for a variety of sensors, including the ALPIDEs~\cite{Dannheim:2020yna,Santra:2022sjw}.

The digitization is done per BX for a data sample containing the signal contribution alone and another sample containing the background contribution alone.
In principle, the digitization should be done on a mixture of signal and background. 
However, the chosen approach provides more flexibility for the subsequent analysis, while having negligible impact on the clusters' characteristics.
Subsequently, the collection of all clusters is handed over to the track-reconstruction algorithms discussed below.
To summarize, the different FullSim samples used in the subsequent analysis are:
\begin{itemize}
    \item \textit{FullSim NBW signal}: Total of 72 BXs simulated with $a_0=10$, yielding $\sim 26k$ positrons (processed positron-by-positron).
    Includes the positrons themselves and all secondary particles that may be produced by these single positrons, e.g., in the vacuum exit window or in the detector layers themselves.
    \item \textit{FullSim NCS background}: Total of 12 BXs with $a_0=10$ (processed BX-by-BX). Includes the secondary fluxes originating from all primary electrons undergoing the NCS process ($\sim 1\%$ of the beam) and the NCS photons.
    \item \textit{FullSim Dump background}: Total of $\sim 30\%$ of one BX. Includes the secondary fluxes originating from all primary beam electrons that did not interact with the laser pulse for $a_0=10$ ($\sim 99\%$ of the beam).
\end{itemize}

The sizes of the three FullSim samples are chosen such that we obtain enough statistics to (i) study the signal, and (ii) to reliably generate much larger, fast simulation background samples, as discussed below, while keeping the computational costs manageable.
The number of the signal positrons per BX is properly re-scaled for evaluation of the detector performance, as discussed below.  

\subsection{Fast simulation}
\label{sec:fastsim}
The FullSim chain discussed above is computationally highly intensive to run, even for only one full BX ($\sim 0.6\textsf{--}1.2\times 10^{10}$ electrons per bunch).
Most of the processing time is taken by the \GEANT step, primarily due to the highly detailed simulation of the electromagnetic and hadronic showers generated in the dump by the primary beam electrons and the NCS ones that reach the dump.
However, this step is unavoidable in order to get a realistic description of the background.

Furthermore, for a statistically-valid derivation of the detector performance one has to simulate much more than one BX, particularly if the expected signal rate is extremely small, as discussed above.
If we assume an overall realistic detector acceptance times efficiency of $\sim\mathcal{O}(10\%)$, the rate of signal positrons at $a_0$ of $4 \textsf{--} 5$ would be $\sim 10^{-3}\textsf{--}10^{-2}$ positrons per BX.
Hence, to claim the necessary background rejection level of $10^{-4}$ tracks per BX, a simulation of at least $10^5$ BXs is required. 
This data size corresponds to a (typical) $\sim2$~hours run at 10~Hz collision rate in the actual experiment.
At present, a simulation of this scale is impossible to realize within a reasonable time frame, even on a large high-performance computing cluster.
This is a known problem and it is common for similar existing and future experiments.
While alternative methods to generate large simulation samples are being explored elsewhere (e.g., with generative machine-learning models), in this study we consider simple toy datasets generated by sampling from the multidimensional FullSim distributions.
The method of generating large statistics background samples is discussed in Sec.~\ref{sec:background_fastsim} since it is tied to the reconstruction framework outlined there.
The different large statistics FastSim samples used in the subsequent analysis are generated from the FullSim equivalent samples as discussed in Sec.~\ref{sec:datasets}.


\section{Expected performance}
\label{sec:reconstruction}
We now turn to the discussion of the expected performance of the E320 tracker.
In particular, we focus on the question of whether it is possible to claim an unambiguous measurement of single positrons at a rate $\sim\mathcal{O}(10^{-3}\textsf{--}10^{-2})$ per BXs (expected for $a_0$ of $4 \textsf{--} 5$, after folding in different losses) in the presence of a large background and, if so, at what accuracy.
This is a key question for one of the main E320 physics measurements: the detection of NBW positrons in the tunneling regime.

To address this question, we employ the ACTS (A Common Tracking Software) framework~\cite{ai2022common}, a modular, experiment-independent toolset designed for high-energy physics applications.
The ACTS functionalities make for a powerful and portable framework for accurate and efficient particle tracking, particularly in experiments like E320, where minimizing false positives in high-background conditions is critical.

\subsection{Detector Implementation in ACTS}
\label{sec:acts_detector}
\begin{figure*}[pos=!ht]
\centering
\begin{overpic}[width=0.8\textwidth]{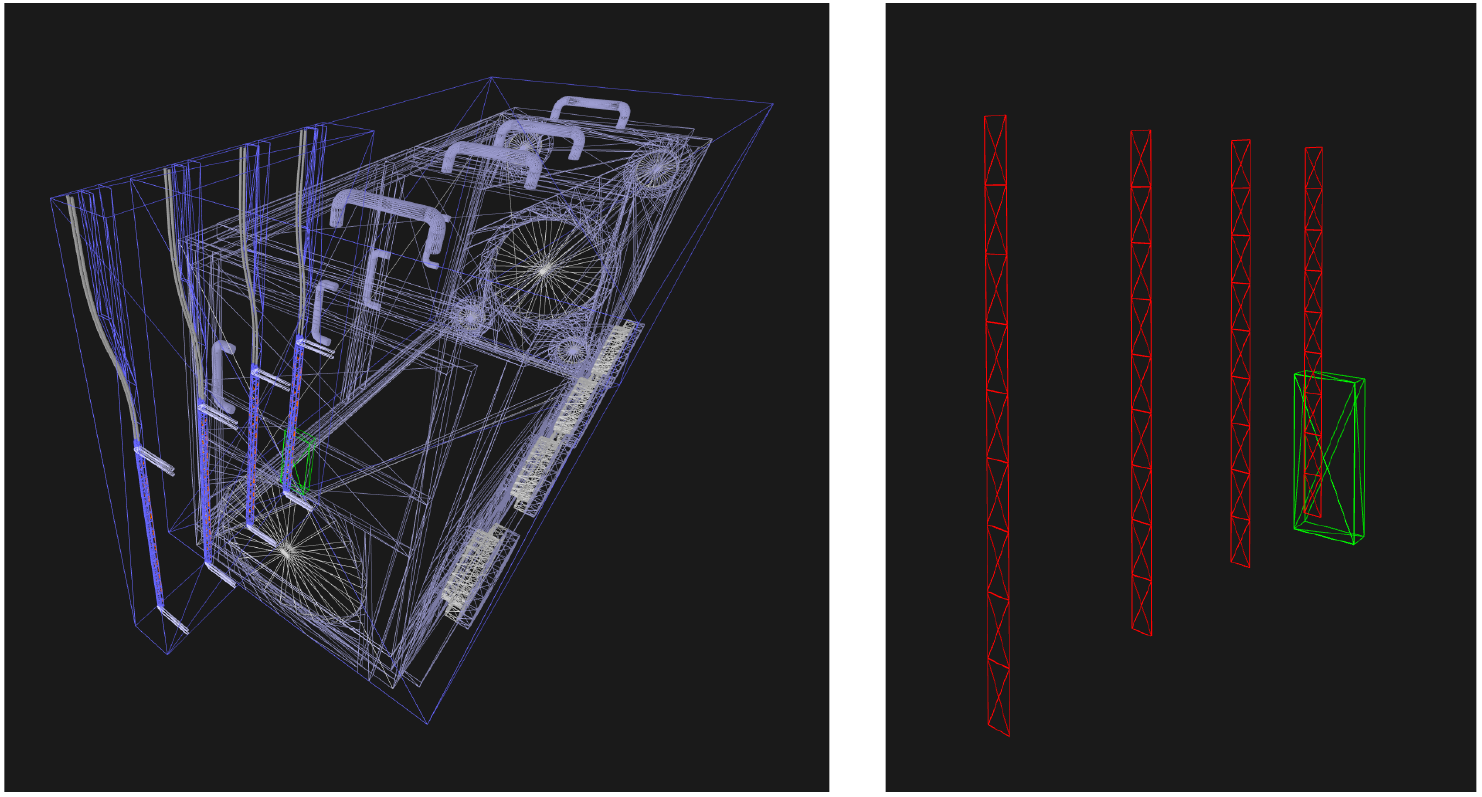}\end{overpic}
\caption{
\GEANT and ACTS detector representation.
Left: \GEANT implementation with the vacuum chamber and the full tracking detector included.
Right: \GEANT volumes included in the ACTS implementation with only the vacuum exit window and the ALPIDE chips present.}
\label{fig:acts_detector}
\end{figure*}

Track reconstruction in ACTS begins with the implementation of the detector geometry based on its \GEANT model.
Only the \GEANT volumes that are sensitive or that may significantly impact the signal positron propagation, for example, dense material regions, are implemented as infinitely-thin surfaces.
For the E320 tracking detector, 36 sensitive surfaces corresponding to ALPIDE chips and one passive surface representing the vacuum exit window, which contributes to the scattering of incoming particles, are implemented.
A comparison between the \GEANT geometry and the \GEANT volumes implemented in ACTS is shown in Fig.~\ref{fig:acts_detector}.
Besides these volumes, the magnetic field maps, derived directly from \GEANT, are also integrated into the ACTS geometry.

The material properties of the tracking detector (both the sensitive and passive components) are transferred from \GEANT to the ACTS implementation using a dedicated material-mapping algorithm.
In essence, each surface of the ACTS detector description is assigned an effective material distribution, which is obtained by averaging properties of the material regions encountered along signal-like track trajectories in \GEANT.
This step is important to ensure that ACTS properly accounts for multiple scattering and energy loss.
A more detailed description of this procedure, along with an evaluation of its performance, is provided in App.~\ref{app:material_mapping}.

\subsection{FastSim implementation with ACTS}
\label{sec:background_fastsim}
As discussed in Sec.~\ref{sec:fastsim}, the fast simulation is integrated into the reconstruction pipeline.
The FastSim begins by sampling the features (positions and momenta) of the background particles at the point of their initial interaction with the tracking detector (not necessarily at the first tracking layer), based on the distributions of FullSim background particles.
The FastSim particles are then propagated by ACTS from the sampled initial states, accounting for multiple scattering and stochastic energy loss via the Bethe-Bloch and Bethe-Heitler processes.
All the particles that produce hits in the layers here are assumed to be electrons.
The pixel hits are predominantly created by electrons and positrons produced everywhere along the beamline, as well as in the detector layers themselves, via conversion of the many background photons.
Since the detector is placed outside of the magnetic field volumes, the positrons behave identically to the electrons.
Charged particles other than electrons (muons, pions, protons) come in much smaller numbers (effectively $<1\%$ of the electrons and positrons). Specifically, for 30\% of one BX in FullSim, we observe one muon (crossing three layers), three protons (each crossing just one layer) and one charged pion (crossing just one layer).
Spatial points are recorded where the propagated particles intersect the detector's sensitive surfaces.
These points represent the geometrical centers of pixel clusters.

As explained in Sec.~\ref{sec:simulation}, there are three types of background that are treated separately.
Within the FastSim framework, we generate the NCS-driven background and the dump-driven one.
The third background type is due to all possible kinds of secondary particles resulting from the interaction of signal NBW positrons with, e.g., the vacuum exit window or the detector modules.
It is taken directly from the FullSim since the computational cost of producing it in large sample sizes is relatively low.

The two FastSim samples are generated from joint feature distributions of the FullSim.
However, the sparsity of the FullSim distributions (given the sample sizes discussed in Sec.~\ref{sec:simulation}) does not allow for a direct multidimensional sampling.
To overcome this problem, adaptive kernel density estimation (KDE) techniques~\cite{abramson1982bandwidth, silverman2018density, sheather1991reliable} are employed.
The use of KDE techniques ensures comprehensive phase-space coverage, preservation of dominant correlations in the underlying distributions, and efficient sampling of the initial particle parameters at a low computational cost.
Though, in principle, KDE distributions for both the NCS and the dump backgrounds can cover the full six-dimensional phase space necessary for the generation of samples, the low sample size of the underlying distributions limits the maximum dimensionality of the kernels.
Thus, the most dominant correlations between the features are identified and used in the KDE construction.
For the NCS and dump backgrounds, a KDE estimation is performed in the $z\textsf{--}\phi\textsf{--}\theta\textsf{--}E$ and $\phi\textsf{--}\theta\textsf{--}E$ subspaces, respectively.

\begin{figure*}[pos=!ht]
\centering
\begin{overpic}[width=0.8\textwidth, trim={0mm 0mm 0mm 0mm}, clip]{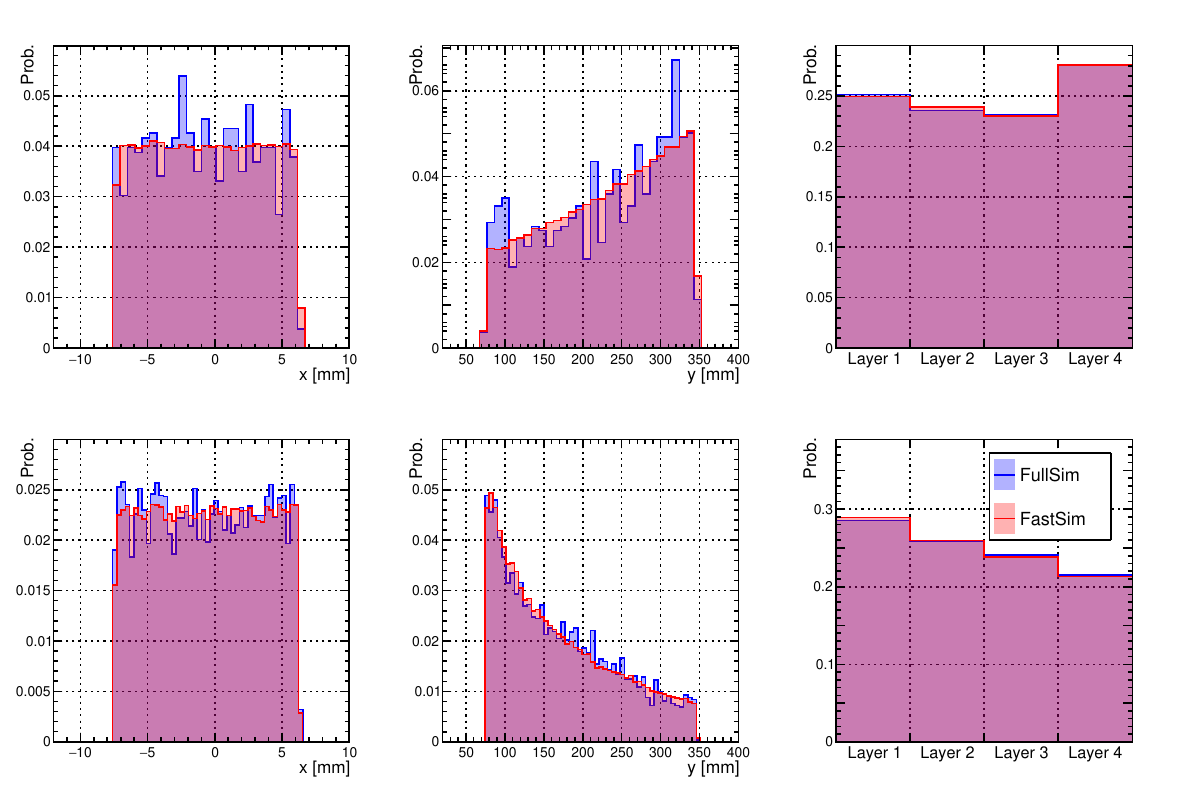}\end{overpic}
\caption{
Comparison of the initial positions of the FullSim and the FastSim background particles for the dump (top row) and the NCS (bottom row) backgrounds.
Left: initial $x$-position of the particles.
Middle: initial $y$-position of the particles.
Right: initial tracking layer of the particles.
}
\label{fig:positions_before}
\end{figure*}

The features omitted from the KDE estimations are sampled independently in the following way.
For both the NCS and the dump backgrounds, the $x$-coordinate is drawn from a uniform distribution, and the $y$-coordinate is sampled from an empirical power-law distribution fitted to the respective FullSim distribution (see Fig.~\ref{fig:positions_before}). 
For the dump background, the $z$-coordinate is sampled from a discrete distribution defined by the positions of the tracking layers along the $z$-axis.

\begin{figure*}[pos=!ht]
\centering
\begin{overpic}[width=0.8\textwidth, trim={0mm 0mm 0mm 0mm}, clip]{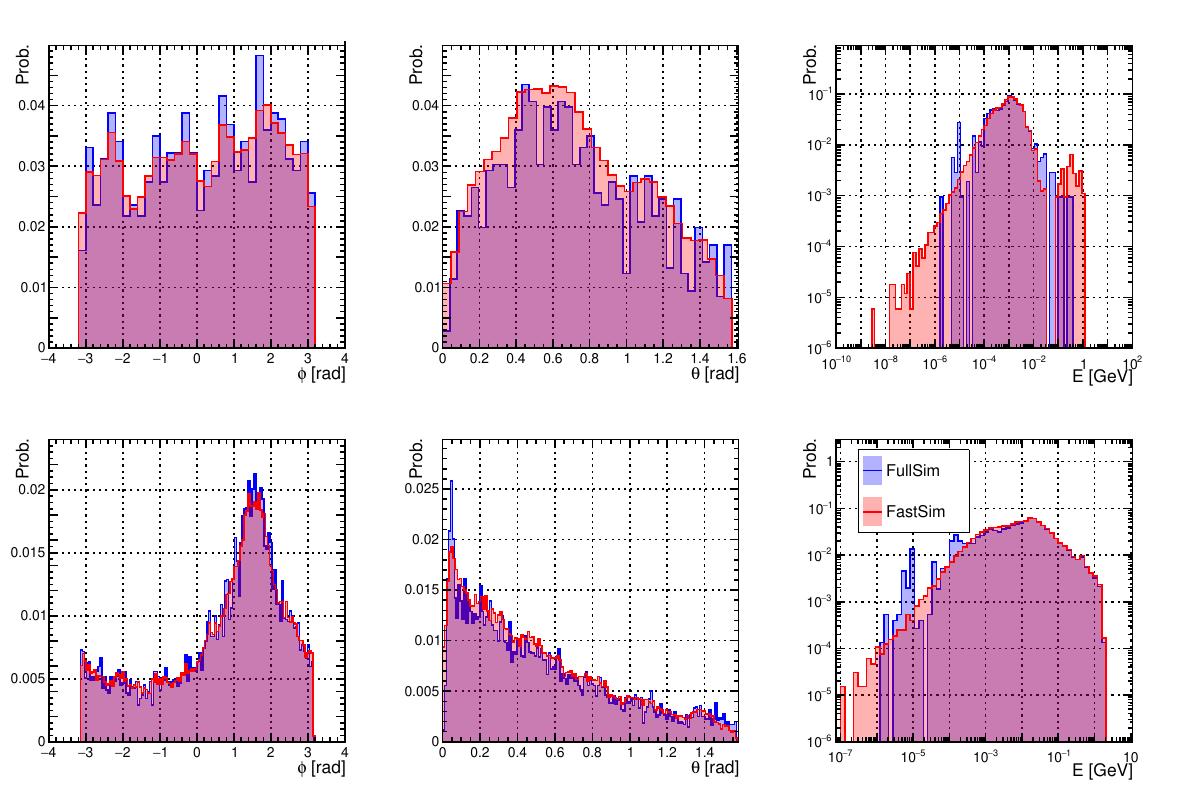}\end{overpic}
\caption{
Comparison of the initial momenta of the FullSim and the FastSim background particles for the dump (top row) and the NCS backgrounds (bottom row).
Left: $\phi$-angle of the particles' initial momenta.
Middle: $\theta$-angle of the particles' initial momenta.
Right: initial energy of the particles.}
\label{fig:kinematics_before}
\end{figure*}

\begin{figure*}[pos=!ht]
\centering
\begin{overpic}[width=0.9\textwidth, trim={0mm 0mm 0mm 0mm}, clip]{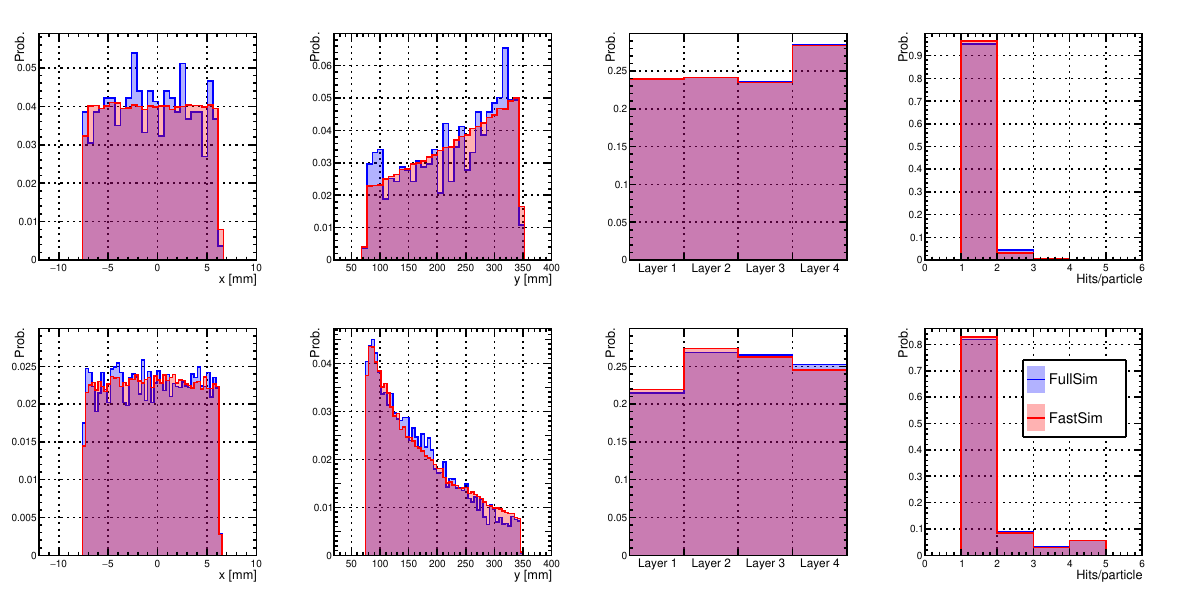}\end{overpic}
\caption{
Comparison of the final detector occupancy and the number of points (clusters) per generated particle in the FullSim and the FastSim for the dump (top row) and the NCS backgrounds (bottom row). 
First column from left: final occupancy in the detector along the $x$-axis.
Second column: final occupancy in the detector along the $y$-axis.
Third column: final occupancy in the detector's tracking layers.
Fourth column: number of generated points (clusters) per particle distributions.
}
\label{fig:multiplicity_after}
\end{figure*}

To complete the FastSim procedure, the sampled features are used as a starting point for the particle propagation, through which all intersection points with the detector tracking layers are recorded.
The size of the cluster represented by each point is sampled directly from the FullSim cluster-size distribution.
The correlation of the cluster size with its position (mostly along the $y$-axis in the effective area covered by the NBW signal) is negligible (see Appendix~\ref{app:cluster_size_sampling}).
This concludes the full information needed for the subsequent tracking.
The distributions used for evaluating the FastSim performance are:
\begin{itemize}
    \item the spatial positions at the beginning of the propagation (see Fig.~\ref{fig:positions_before}),
    \item the energy and the momentum direction at the beginning of the propagation (see Fig.~\ref{fig:kinematics_before}),
    \item the spatial positions and the number of points (clusters) per particle at the end of the propagation (see Fig.~\ref{fig:multiplicity_after}).
\end{itemize}

A comparison of the pairwise correlations between features of the FullSim and the FastSim particles at the initial state is shown in App.~\ref{app:fast_sim_inital_corr}.

Note that the distributions of points per particle in Fig.~\ref{fig:multiplicity_after} are directly tied to the complexity of background rejection in the reconstruction.
Naively, background particles that are associated with more points (clusters) in the detector layers pose greater challenges for the reconstruction algorithms.
The correlations between these points (clusters) increase the difficulty of background rejection.
Therefore, a good agreement between the FullSim and the FastSim distributions ensures that the conditions for track reconstruction are similar in both cases.

The similarity between the FastSim and FullSim distributions is only evaluated qualitatively from Fig.~\ref{fig:positions_before},~\ref{fig:kinematics_before} and~\ref{fig:multiplicity_after}.
This qualitative similarity is sufficient for evaluating the feasibility of our approach with respect to the E320 measurement goals.
Therefore, it is only important that the overall trends across both the initial position and momentum, as well as the final detector occupancy, are in agreement.
The small discrepancies observed when comparing the dump-background distributions can be attributed to the limited statistics of the FullSim sample used for the KDE construction and the fits of the non-KDE features.
The number of points per particle distributions are in good agreement between the FastSim and the FullSim, with the differences fully explained by the statistical errors of the bin entries.

To avoid underestimations in the subsequent analysis, we conservatively generate a slightly larger number of FastSim particles per BX for all background scenarios. 
While the agreement between FullSim and FastSim is not perfect in a small subset of the kinematic features, the overall behavior is well reproduced and most importantly there are no truncated regions (i.e. absent in FastSim compared to FullSim) in any of the features. 
Therefore, we can be confident that with a very large sample of BXs the subsequent tracking algorithm would encounter all edge scenarios. 
We recall that the purpose of the FastSim approach is to enable a generation of very large-scale samples of BXs ($\mathcal{O}(10^6)$) with a large number of particles ($\sim4\times 10^3\textsf{--}4\times 10^4$) per BX for the subsequent tracking estimations. 
The same sample sizes are strictly impossible to obtain with FullSim.

For completeness, Fig.~\ref{fig:simulation_pipeline} in Appendix~\ref{app:simflow} shows a schematic representation of the simulation pipeline, including both FullSim and FastSim steps.

\subsection{Datasets}
\label{sec:datasets}
Two datasets are analyzed in this work.
The first one, which consists of pure background events, is designed to study the background rejection capabilities of the reconstruction chain.
A total of $10^6$ BXs are generated following the procedure outlined in Sec.~\ref{sec:background_fastsim}.
For each BX, 660 background particles from the NCS KDE and 3450 particles from the dump KDE are generated.
These numbers exceed the 627 and 3303 particles per BX estimated from the FullSim datasets by their respective error estimates. 
For the NCS, the error is estimated as the standard deviation of the 12 FullSim BXs, while for the dump case this is impossible since we have only 30\% of one BX simulated. 
Hence the error for it is estimated by employing binomial statistics considerations.
This is done to ensure that the results provide a conservative estimate of the rejection performance.
Additionally, recall that the NCS background is conservatively determined for $a_0=10$.

The second dataset combines signal and background particles.
The signal NBW positrons and the secondary particles they produce are obtained from the FullSim with a total of $\sim2.6\times 10^4$ positrons (as discussed in Sec.~\ref{sec:fullsim}).
For reconstruction, the signal sample is split into individual positrons and their secondaries.
The number of positrons having at least one pixel cluster in the detector is $\sim 1.85\times 10^4$ ($\sim 71\%$ of the total), where the rest are completely missing the detector.
For each such positron, a fresh randomized copy of the FastSim background sample (NCS and dump) is embedded.
This is done using the same generation parameters as for the pure background dataset.
The combination of a signal positron, its secondaries, and the NCS and the dump backgrounds constitutes a single BX for reconstruction.

The points (FastSim background) and the clusters (FullSim NBW signal and its associated backgrounds) in the datasets listed above are transformed into particle position ``measurements'' and their uncertainties.
The latter are given by the standard error of the mean, calculated as $L_{x(y)}/\sqrt{12 N}$, where $L_{x(y)}$ is the size of the ALPIDE pixel in $x$ ($y$), and $N$ is the number of pixels in the cluster. 

\begin{figure*}[pos=!ht]
\centering
\begin{overpic}[width=0.9\textwidth]{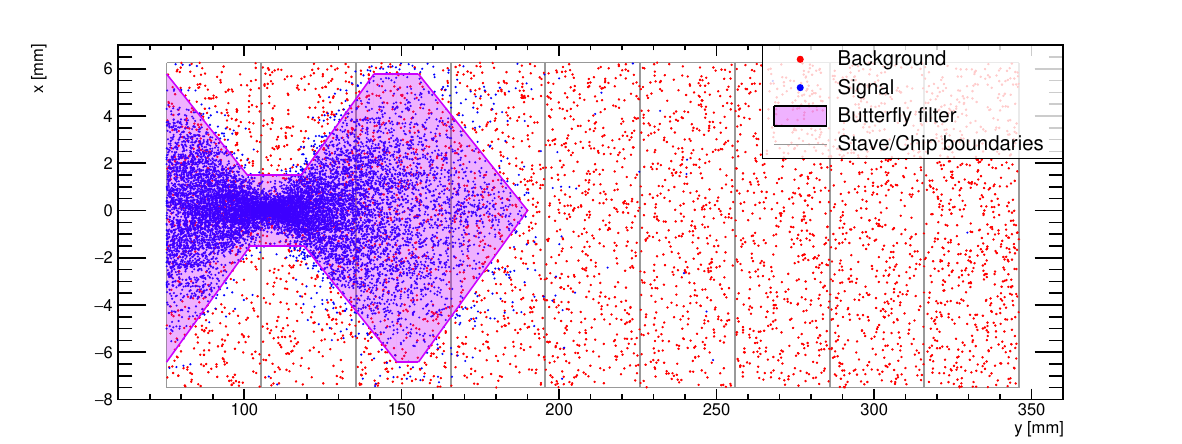}\end{overpic}
\caption{
The butterfly-shaped geometrical cut.
For illustration, measurements belonging to 3,100 FullSim signal positrons (equivalent to approximately 10 times the expectation for $a_0=10$) and one BX of FastSim background particles are shown in the global coordinate system. The plot is cumulative for measurements from all four detector layers in order to capture the effect of the small inclination of the particles’ trajectories imposed by the dipole magnet.
The shaded violet area shows the butterfly-like pattern in the $x\textsf{--}y$ plane where measurements are retained.
The cut is the same between the four layers. 
The stave and the chip boundaries within it are indicated.}
\label{fig:hourglass}
\end{figure*}

All measurements undergo filtering based on their positions in the $x\textsf{--}y$ plane.
Due to the chromaticity of the quadrupoles in the spectrometer beamline, the NBW positrons have a characteristic ``butterfly-shaped'' distribution in the detector plane (energy-angle correlation).
This is exploited by applying a spatial cut as shown in Fig.~\ref{fig:hourglass}.
It significantly reduces the presence of background measurements, while keeping the signal measurements intact: $\sim 78\%$ of the background measurements are removed, while $\sim 96\%$ of the signal measurements are retained.
All passing measurements are subsequently embedded in the corresponding sensitive surfaces of the detector in ACTS for the next tracking step.

The performance of the reconstruction algorithms discussed below is evaluated within the detector acceptance.
The acceptance includes the butterfly cut discussed above, which effectively restricts the energy range of signal positrons to $1.5\textsf{--}4$~GeV for the dipole settings that are simulated here.
The lower bound of this range approximately corresponds to the top edge of the dipole exit flange, while the upper bound corresponds to the bottom edge of the sensitive area at the first detector layer.
The ALPIDE chips' detection efficiency for minimum ionizing particles ($\gtrsim99\%$, typically~\cite{AGLIERIRINELLA2017583,MAGER2016434,Abelevetal:2014dna}) is also included in the acceptance definition. The final observable performance metrics of interest -- such as the track reconstruction rate -- are evaluated for different cases characterized by the number of signal measurements in each detector layer. 

Next, we describe how one can simulate the detector performance for an arbitrary value of $a_0$.
To this end, all calculations involving the signal are done first with the ``exactly one positron per BX'' assumption, where each of these positrons must have \textit{at least one measurement in the detector volume}.
As discussed earlier, this amounts to $\mathcal{A}\sim 71\%$ of the generated positrons and to a good approximation it is independent of $a_0$ (at least in the parameter range relevant for E320).
This choice is denoted as  ``$1e^+/{\rm BX}$''.

The rate for a fixed $a_0$, $\Gamma(a_0)$, is obtained by scaling the $\Gamma(1e^+/{\rm BX})$ scenario via $\Gamma(a_0)=\Gamma_{\rm prod}(a_0)\times \Gamma(1e^+/{\rm BX})\times \mathcal{A}$, where $\Gamma_{\rm prod}(a_0)$ is the positron‐production rate at the IP shown in Fig.~\ref{fig:rate}.
In the remainder of this study, we quote only the $\Gamma(a_0=5)$ rates, for which the rate of the positron production at the IP is $\Gamma_{\rm prod}(a_0=5) = 0.168$ positrons per BX.

\subsection{Position to Kinematics mapping}
\label{sec:acts_kinematics}
Track finding and track fitting algorithms, based on the propagation of a particle through the detector, require an initial guess of the track parameters at the IP (production vertex and four-momentum).
The preceding step, called ``track seeding'', also requires an estimate of the particle's momentum at the first tracking layer (FTL) of the setup.
The inference of those parameters is done by mapping a set of known positions in the detector to the corresponding known parameters of signal-like particles.
The map object storing the correspondence, $\mathcal{M}$, is filled using ACTS in a particle-gun mode. 
We generate particles at the IP and propagate them to the FTL, where each of the nine sensors' surfaces has an $x\textsf{--}y$ grid imposed onto it.
The grid cells intersected by the particles are identified, and the corresponding IP/FTL track parameters are stored within them.
The resulting collections of the tracks' parameters within each cell are averaged.
Thus, $\mathcal{M}$ maps the $x\textsf{--}y$ measurement positions in the FTL to corresponding parameters of the tracks at the IP and the FTL.

\begin{figure*}[pos=!ht]
\centering
\begin{subfigure}{0.45\textwidth}
\begin{overpic}[width=1.0\linewidth]{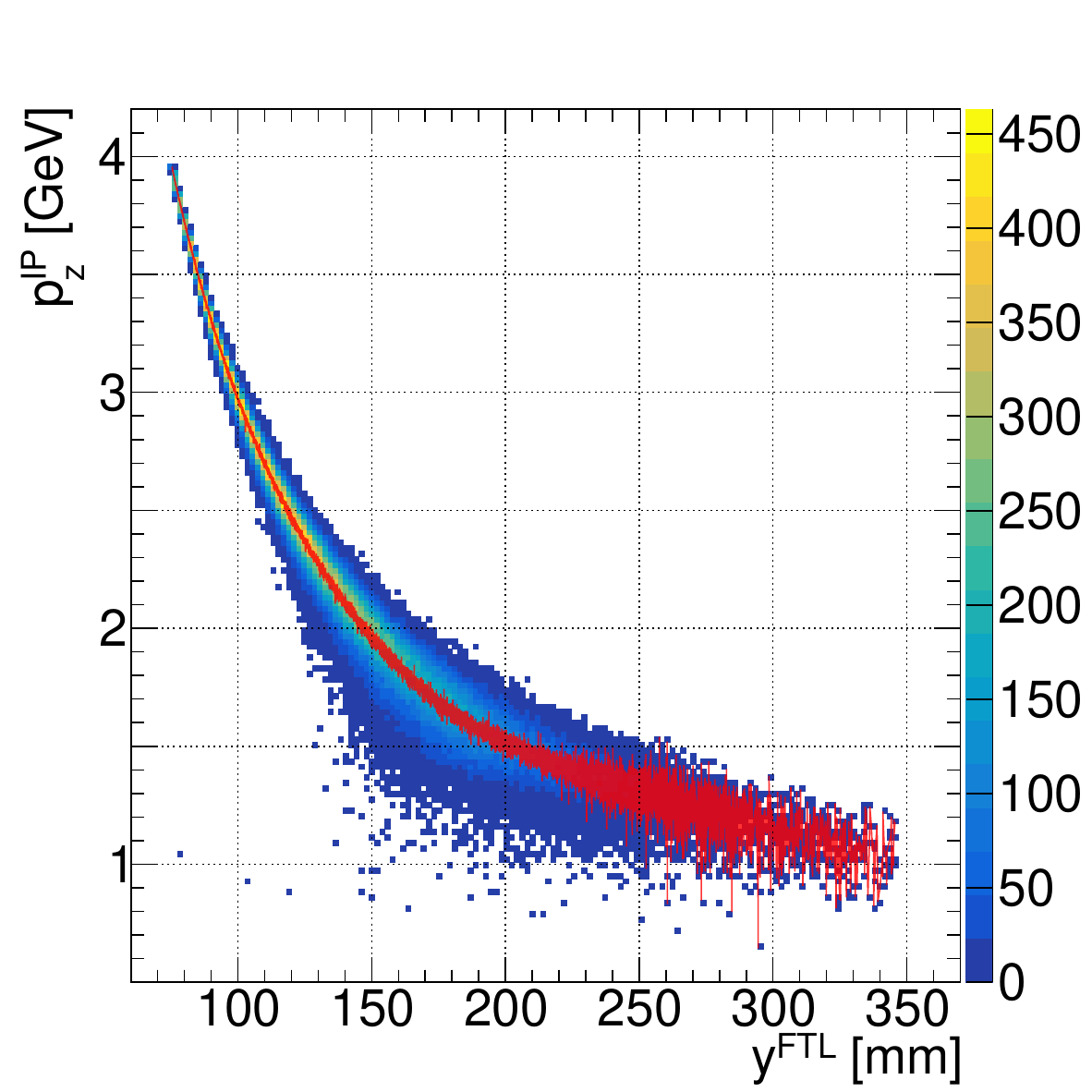}\end{overpic}
\end{subfigure}
\begin{subfigure}{0.45\textwidth}
\begin{overpic}[width=1.0\linewidth]{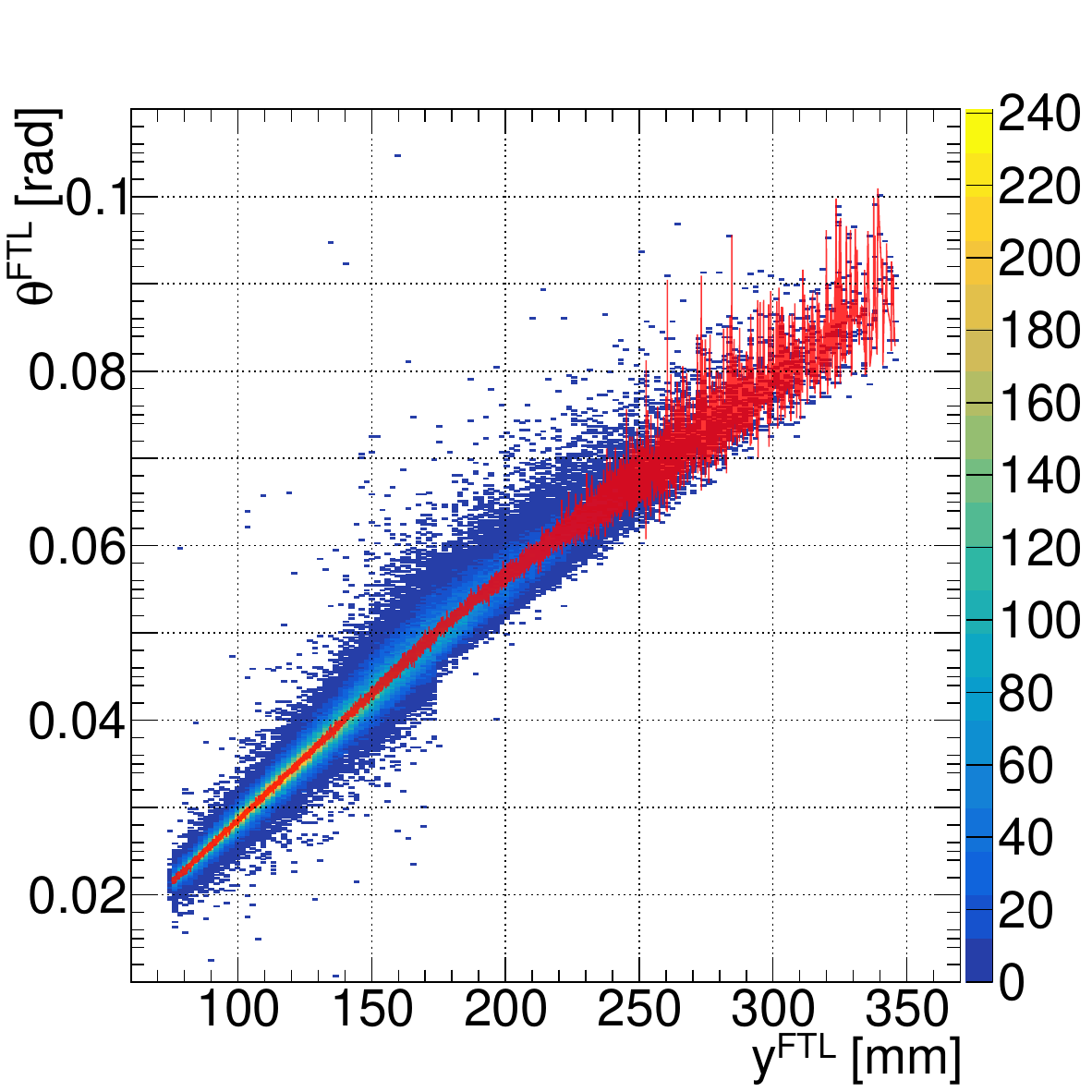}\end{overpic}
\end{subfigure}
\caption{
Two dimensional projections of the six-dimensional $\mathcal{M}$ map: $y^{\rm FTL}\to p_z^{\rm IP}$ (left) and $y^{\rm FTL}\to \theta^{\rm FTL}$ (right).
The red line shows the average, which is later used to infer the IP and FTL momenta from the position at the FTL.}
\label{fig:kinematics_map}
\end{figure*}

The IP parameters used to build $\mathcal{M}$ are taken from a large ($2\times 10^5$) dedicated truth-only sample of NBW positrons generated with \PTARMIGAN. 
Each grid imposed on the FTL surfaces is binned only along the $y$-axis, using 1000 bins.
The binning in the $x$-direction is omitted to avoid the prohibitively larger sample size required to sufficiently populate a two-dimensional grid.
This simplified estimation of the initial guess is effectively ignoring the $x$-component of the momentum.
This, however, has little impact on the reconstruction performance discussed below, since the latter is mostly governed by the dominant $z$-component of the IP momentum.
Fig.~\ref{fig:kinematics_map} illustrates two example projections of $\mathcal{M}$: $y^{\rm FTL} \to p_z^{\rm IP}$ and $y^{\rm FTL} \to \theta^{\rm FTL}$.
The (small) spread seen in these 2D distributions is due to material effects, as well as the non-trivial transport imposed by the quadrupoles given the non-vanishing transverse position and momenta at the IP.
The average (in the full dimensionality of $\mathcal{M}$) is used to predict the seed parameters as discussed next in Sec.~\ref{sec:acts_seeding}.
Additional details on the mapping procedure are provided in App.~\ref{app:kinematics_guess}.

\subsection{Seeding}
\label{sec:acts_seeding}
Seeding is the initial step of track reconstruction.
It aims to reduce the combinatorial complexity of the reconstruction process by grouping measurements into small subsets.
Each subset potentially contains all measurements corresponding to a signal particle with a specific momentum.
The track hypothesis for a given momentum is then tested exclusively against the measurements within the corresponding subset.
This eliminates the need to consider all measurements recorded by the detector for a given event.

\begin{figure*}[pos=!ht]
\centering
\begin{overpic}[width=0.8\textwidth]{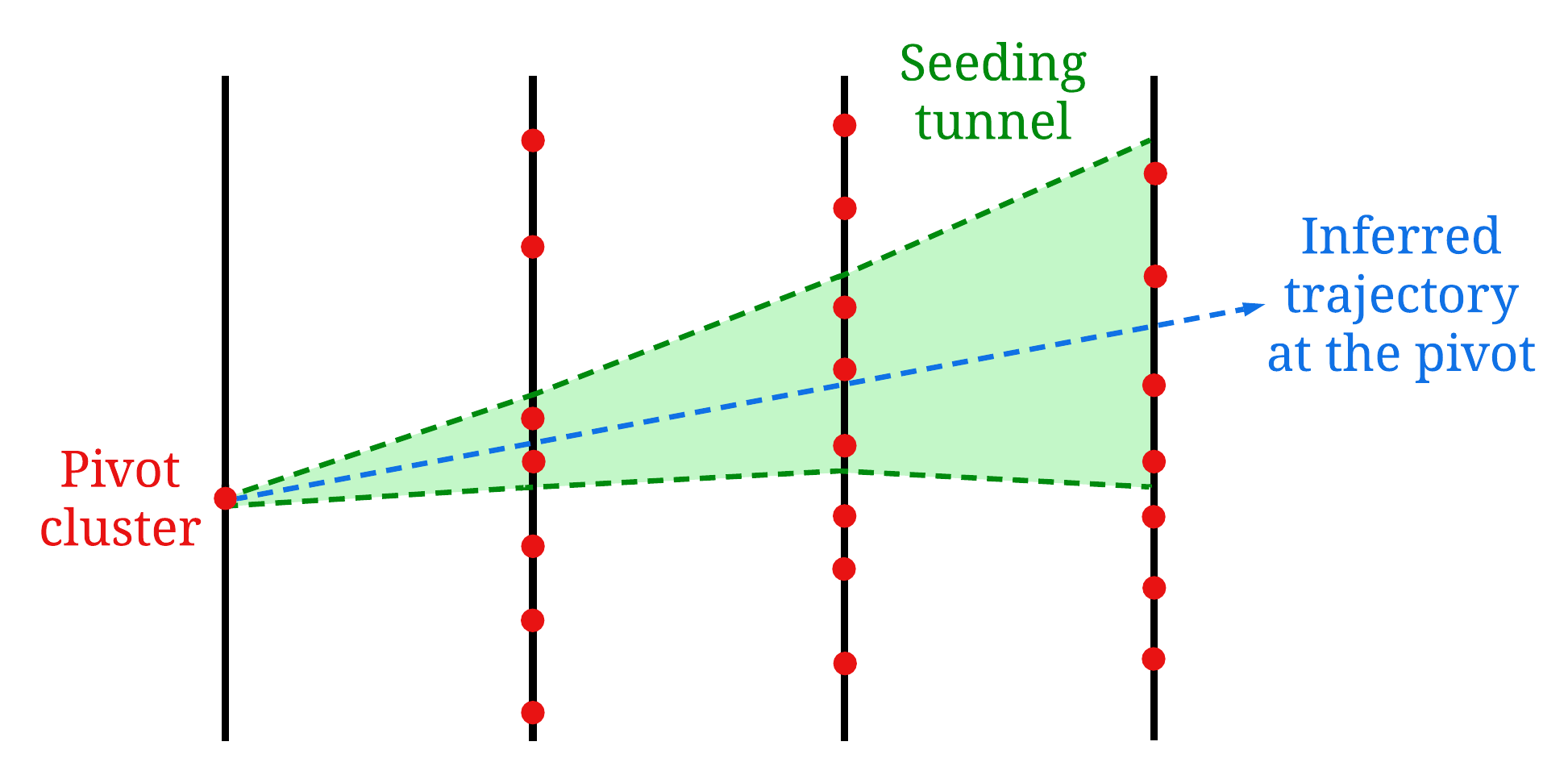}\end{overpic}
\caption{
A schematic illustration of the seeding algorithm. Measurements falling within the seeding tunnel around the linear track estimate are considered part of the seed.
The size of the rectangular bounds is growing with $z$.}
\label{fig:seeding}
\end{figure*}

The seeding algorithm used in this study is performed in two steps.
In the first step, a so-called pivot measurement is selected from the first tracking
layer (FTL).
Based on the pivot's position, estimates of the particle's vertex, momentum at the IP, and momentum at the FTL are taken from the $\mathcal{M}$ map defined earlier.
In the second step, a straight line is drawn from the pivot's position at the FTL through the rest of the tracking layers along the inferred direction of the particle's trajectory at the pivot.
Rectangular boundaries are constructed around the intersections of this line with the sensitive surfaces.
Each boundary lies within the plane of its respective surface and it grows with $z$.
Measurements that fall within these rectangular bounds are associated with the estimated IP parameters of the particle.
These measurements, together with the IP parameters, are referred to as a seed.
These steps are repeated for all pivot measurements, resulting in a collection of seeds for a given BX.
The algorithm is shown diagrammatically in Fig.~\ref{fig:seeding}.

The boundary widths along the $y$-axis of $\pm 350~\mu{\rm m}$, $\pm 400~\mu{\rm m}$, and $\pm 450~\mu{\rm m}$ are selected for the second, third, and fourth tracking layers, respectively.
This increase accounts for the growing discrepancy between the inferred and true particle trajectory, resulting from effects such as multiple scattering and the propagation of momentum-estimation errors at the pivot into spatial uncertainties in the downstream tracking layers.
The width in the $x$-direction is chosen to ensure full coverage of the chip along the $x$-axis, due to the specific construction of the $\mathcal{M}$ map object discussed in Sec.~\ref{sec:acts_kinematics}, which is insensitive to the initial momentum along $x$.
These boundaries effectively define a 3D seed ``tunnel'' encapsulating all seed measurements.\\

\begin{figure*}[pos=!ht]
\centering
\begin{subfigure}{0.32\textwidth}
\begin{overpic}[width=1.0\linewidth]{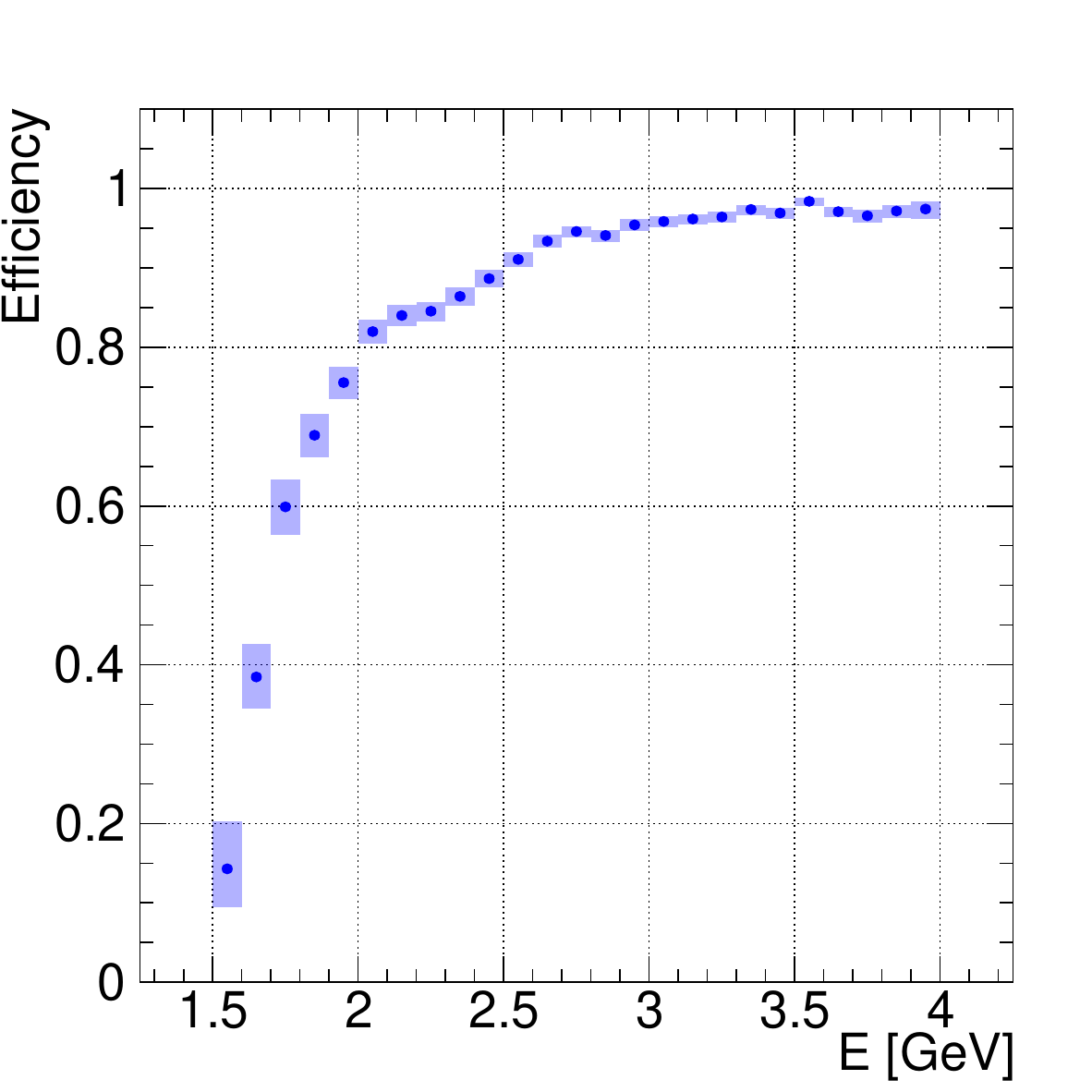}\end{overpic}
\end{subfigure}
\begin{subfigure}{0.32\textwidth}
\begin{overpic}[width=1.0\linewidth]{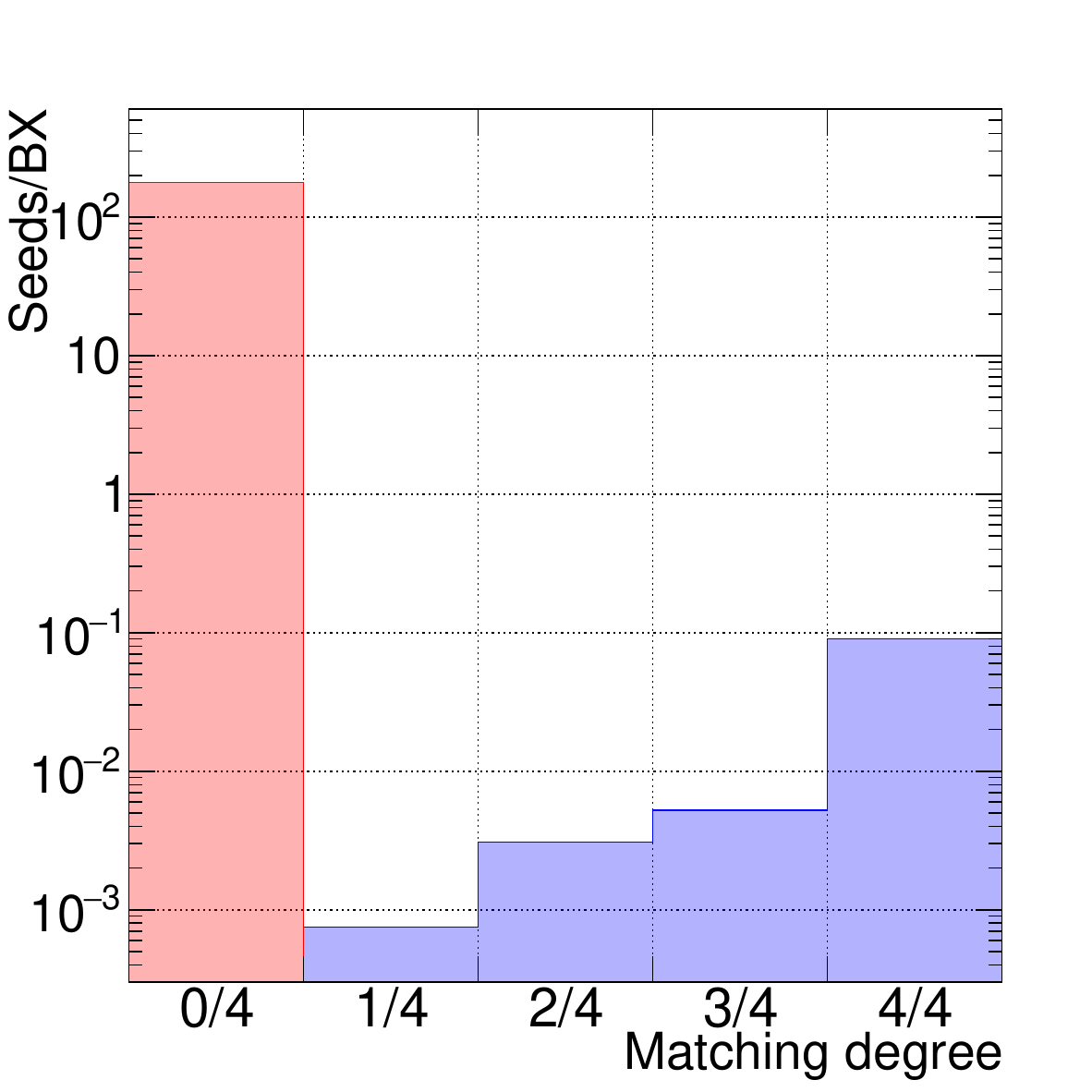}\end{overpic}
\end{subfigure}
\begin{subfigure}{0.32\textwidth}
\begin{overpic}[width=1.0\linewidth]{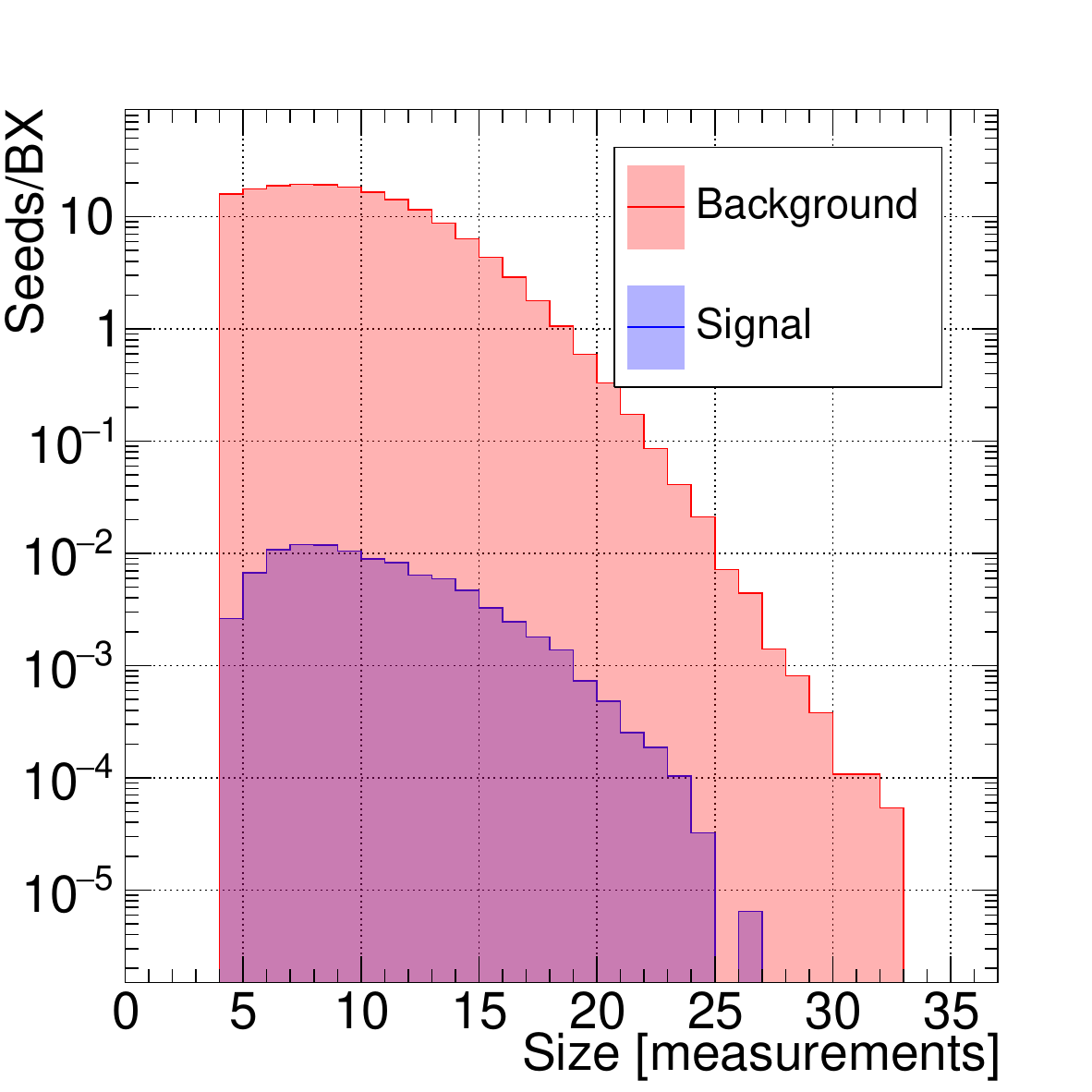}\end{overpic}
\end{subfigure}
\caption{
Performance of the seeding algorithm, measured within the acceptance.
Top left: the seeding efficiency as a function of the true FullSim NBW positrons energy at the IP.
Top right: the seeds matching degree distribution.
Bottom: the seed size distribution for signal and background. The plots per BX are scaled to $a_0=5$ and are given within the considered acceptance.}
\label{fig:seeding_performance}
\end{figure*}

Fig.~\ref{fig:seeding_performance} shows the seeding performance.
To quantify this performance we introduce the seed ``matching degree'' concept, which is defined as follows: let the seed pivot be associated with a signal positron that produces $N$ measurements in the detector.
Further, let $M$ of these measurements fall inside the seed tunnel.
Then the matching degree of this seed is $M/N$.
If the pivot is not associated with a signal positron, the seed matching degree is set to zero.
Within the acceptance, the matching degree narrows down to 0/4, 1/4, ..., 4/4.
The seeding efficiency is measured within the acceptance in bins of the signal positron’s true energy.
It is defined as the fraction of fully matched (matching degree 4/4) seeds out of all positrons in a given energy bin.

The seed size refers to the number of measurements falling within the seed tunnel from all layers.
Each layer must have at least one measurement in the tunnel to make a seed and minimum seed size of four is imposed for all subsequent reconstruction steps.
The efficiency of the seeding algorithm within the acceptance is $\sim 91\%$.
As expected, it decreases for lower-energy positrons ($\sim1.5\textsf{--}2.5$~GeV).
Capturing these positrons would require expanding the tunnel boundaries along the $y$-direction.
However, this would increase the seed sizes and, in turn, the computational complexity of later stages in the reconstruction workflow.
We remind that Fig.~\ref{fig:seeding_performance} and all figures in the subsequent analysis are scaled to the $a_0=5$ scenario.
This applies for all distributions below that are normalized to the number of BXs.
With the collection of seeds (in a given BX), we can now proceed to run the track finding and fitting algorithms.

\subsection{Track Finding}
\label{sec:acts_track_finding}
The track finding procedures in ACTS rely on the Kalman Filter (KF) and Combinatorial KF (CKF) algorithms~\cite{ai2022common}.
As shown in Fig.~\ref{fig:seeding_performance}, the number of possible measurement combinations within a single seed may still be too large to conduct an exhaustive search for the combination of measurements most compatible with a signal positron.
To mitigate this, a further filtering step is introduced, where ``track candidates'' are generated from the seeds using the CKF algorithm.
The initial track parameters (position and momentum at the IP) are taken from the seed.
The track state is then propagated from the IP through the experimental setup using the KF algorithm, accounting for material effects and magnetic fields.

\begin{figure*}[pos=!ht]
\centering
\begin{overpic}[width=0.8\textwidth]{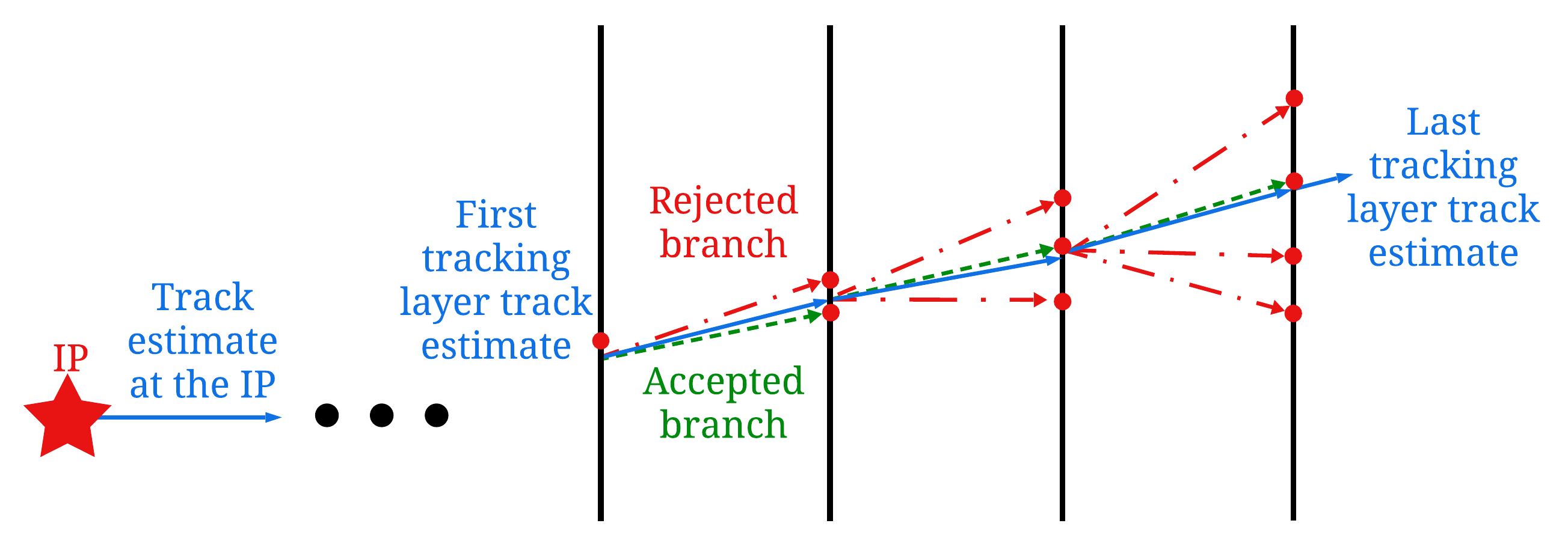}\end{overpic}
\caption{
An illustration of the CKF algorithm.
The dot-dashed red branches are rejected based on the $\chi^2$ estimate, while the dashed green branches correspond to accepted measurements.
The solid blue lines represent track estimates at a given tracking layer post Kalman update at this layer.}
\label{fig:ckf}
\end{figure*}

At each sensitive surface, the corresponding measurements are retrieved (within the seed tunnel only), and the track branches through them.
The branches are evaluated for compatibility with the track hypothesis based on the $\chi^2$ of the surface measurements relative to the KF intersection estimate.
The measurement with the lowest $\chi^2$ is selected, followed by a Kalman update before continuing the propagation.
Once the propagation is complete, the measurements from accepted branches are grouped into a track candidate for further fitting with the KF.
A schematic representation of the algorithm is shown in Fig.~\ref{fig:ckf}.

\begin{figure*}[pos=!ht]
\centering
\begin{subfigure}{0.39\textwidth}
\begin{overpic}[width=1.0\linewidth]{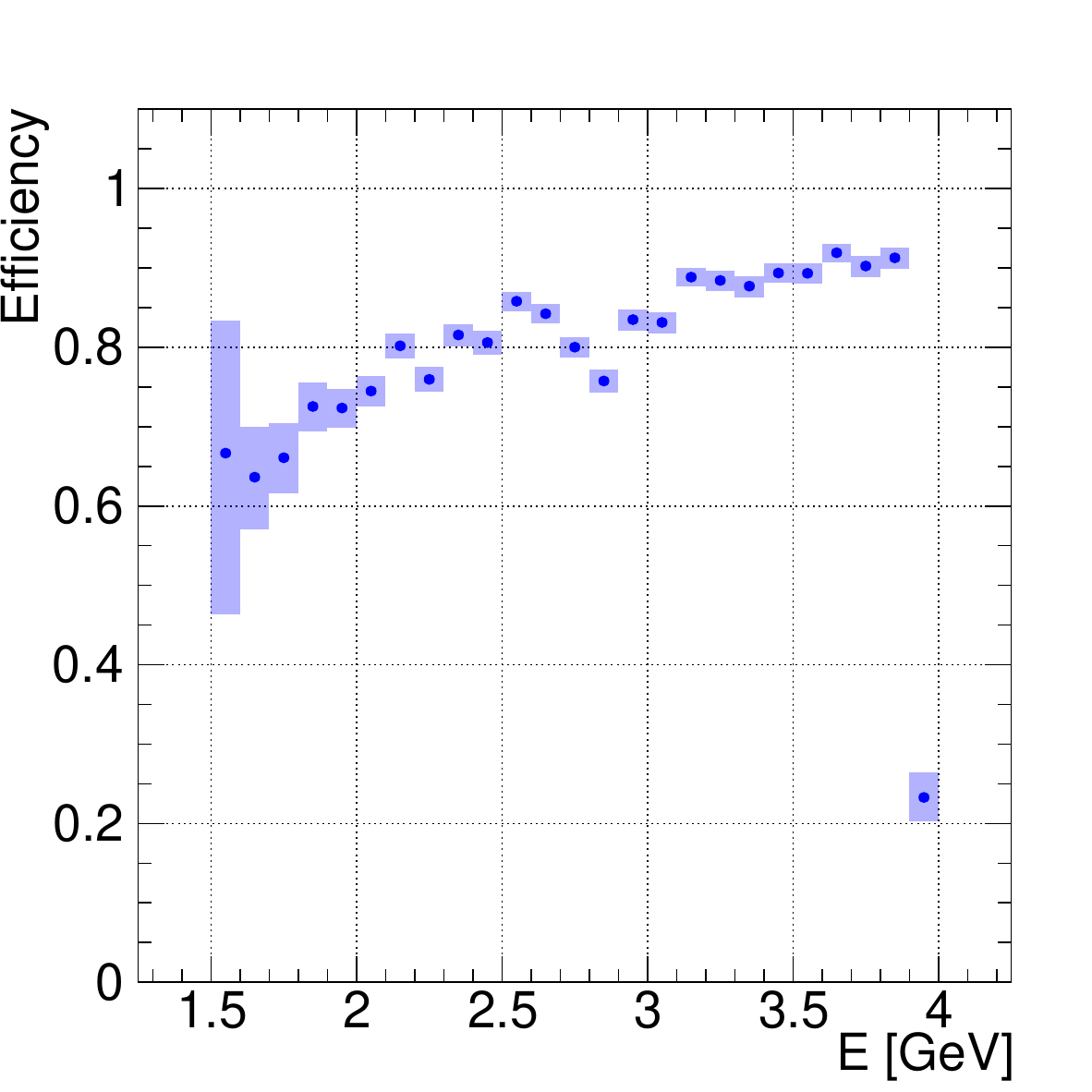}\end{overpic}
\end{subfigure}
\begin{subfigure}{0.39\textwidth}
\begin{overpic}[width=1.0\linewidth]{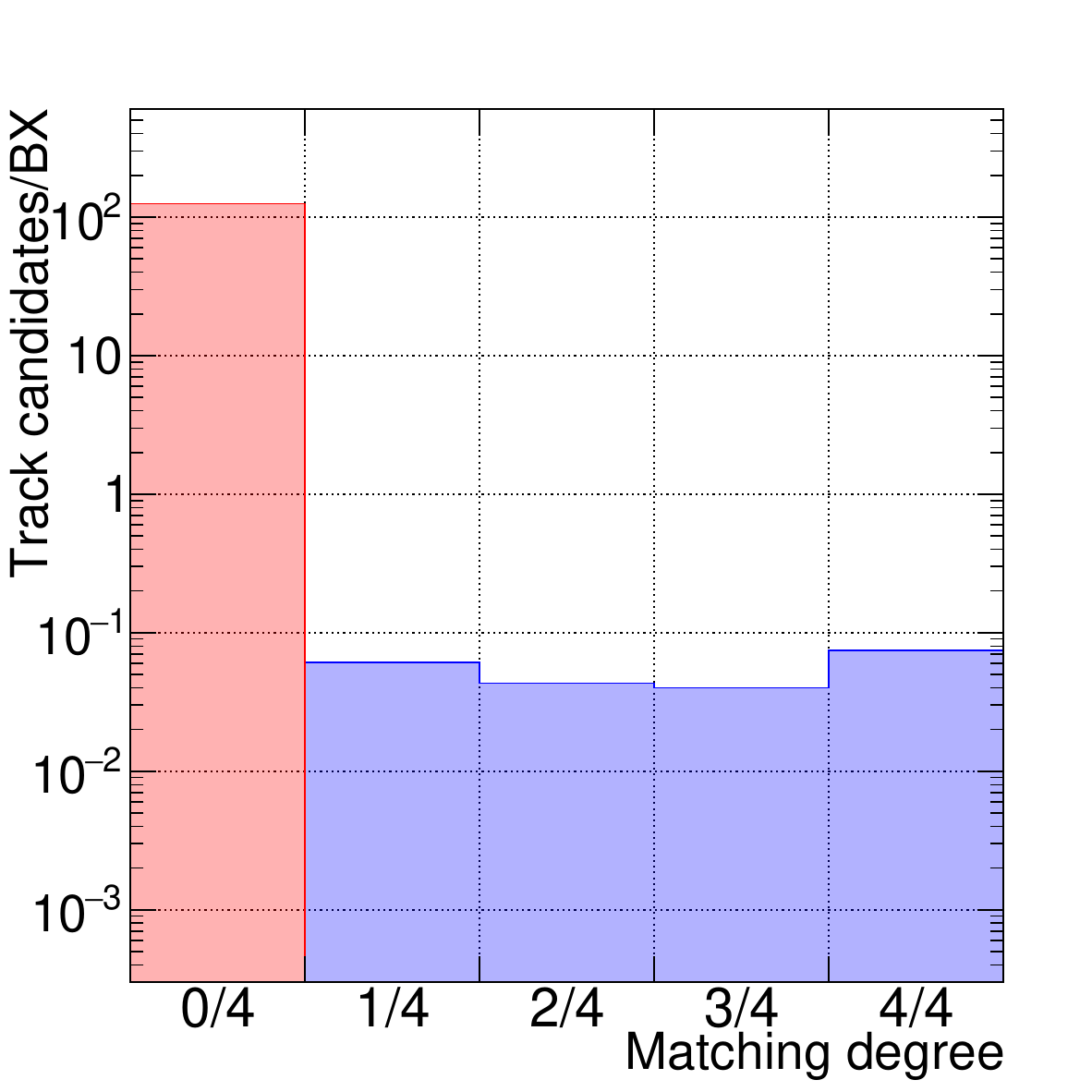}\end{overpic}
\end{subfigure}
\caption{
The CKF algorithm performance.
Left: the CKF signal efficiency as a function of the true FullSim NBW positrons’ energy at the IP.
Right: the distribution of matching degrees for track candidates. The plot per BX are scaled to $a_0=5$ and are given within the considered acceptance.}
\label{fig:ckf_performance}
\end{figure*}

To limit the computational overhead, a maximum of 1000 propagation steps is imposed. 
Tracks failing to reach the last layer within this limit are excluded from further analysis.
Fig.~\ref{fig:ckf_performance} shows the track finding algorithm performance.
The track candidate matching degree is defined analogously to the seed matching degree, with one difference.
Instead of tying the definition of $M$ and $N$ to the pivot measurement (such that the matching degree is automatically set to 0/4 if the pivot is not a signal measurement), $M$ simply represents the number of signal positron measurements and $N=4$.
In practice, the matching degree takes the same values as in the seeding case.

The track finding signal efficiency is defined as the ratio of track candidates with a matching degree of 4/4 to the number of fully matched seeds.
Approximately 82\% of all signal tracks fully captured by the seeding are also retained by the CKF filtering process.
The efficiency drops observed near $\sim2.2$~GeV and $\sim2.8$~GeV are attributed to positrons of these energies passing close to the gaps between the sensors of the stave (in at least one layer, see Fig.~\ref{fig:acts_detector}). 
A sharp efficiency decline is also observed for positrons with energies above 3.9~GeV. 
Similarly, this is attributed to the positrons passing close to the bottom edge of the lowest sensor of the first tracking layer.
It should be noted that the observed efficiency drops can be avoided by performing an exhaustive search over all possible track candidates. 
That is, every possible combination of clusters across the detector layers is considered as a track candidate. 
However, such an exhaustive search comes at a substantial computational cost. 
Specifically, for the same output of the seeding step, the number of track candidates increases by a factor of approximately 30 compared to the CKF output. 
This increase, propagating to the computational cost of subsequent track fitting and post-reconstruction analysis, is prohibitive given the expected dataset sizes in E320 and the not unlimited computing resources available.

At this stage in the pipeline, the dominant fraction of the accepted candidates consists entirely of background measurements (matching degree 0/4).
In addition, there is a non-negligible contribution from the matching degrees 1/4, 2/4, and 3/4.
The latter are due to the fact that measurement-sharing between track candidates is allowed in this study.
That is, if the pixel (and thus cluster) occupancy is high enough in the detector, one may end up with multiple track candidates sharing the same cluster.
For example, in the same BX there can be a 4/4 candidate and a 1/4 candidate, where the latter has three background clusters and one cluster picked up from the signal and belonging to the former 4/4 track.
These candidates are expected to be incompatible with signal kinematics after fitting, which is discussed next in Sec.~\ref{sec:acts_track_fitting}.
Thus, they are expected to be rejected in subsequent analysis.

\subsection{Track fitting}
\label{sec:acts_track_fitting}
The track candidates are fitted using the KF algorithm.
It involves standard propagation steps with Kalman updates applied at each measurement encounter.
\begin{figure*}[pos=t]
\centering
\begin{overpic}[width=0.8\textwidth]{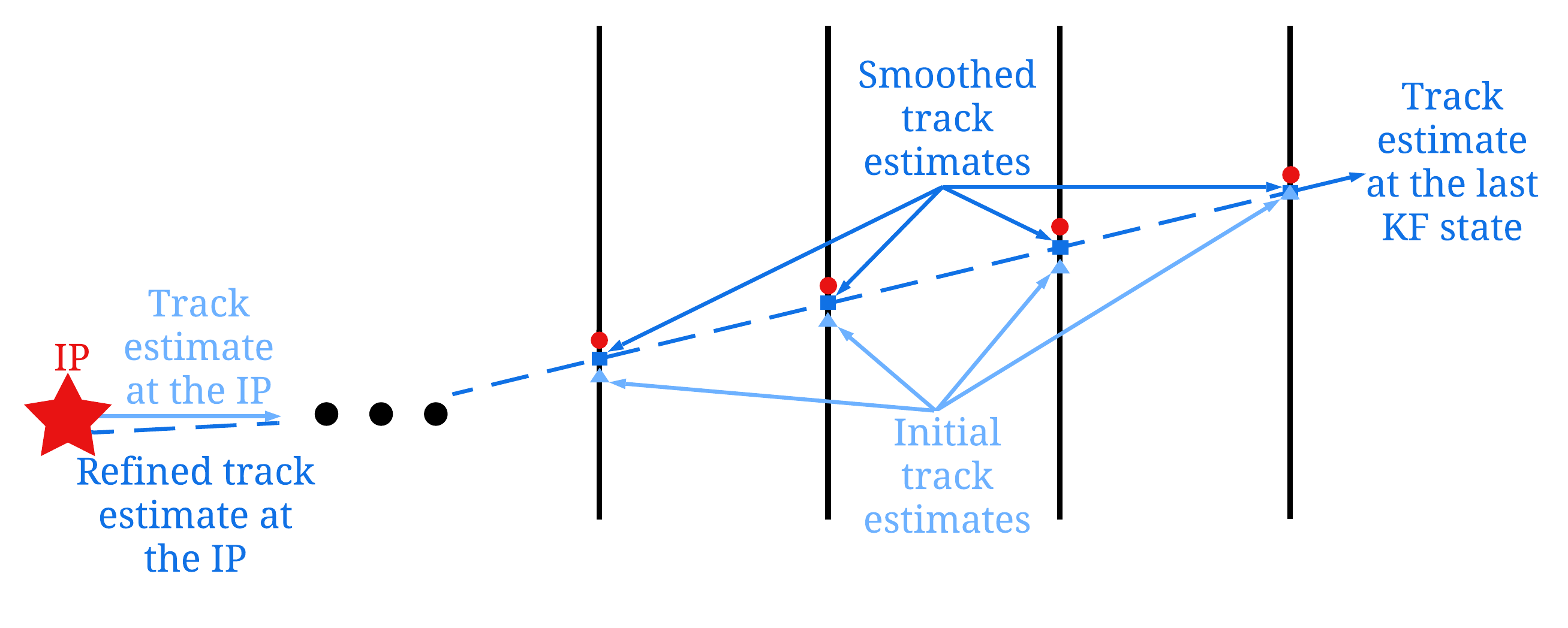}\end{overpic}
\caption{
An illustration of the KF algorithm smoothing concept.
The track candidate measurements are shown as red points.
The pre-smoothing track estimates are shown as light blue triangles, while the smoothed estimates are shown as dark blue squares.
The track estimate at the last layer remains unchanged, as all available information is already incorporated during inference.}
\label{fig:kf_smoothing}
\end{figure*}
Upon reaching the last layer of the detector, a smoothing step is performed, where the track state is propagated backward to the IP.
During this back-propagation pass, the track estimates are updated at each tracking layer, incorporating information from all estimates in the downstream (as counted along the beamline) layers.
The smoothed kinematic estimate at the IP is then used to refine the initial track parameters.
This process is illustrated in Fig.~\ref{fig:kf_smoothing}.

With the same limit on the number of propagation steps as for the CKF stage, the efficiency of the KF fitting algorithm with respect to the CKF step is 100\%, requiring only that the KF fit is successful. Further requirements on the track parameters and fit quality are applied post-reconstruction as discussed in Sec.~\ref{sec:acts_track_analysis}.
All track candidates identified at the CKF stage are successfully fitted and propagated to the post-reconstruction analysis. 
Additionally, the matching degree distribution remains identical to that of the track-finding stage.

\begin{figure*}[pos=!ht]
\centering
\begin{overpic}[width=0.8\textwidth]{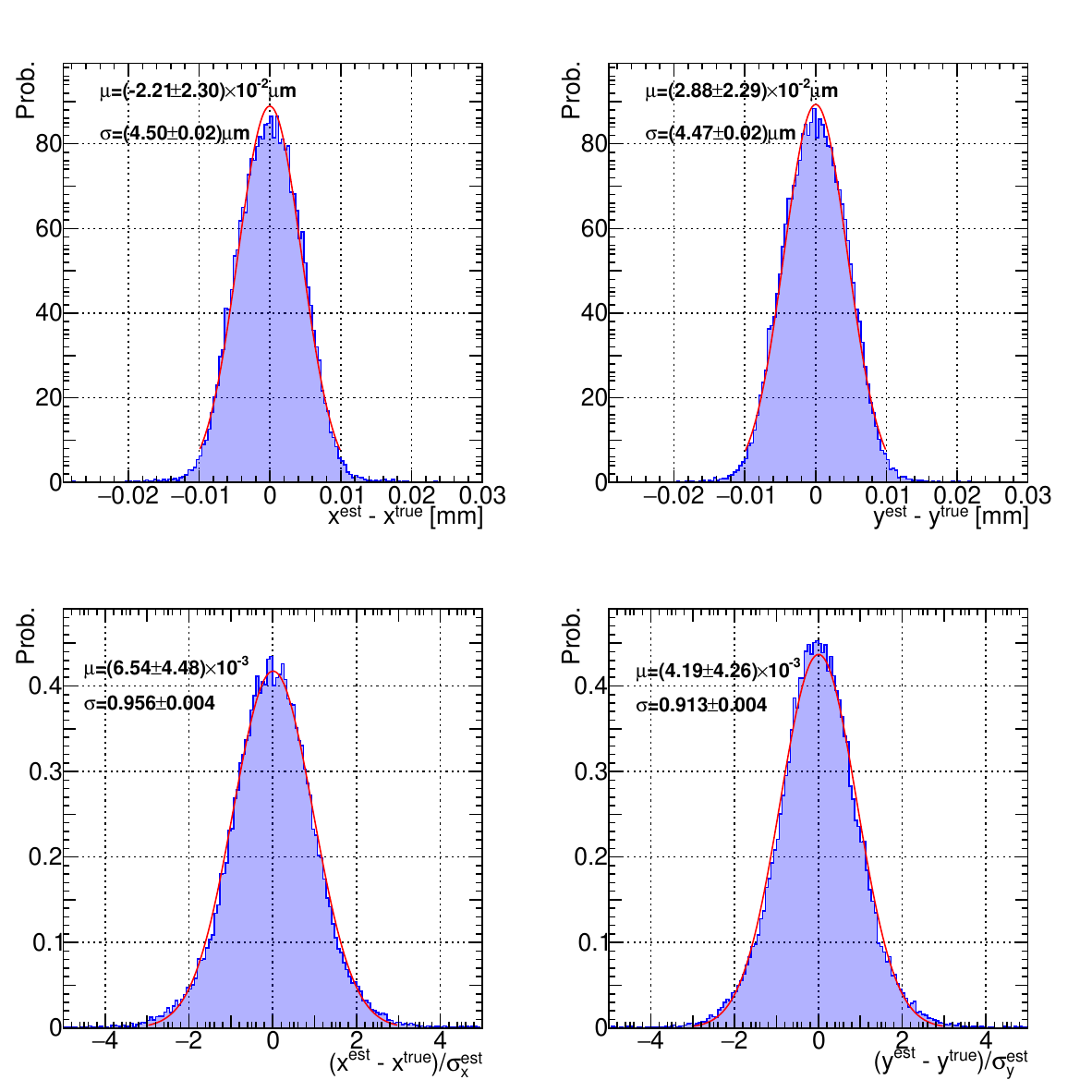}\end{overpic}
\caption{
The residuals and pulls of the fitted tracks, where superscript ``est.'' signifies the estimated quantities.
The Gaussian distribution fits are shown in red.
Top: $x$ and $y$ residuals distributions.
Bottom: $x$ and $y$ pulls distributions.
The data in the plots are given for the global coordinates and inclusively for all layers together. 
The per-layer estimations yield similar characteristics and hence the data are merged to enhance statistics.}
\label{fig:kf_res_pulls}
\end{figure*}

Fig.~\ref{fig:kf_res_pulls} shows the track residuals and pulls distributions.
The standard deviations of the two residual distributions (both axes) indicate a spatial resolution slightly below the expected $5~\mu{\rm m}$ resolution of the ALPIDE sensors and the pull distributions are very close to normal, with standard deviations slightly below 1.
The source of these small incompatibilities are traced to slight overestimation of the measurements (clusters) uncertainties.
This overestimation can be attributed to residual mismodelling of the material in ACTS (see App.~\ref{app:material_mapping}) and the digitization step.

\begin{figure}[pos=!ht]
\centering
\begin{overpic}[width=0.48\textwidth]{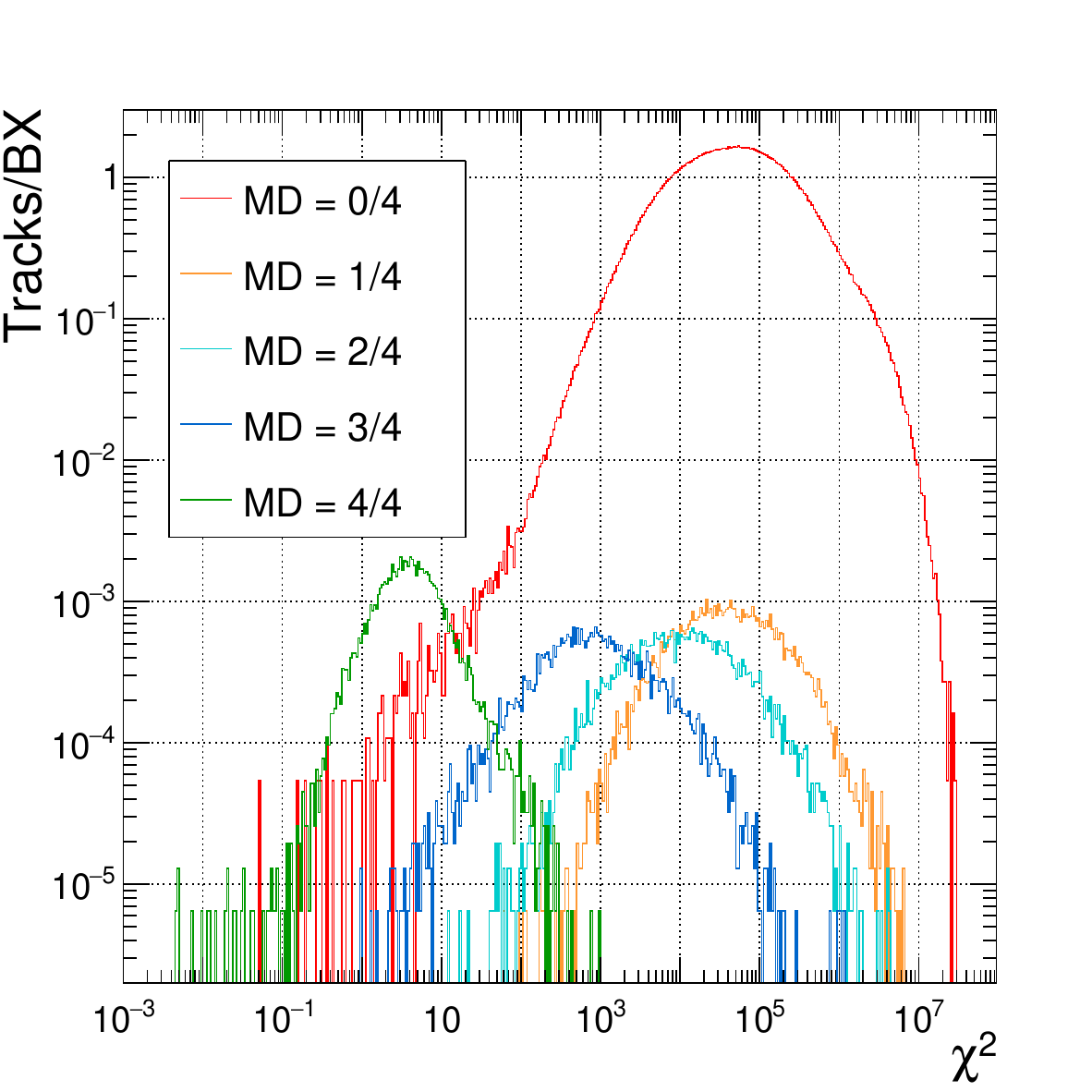}\end{overpic}
\caption{
The $\chi^2$ distributions for fitted tracks of different matching degrees. The distribution is normalized to the $a_0=5$ scenario.}
\label{fig:kf_chi2}
\end{figure}

Fig.~\ref{fig:kf_chi2} presents the $\chi^2$ distributions for fitted tracks categorized by their matching degree in the KF step.
The majority of non-fully matched tracks (matching degree below 4/4) are peaking at very high values, around $\chi^2 \sim 10^3\textsf{--}10^5$, while the fully matched ones are peaking below $\chi^2 \sim 10$.
However, some non-fully matched tracks are still found in the $\chi^2 < 10$ range.
These tracks closely resemble the signal positrons and present the greatest challenge for rejection.

\subsection{Track Selection}
\label{sec:acts_track_analysis}
A series of selection cuts is applied to eliminate poorly reconstructed or non-signal-like tracks.
The cuts are primarily designed to suppress the background rate relative to the signal.
That is, the number of background tracks reconstructed per BX should be reduced to at least an order of magnitude below the expected number of signal tracks reconstructed per BX.

Using the KF-fitted tracks' parameters at the IP, the following selection criteria are applied.
The tracks' $\chi^2$ (as seen in Fig.~\ref{fig:kf_chi2}) must be below 5 and its momentum along the $x$-axis at the IP must be in $|p^{\rm IP}_x|<5.5~{\rm MeV}$ (according to the expected beam divergence and the transverse momentum transfer from the laser~\cite{PhysRevD.93.085028}).
The other cuts listed below are simply chosen based on maximizing the signal-to-background ratio in terms of tracks counting.
The track vertex ($x,y$) significance along $x$ and $y$ must be in $|x^{\rm IP}/\sigma_{x^{\rm IP}}|<0.45$ and in $|y^{\rm IP}/\sigma_{y^{\rm IP}}|<0.5$.
The track azimuth and polar angle significances must be in $|\phi^{\rm IP}/\sigma_{\phi^{\rm IP}}|<1$ and in $|\theta^{\rm IP}/\sigma_{\theta^{\rm IP}}|<1.2$.
Finally, the track's energy must be in the range between 2.0 and 3.9~GeV.
This range roughly corresponds to the vertical aperture between the top edge of the flange of the vacuum pipe close to the dipole exit and the bottom edge of the lowest sensor at the FTL.
This aperture is defined by the specific magnets' settings (fixed during the NBW runs).
While tracks can be reconstructed even below this range, their quality is significantly reduced by scattering in the material and hence they are removed.

\begin{figure*}[pos=!ht]
\centering
\begin{overpic}[width=0.8\textwidth]{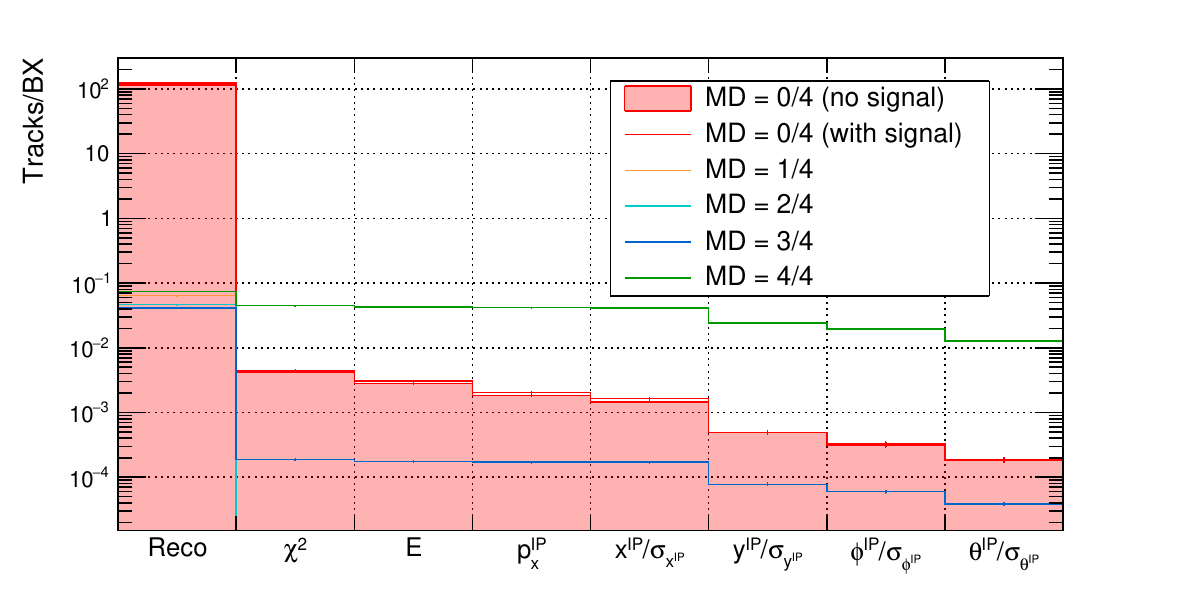}\end{overpic}
\caption{The cutflow of the post-reconstruction analysis chain for pure background and signal-plus-background samples.
The statistical uncertainties are indicated. The cutflow is normalized to the $a_0=5$ scenario.}
\label{fig:cutflows}
\end{figure*}

Fig.~\ref{fig:cutflows} presents the cutflows for the pure-background and signal-plus-background samples.
To evaluate the impact of the signal presence on the background reconstruction rate, a sample of $\sim 1.85 \times 10^4$ FullSim NBW positrons is embedded (following the same procedure as outlined in Sec.~\ref{sec:datasets}) into the detector 50 times.
The FastSim background is generated independently for each NBW positron, irrespective of repetition.
Consequently, the signal-plus-background cutflow is assessed on a dataset of $\sim10^6$ BXs, where each BX contains a uniquely sampled background, and NBW positrons are reintroduced every $\sim1.85 \times 10^4$ BXs.

As mentioned at the end of Sec.~\ref{sec:datasets}, for evaluation of the reconstruction rates we relax the acceptance requirement 
on the number of signal measurements in the detector from exactly four to at least one.
The definition of the matching degree remains effectively the same as for the track-finding stage (see Sec.~\ref{sec:acts_track_finding}).
The impact of this acceptance relaxation on the reconstruction rates, discussed below, is found to be negligible. 

For comparison, the cutflow estimated on the pure-background sample is shown alongside the signal-plus-background cutflow.
The reconstruction rate for pure background tracks is $(1.87\pm0.13)\times10^{-4}$ tracks per BX and $(1.82\pm0.13)\times10^{-4}$ tracks per BX in the signal-plus-background case (matching degree 0/4).
There are no tracks with matching degrees 1/4 and 2/4 that survive to the end of the selection process.
The rate for matching degree 3/4, assuming $a_0=5$, is $(3.84\pm0.22)\times10^{-5}$ tracks per BX.
Finally, the rate for signal (matching degree 4/4) under the same $a_0$ assumption is $(12.80\pm0.04)\times10^{-3}$ tracks per BX.
This is almost two orders of magnitude higher than the pure background and the matching degree 0/4 background (in the signal-plus-background sample) rates.

\begin{figure}[pos=!ht]
\centering
\begin{overpic}[width=0.48\textwidth]{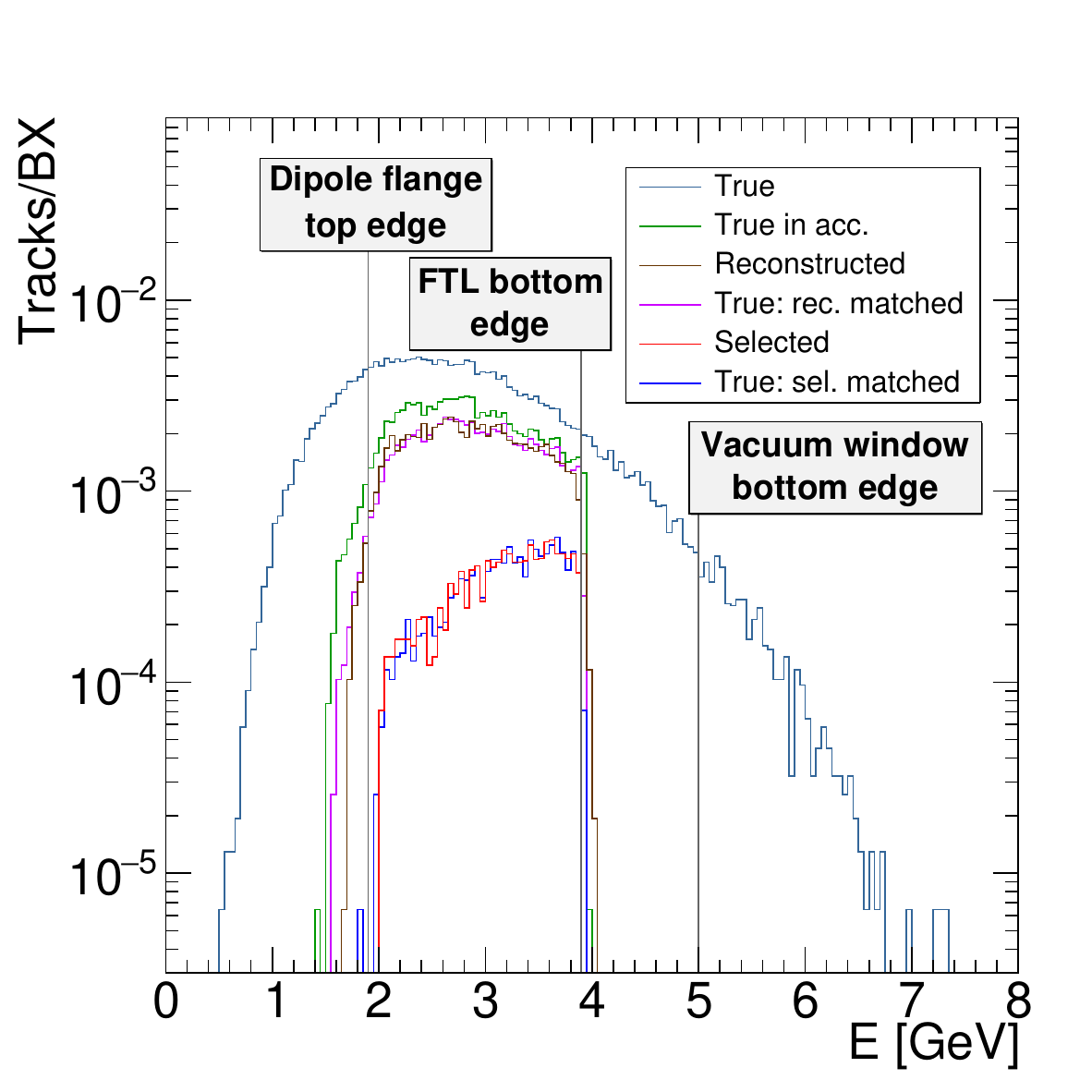}\end{overpic}
\caption{A comparison of the true and the reconstructed NBW positron energy before and after the cuts are applied, including the uncut true energy spectrum as given at the IP by
\PTARMIGAN.
The vertical lines mark the physical cutoffs (the dipole flange, the sensor's bottom edge, and the vacuum exit window's bottom edge). The distribution carries the $a_0=10$ scenario and is normalized to the $a_0=5$ one. The shapes for $a_0=5$ are only slightly harder (see Appendix~\ref{app:a0510signal}).}
\label{fig:e_spectrum}
\end{figure}

Fig.~\ref{fig:e_spectrum} shows the comparison between the true and the estimated spectrum of the NBW positrons before and after application of the cuts.
The estimated energies are in agreement with the true ones in general, particularly after the cuts.
The complete ``true'' energy spectrum at the IP is also shown without any selection as a reference to illustrate the overall losses.
We note that although these spectra are normalized to the $a_0=5$ scenario as discussed earlier, their shapes corresponds to the $a_0=10$ simulation.
Generally, the NBW spectra for the former (or any $4 \lesssim a_0<10$ scenario) would be slightly harder, but the evolution seen in Fig.~\ref{fig:e_spectrum} will be effectively the same.

\begin{figure}[pos=!ht]
\centering
\begin{overpic}[width=0.48\textwidth]{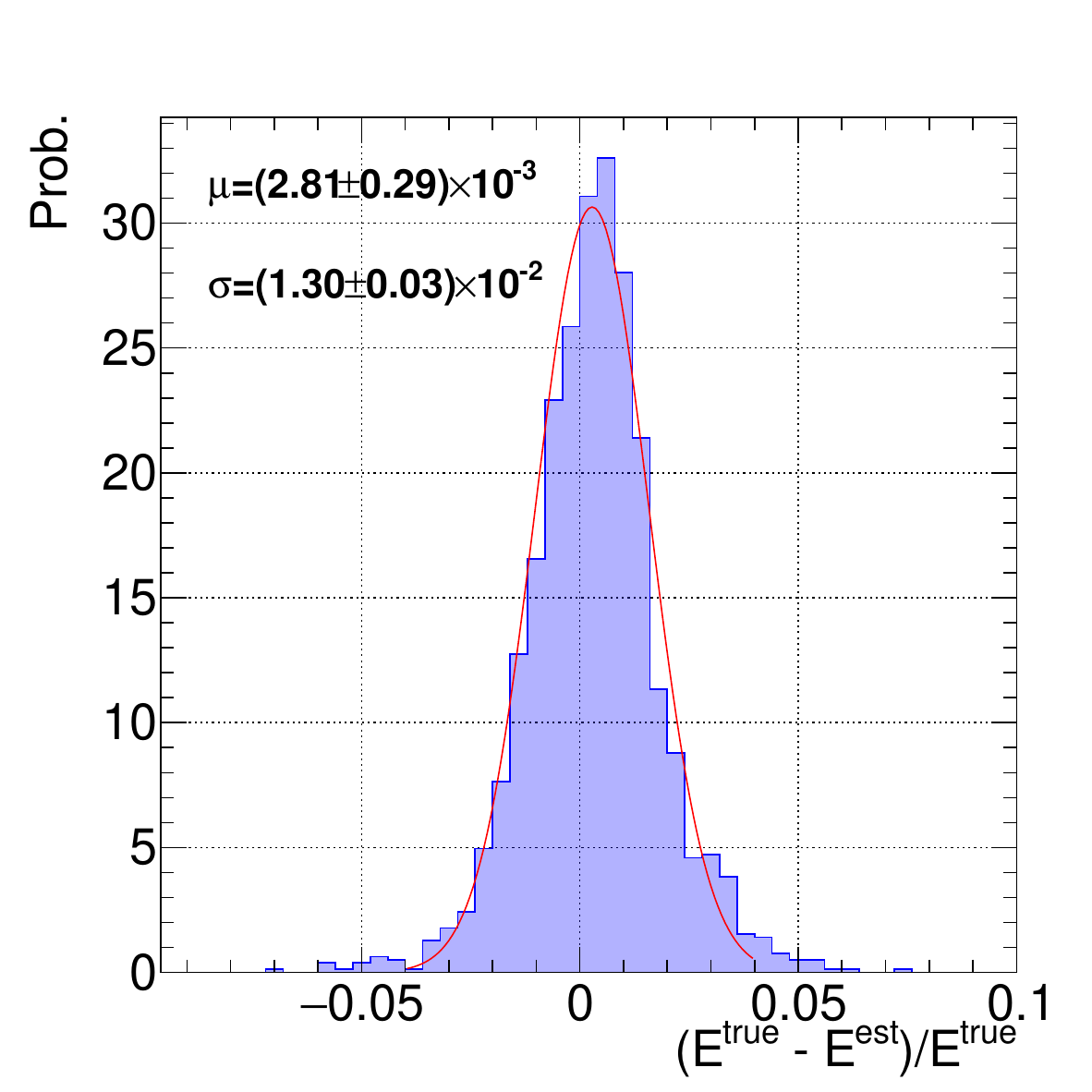}\end{overpic}
\caption{The energy response for reconstructed NBW positron energy after the cuts are applied with a Gaussian fit. Only the 4/4 matching degree tracks are considered.}
\label{fig:e_response}
\end{figure}

To complete the discussion, and since some of the distributions seen in Fig.~\ref{fig:e_spectrum} are given at the truth level, the reconstructed energy response for the matching degree 4/4 tracks passing the cuts is shown in Fig.~\ref{fig:e_response}.
The contribution of the 3/4 category is negligible and follows the same trend.
A Gaussian fit to the response shape is performed and the energy resolution, as estimated from the quadrature of the mean and width of the Gaussian shape, is $\sim 1.33\%$.

\subsection{Uncertainties}
\label{sec:uncertaintis}
This study is focused on the question whether it is possible to unambiguously measure single positrons amid large backgrounds per BX.
Hence, we need to also consider cases of potential  underestimation or overestimation of the background, as discussed below.
Other sources of uncertainties, such as the residual misalignment of the detector elements to themselves and to the beamline elements, the magnetic fields knowledge, etc., are not considered here.
The impact of these other sources is expected to be less significant, and we defer their discussion to future work.

\begin{figure}[pos=!ht]
\centering
\begin{overpic}[width=0.48\textwidth]{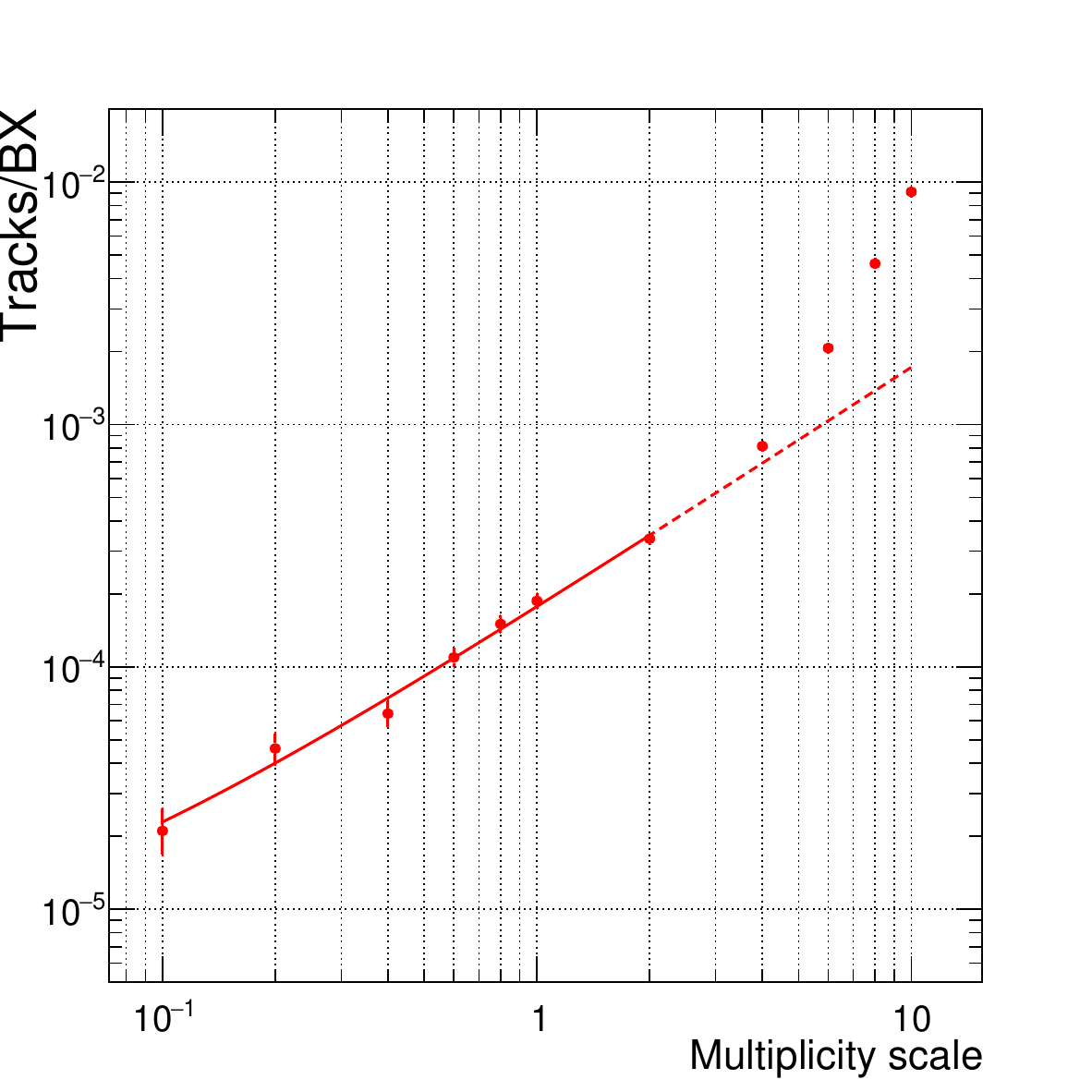}\end{overpic}
\caption{
The background rate (number of background tracks per BX) as a function of the background particles scaling factor.
The latter is a multiplier of the numbers of particles generated with the FastSim discussed in Sec.~\ref{sec:datasets}.
A linear fit in the range between 0.1 and 2 is shown as a solid red line.
The fit extrapolation in the range between 2.0 and 10.0 is shown as a dashed red line.
The point where the scaling factor is equal to 1 corresponds to the conservative $a_0=10$  background scenario.}
\label{fig:bkg_rate_with_scale}
\end{figure}

To evaluate the robustness of the reconstruction and selection pipeline under scenarios, where the background is significantly underestimated or overestimated, additional background-only samples are generated.
Each such sample contains $10^6$ BXs, where, using the same underlying distributions, the number of background particles in each BX (see Sec.~\ref{sec:datasets}) of a given sample is scaled by a factor between 0.1 and 10.
These samples are analyzed exactly as the nominal one, and the background rate is plotted against the scaling factor.
The result is shown in Fig.~\ref{fig:bkg_rate_with_scale}. 
The background rate (number of pure background reconstructed tracks per BX) varies from $2.1 \times 10^{-5}$ to $9.1 \times 10^{-3}$.
The scaling is linear in the range of scaling factors up to $\sim 3$, as shown by the linear fit.
The goodness of fit for the linear assumption up to a scale factor of 2 is $\chi^2/N_{\rm DoF}=0.78$, whereas the corresponding fit up to a scale factor of 4 is significantly worse.
The background reconstruction rate at 10 times the nominal case is still lower (by a factor of $\sim 0.7$) than the signal reconstruction rate under the assumption of $a_0=5$.
This indicates that the tracking pipeline remains effective even under significantly elevated background conditions, well beyond the already conservative baseline estimates taken here.

\section{Outlook}
\label{sec:outlook}
We present new hardware and software developments for SF-QED experiments, reflecting a paradigm shift toward advanced particle-detection techniques in current and future studies.
Specifically, we outline a strategy for unambiguous detection of single NBW positrons in E320, using a systematically conservative estimate of the overwhelming background.
At the already achievable $a_0=5$ benchmark, the E320 experiment is expected to produce $\sim$0.168 NBW $e^+e^-$ pairs per collision, out of them $\sim71\%$ positrons per collision will have at least one pixel cluster the detector, and $\sim 1.28\times 10^{-2}$ of them will be reconstructed and selected as good tracks (with an energy resolution of $\sim 1.33\%$). 
The reconstruction efficiency ($\sim 65\%$) combined with the relatively low selection efficiency ($\sim 15\%$) allows to still measure the signal, while benchmarking the error on the signal rate estimate due to the overwhelming background particles at only $\leq 1.87\times 10^{-4}$. 
Thus, with an effective collision rate of 5~Hz, this will lead to a statistically significant number of NBW positrons even in a single $\lesssim 1$~hour run.
Besides demonstrating that the rate of background positron-like tracks can be kept well below the expected signal rate for characteristic laser intensities of $a_0=4\textsf{--}5$ in upcoming E320 campaigns, we show that this statement still holds even if the background fluxes are significantly underestimated.

Although not discussed in detail here, the same approach applies to much larger NBW positron multiplicities per collision.
This aspect is partially covered in the LUXE technical design report~\cite{LUXETDR}, and the algorithms described here are also being adopted by LUXE.

Electron–laser collision data are already being collected in E320 using a prototype tracking detector based on the same technology, albeit with smaller acceptance than the detector shown in Figs.~\ref{fig:detector_default} and~\ref{fig:detector_tilted}.
While tracking detectors have been a common practice in high-energy physics for decades, this is the first time such a solution has been used in a SF-QED experiment.
Given the rapidly expanding landscape of dedicated SF-QED experiments worldwide -- with several multi-PW facilities already online -- we expect this or similar solutions to be adopted elsewhere.

\section*{Acknowledgments}

We thank the entire E320 collaboration as well as the FACET-II facility for their input and support.
FACET-II is supported by the U.S. Department of Energy under Contract No. DE-AC02-76SF00515.

We also thank the colleagues from the ALICE ITS project for useful discussions and help
related to the ALPIDE sensors. We would like to especially thank Luciano Musa, Gianluca Aglieri
Rinella, Antonello Di Mauro, Magnus Mager, Corrado Gargiulo, Felix Reidt, Ivan Ravasenga, Ruben
Shahoyan and Walter Snoeys.
We also wish to thank Simon Spannagel and Paul Sch\"{u}tze for the
detailed and dedicated help and for the useful discussions about Allpix$^2$.
Finally, we want to thank the ACTS developers Andreas Salzburger, Paul Gessinger, and Andreas Stefl for their help and support with implementation and extension of the track reconstruction pipeline.

The work of Noam Tal Hod's group  at Weizmann is supported by a research grant from the Estate of Dr.\ Moshe Gl\"{u}ck, the Minerva foundation with funding from the Federal German Ministry for Education and Research, the ISRAEL SCIENCE FOUNDATION (grant No. 708/20 and 1235/24), the Anna and Maurice Boukstein Career Development Chair, the Benoziyo Endowment Fund for the Advancement of Science, the Estate of Emile Mimran, the Estate of Betty Weneser, a research grant from the Estate of Gerald Alexander, a research grant from the Potter's Wheel Foundation, a research grant from Adam Glickman and the Sassoon \& Marjorie Peress Legacy Fund, the Deloro Center for Space and Optics, and in part by the Krenter-Perinot center for High-Energy particle physics.

S.\ Meuren is supported by the Agence Nationale de la Recherche (ANR) under the Chaire de Professeur Junior (CPJ) program. Furthermore, he is grateful for support from the visitor program of the Stanford PULSE institute.

D.\ A.\ Reis is supported by the U.S. Department of Energy Office of Science, Office of Fusion Energy Sciences under award DE-SC0020076.



\bibliography{bibliography.bib}

@article{ATLAS:2020hii,
    author = "{ATLAS Collaboration}",
    collaboration = "ATLAS",
    title = "{Measurement of light-by-light scattering and search for axion-like particles with 2.2 nb$^{-1}$ of Pb+Pb data with the ATLAS detector}",
    eprint = "2008.05355",
    archivePrefix = "arXiv",
    primaryClass = "hep-ex",
    reportNumber = "CERN-EP-2020-135",
    doi = "10.1007/JHEP03(2021)243",
    journal = "JHEP",
    volume = "03",
    pages = "243",
    year = "2021",
    note = "[Erratum: JHEP 11, 050 (2021)]"
}

@article{CMS:2024bnt,
    author = "{CMS Collaboration}",
    collaboration = "CMS",
    title = "{Measurement of light-by-light scattering and the Breit-Wheeler process, and search for axion-like particles in ultraperipheral PbPb collisions at $\sqrt{s_\mathrm{NN}}$ = 5.02 TeV}",
    eprint = "2412.15413",
    archivePrefix = "arXiv",
    primaryClass = "nucl-ex",
    reportNumber = "CMS-HIN-21-015, CERN-EP-2024-284",
    month = "12",
    year = "2024"
}

@article{PhysRevLett.94.114801,
  title = {Clocking Femtosecond X Rays},
  author = {Cavalieri, A. L. and Fritz, D. M. and Lee, S. H. and Bucksbaum, P. H. and Reis, D. A. and Rudati, J. and Mills, D. M. and Fuoss, P. H. and Stephenson, G. B. and Kao, C. C. and Siddons, D. P. and Lowney, D. P. and MacPhee, A. G. and Weinstein, D. and Falcone, R. W. and Pahl, R. and Als-Nielsen, J. and Blome, C. and D\"usterer, S. and Ischebeck, R. and Schlarb, H. and Schulte-Schrepping, H. and Tschentscher, Th. and Schneider, J. and Hignette, O. and Sette, F. and Sokolowski-Tinten, K. and Chapman, H. N. and Lee, R. W. and Hansen, T. N. and Synnergren, O. and Larsson, J. and Techert, S. and Sheppard, J. and Wark, J. S. and Bergh, M. and Caleman, C. and Huldt, G. and van der Spoel, D. and Timneanu, N. and Hajdu, J. and Akre, R. A. and Bong, E. and Emma, P. and Krejcik, P. and Arthur, J. and Brennan, S. and Gaffney, K. J. and Lindenberg, A. M. and Luening, K. and Hastings, J. B.},
  journal = {Phys. Rev. Lett.},
  volume = {94},
  issue = {11},
  pages = {114801},
  numpages = {4},
  year = {2005},
  month = {3},
  publisher = {American Physical Society},
  doi = {10.1103/PhysRevLett.94.114801},
  url = {https://link.aps.org/doi/10.1103/PhysRevLett.94.114801}
}

@inbook{doi:10.1142/9789811279461_0008,
	author = {L. V. Keldysh},
	booktitle = {Selected Papers of Leonid V Keldysh},
	doi = {10.1142/9789811279461_0008},
	eprint = {https://www.worldscientific.com/doi/pdf/10.1142/9789811279461_0008},
	pages = {56-63},
	title = {Ionization in the field of a strong electromagnetic wave},
	url = {https://www.worldscientific.com/doi/abs/10.1142/9789811279461_0008}}

@article{PhysRevD.106.013010,
  title = {High-resolution modeling of nonlinear Compton scattering in focused laser pulses},
  author = {Nielsen, C. F. and Holtzapple, R. and King, B.},
  journal = {Phys. Rev. D},
  volume = {106},
  issue = {1},
  pages = {013010},
  numpages = {15},
  year = {2022},
  month = {7},
  publisher = {American Physical Society},
  doi = {10.1103/PhysRevD.106.013010},
  url = {https://link.aps.org/doi/10.1103/PhysRevD.106.013010}
}

@article{RevModPhys.94.045001,
  title = {Charged particle motion and radiation in strong electromagnetic fields},
  author = {Gonoskov, A. and Blackburn, T. G. and Marklund, M. and Bulanov, S. S.},
  journal = {Rev. Mod. Phys.},
  volume = {94},
  issue = {4},
  pages = {045001},
  numpages = {63},
  year = {2022},
  month = {10},
  publisher = {American Physical Society},
  doi = {10.1103/RevModPhys.94.045001},
  url = {https://link.aps.org/doi/10.1103/RevModPhys.94.045001}
}

@article{PhysRevD.107.096004,
  title = {Locally monochromatic two-step nonlinear trident process in a plane wave},
  author = {Tang, S. and King, B.},
  journal = {Phys. Rev. D},
  volume = {107},
  issue = {9},
  pages = {096004},
  numpages = {14},
  year = {2023},
  month = {5},
  publisher = {American Physical Society},
  doi = {10.1103/PhysRevD.107.096004},
  url = {https://link.aps.org/doi/10.1103/PhysRevD.107.096004}
}

@article{PhysRevD.101.056017,
  title = {Trident process in laser pulses},
  author = {Dinu, Victor and Torgrimsson, Greger},
  journal = {Phys. Rev. D},
  volume = {101},
  issue = {5},
  pages = {056017},
  numpages = {19},
  year = {2020},
  month = {3},
  publisher = {American Physical Society},
  doi = {10.1103/PhysRevD.101.056017},
  url = {https://link.aps.org/doi/10.1103/PhysRevD.101.056017}
}

@article{PhysRevLett.105.080401,
  title = {Complete QED Theory of Multiphoton Trident Pair Production in Strong Laser Fields},
  author = {Hu, Huayu and M\"uller, Carsten and Keitel, Christoph H.},
  journal = {Phys. Rev. Lett.},
  volume = {105},
  issue = {8},
  pages = {080401},
  numpages = {4},
  year = {2010},
  month = {8},
  publisher = {American Physical Society},
  doi = {10.1103/PhysRevLett.105.080401},
  url = {https://link.aps.org/doi/10.1103/PhysRevLett.105.080401}
}

@article{di_piazza_extremely_2012,
	title	= {{E}xtremely high-intensity laser interactions with fundamental quantum systems},
	volume	= {84},
	doi	= {10.1103/RevModPhys.84.1177},
	journal	= {Rev. Mod. Phys.},
	author	= {Di Piazza, A. and M\"uller, C. and Hatsagortsyan, K. Z. and Keitel, C. H.},
	year	= {2012},
	pages	= {1177--1228},
}

@article{PhysRevD.93.085028,
  title = {Semiclassical picture for electron-positron photoproduction in strong laser fields},
  author = {Meuren, Sebastian and Keitel, Christoph H. and Di Piazza, Antonino},
  journal = {Phys. Rev. D},
  volume = {93},
  issue = {8},
  pages = {085028},
  numpages = {17},
  year = {2016},
  month = {4},
  publisher = {American Physical Society},
  doi = {10.1103/PhysRevD.93.085028},
  url = {https://link.aps.org/doi/10.1103/PhysRevD.93.085028}
}

@article{PhysRevD.91.013009,
  title = {Polarization-operator approach to pair creation in short laser pulses},
  author = {Meuren, Sebastian and Hatsagortsyan, Karen Z. and Keitel, Christoph H. and Di Piazza, Antonino},
  journal = {Phys. Rev. D},
  volume = {91},
  issue = {1},
  pages = {013009},
  numpages = {20},
  year = {2015},
  month = {1},
  publisher = {American Physical Society},
  doi = {10.1103/PhysRevD.91.013009},
  url = {https://link.aps.org/doi/10.1103/PhysRevD.91.013009}
}

@article{Ritus,
	author = {Ritus, V.  I. },
	doi = {10.1007/BF01120220},
	id = {Ritus1985},
	journal = {Journal of Soviet Laser Research},
	number = {5},
	pages = {497--617},
	title = {Quantum effects of the interaction of elementary particles with an intense electromagnetic field},
	url = {https://doi.org/10.1007/BF01120220},
	volume = {6},
	year = {1985}
}

@article{Storey:2023wlu,
    author = "Storey, Douglas and others",
    title = "{Status and first results from FACET-II towards the demonstration of plasma wakefield acceleration, coherent radiation generation, and probing strong-field QED}",
    doi = "10.18429/JACoW-IPAC2023-TUPA104",
    journal = "JACoW",
    volume = "IPAC2023",
    pages = "TUPA104",
    year = "2023"
}

@article{mourou_optics_2006,
  title = {Optics in the relativistic regime},
  author = {Mourou, Gerard A. and Tajima, Toshiki and Bulanov, Sergei V.},
  journal = {Rev. Mod. Phys.},
  volume = {78},
  issue = {2},
  pages = {309--371},
  year = {2006},
  month = {4},
  doi = {10.1103/RevModPhys.78.309},
}

@article{yan_high_2017,
  title = {High‐order multiphoton Thomson scattering},
  author = {Yan, Wenchao and Fruhling, Colton and Golovin, Grigory and Haden, Daniel and Luo, Ji and Zhang, P. and Zhao, B. and Zhang, J. and Liu, C. and Chen, Minsi and Chen, Shilei and Banerjee, S. and Umstadter, Donald},
  journal = {Nature Photonics},
  volume = {11},
  number = {8},
  pages = {514--520},
  year = {2017},
  doi = {10.1038/nphoton.2017.100},
}

@article{PhysRevE.48.3003,
  title = {Nonlinear Thomson scattering of intense laser pulses from beams and plasmas},
  author = {Esarey, Eric and Ride, Sally K. and Sprangle, Phillip},
  journal = {Phys. Rev. E},
  volume = {48},
  issue = {4},
  pages = {3003--3021},
  numpages = {0},
  year = {1993},
  month = {10},
  publisher = {American Physical Society},
  doi = {10.1103/PhysRevE.48.3003},
  url = {https://link.aps.org/doi/10.1103/PhysRevE.48.3003}
}

@article{HUNTSTONE2021165210,
	author = {Keenan Hunt-Stone and Robert Ariniello and Christopher Doss and Valentina Lee and Mike Litos},
	doi = {https://doi.org/10.1016/j.nima.2021.165210},
	issn = {0168-9002},
	journal = {Nuclear Instruments and Methods in Physics Research Section A: Accelerators, Spectrometers, Detectors and Associated Equipment},
	pages = {165210},
	title = {Electro-optic sampling beam position monitor for relativistic electron beams},
	url = {https://www.sciencedirect.com/science/article/pii/S0168900221001947},
	volume = {999},
	year = {2021}}

@article{RevModPhys.77.1131,
  title = {The interaction of relativistic particles with strong crystalline fields},
  author = {Uggerh\o{}j, Ulrik I.},
  journal = {Rev. Mod. Phys.},
  volume = {77},
  issue = {4},
  pages = {1131--1171},
  numpages = {0},
  year = {2005},
  month = {10},
  publisher = {American Physical Society},
  doi = {10.1103/RevModPhys.77.1131},
  url = {https://link.aps.org/doi/10.1103/RevModPhys.77.1131}
}

@article{Sauter,
	author = {Sauter, Fritz},
	doi = {10.1007/BF01339461},
	id = {Sauter1931},
	journal = {Zeitschrift f{\"u}r Physik},
	number = {11},
	pages = {742--764},
	title = {{\"U}ber das Verhalten eines Elektrons im homogenen elektrischen Feld nach der relativistischen Theorie Diracs},
	url = {https://doi.org/10.1007/BF01339461},
	volume = {69},
	year = {1931}
}

@article{PhysRevLett.134.085001,
  title = {Experimental Generation of Extreme Electron Beams for Advanced Accelerator Applications},
  author = {Emma, C. and Majernik, N. and Swanson, K. K. and Ariniello, R. and Gessner, S. and Hessami, R. and Hogan, M. J. and Knetsch, A. and Larsen, K. A. and Marinelli, A. and O'Shea, B. and Perez, S. and Rajkovic, I. and Robles, R. and Storey, D. and Yocky, G.},
  journal = {Phys. Rev. Lett.},
  volume = {134},
  issue = {8},
  pages = {085001},
  numpages = {7},
  year = {2025},
  month = {2},
  publisher = {American Physical Society},
  doi = {10.1103/PhysRevLett.134.085001},
  url = {https://link.aps.org/doi/10.1103/PhysRevLett.134.085001}
}

@article{PhysRevAccelBeams.27.051302,
  title = {Wakefield generation in hydrogen and lithium plasmas at FACET-II: Diagnostics and first beam-plasma interaction results},
  author = {Storey, D. and Zhang, C. and San Miguel Claveria, P. and Cao, G. J. and Adli, E. and Alsberg, L. and Ariniello, R. and Clarke, C. and Corde, S. and Dalichaouch, T. N. and Doss, C. E. and Ekerfelt, H. and Emma, C. and Gerstmayr, E. and Gessner, S. and Gilljohann, M. and Hast, C. and Knetsch, A. and Lee, V. and Litos, M. and Loney, R. and Marsh, K. A. and Matheron, A. and Mori, W. B. and Nie, Z. and O'Shea, B. and Parker, M. and White, G. and Yocky, G. and Zakharova, V. and Hogan, M. J. and Joshi, C.},
  journal = {Phys. Rev. Accel. Beams},
  volume = {27},
  issue = {5},
  pages = {051302},
  numpages = {15},
  year = {2024},
  month = {5},
  publisher = {American Physical Society},
  doi = {10.1103/PhysRevAccelBeams.27.051302},
  url = {https://link.aps.org/doi/10.1103/PhysRevAccelBeams.27.051302}
}

@article{PhysRevLett.130.071601,
  title = {Precision Measurement of Trident Production in Strong Electromagnetic Fields},
  author = {Nielsen, Christian F. and Holtzapple, Robert and Lund, Mads M. and Surrow, Jeppe H. and S\o{}rensen, Allan H. and S\o{}rensen, Marc B. and Uggerh\o{}j, Ulrik I.},
  collaboration = {CERN NA63},
  journal = {Phys. Rev. Lett.},
  volume = {130},
  issue = {7},
  pages = {071601},
  numpages = {5},
  year = {2023},
  month = {2},
  publisher = {American Physical Society},
  doi = {10.1103/PhysRevLett.130.071601},
  url = {https://link.aps.org/doi/10.1103/PhysRevLett.130.071601}
}

@article{HeliumLikeUranium,
	author = {Loetzsch, R. and Beyer, H. F. and Duval, L. and Spillmann, U. and Bana{\'s}, D. and Dergham, P. and Kr{\"o}ger, F. M. and Glorius, J. and Grisenti, R. E. and Guerra, M. and Gumberidze, A. and He{\ss}, R. and Hillenbrand, P. -M. and Indelicato, P. and Jagodzinski, P. and Lamour, E. and Lorentz, B. and Litvinov, S. and Litvinov, Yu. A. and Machado, J. and Paul, N. and Paulus, G. G. and Petridis, N. and Santos, J. P. and Scheidel, M. and Sidhu, R. S. and Steck, M. and Steydli, S. and Szary, K. and Trotsenko, S. and Uschmann, I. and Weber, G. and St{\"o}hlker, Th. and Trassinelli, M.},
	doi = {10.1038/s41586-023-06910-y},
	id = {Loetzsch2024},
	journal = {Nature},
	number = {7996},
	pages = {673--678},
	title = {Testing quantum electrodynamics in extreme fields using helium-like uranium},
	url = {https://doi.org/10.1038/s41586-023-06910-y},
	volume = {625},
	year = {2024}
}

@article{Matheron:2024hwy,
    author = "Matheron, Aim\'e and Andriyash, Igor and Davoine, Xavier and Gremillet, Laurent and Pouyez, Mattys and Grech, Mickael and Lancia, Livia and Phuoc, Kim Ta and Corde, S\'ebastien",
    title = "{Self-triggered strong-field QED collisions in laser-plasma interaction}",
    eprint = "2408.13238",
    archivePrefix = "arXiv",
    primaryClass = "physics.plasm-ph",
    month = "8",
    year = "2024"
}

@article{PhysRevLett.132.175002,
  title = {Light-Matter Interaction near the Schwinger Limit Using Tightly Focused Doppler-Boosted Lasers},
  author = {Za\"{i}m, Ne\"{i}l and Sainte-Marie, Antonin and Fedeli, Luca and Bartoli, Pierre and Huebl, Axel and Leblanc, Adrien and Vay, Jean-Luc and Vincenti, Henri},
  journal = {Phys. Rev. Lett.},
  volume = {132},
  issue = {17},
  pages = {175002},
  numpages = {7},
  year = {2024},
  month = {4},
  publisher = {American Physical Society},
  doi = {10.1103/PhysRevLett.132.175002},
  url = {https://link.aps.org/doi/10.1103/PhysRevLett.132.175002}
}

@article{corels_compton_2024,
	author = {Mirzaie, Mohammad and Hojbota, Calin Ioan and Kim, Do Yeon and Pathak, Vishwa Bandhu and Pak, Tae Gyu and Kim, Chul Min and Lee, Hwang Woon and Yoon, Jin Woo and Lee, Seong Ku and Rhee, Yong Joo and Vranic, Marija and Amaro, {\'O}scar and Kim, Ki Yong and Sung, Jae Hee and Nam, Chang Hee},
	doi = {10.1038/s41566-024-01550-8},
	id = {Mirzaie2024},
	journal = {Nature Photonics},
	number = {11},
	pages = {1212--1217},
	title = {All-optical nonlinear Compton scattering performed with a multi-petawatt laser},
	url = {https://doi.org/10.1038/s41566-024-01550-8},
	volume = {18},
	year = {2024}
}

@article{altarelli2007european,
  title={The European X-ray free-electron laser. Technical design report},
  author={Altarelli, Massimo and Brinkmann, Reinhard and Chergui, Majed},
  year={2007}
}

@article{weber2017p3,
  title={P3: An installation for high-energy density plasma physics and ultra-high intensity laser--matter interaction at ELI-Beamlines},
  author={Weber, Stefan and Bechet, S and Borneis, S and Brabec, L and Bu{\v{c}}ka, M and Chacon-Golcher, E and Ciappina, M and DeMarco, M and Fajstavr, A and Falk, K and others},
  journal={Matter and Radiation at Extremes},
  volume={2},
  number={4},
  pages={149--176},
  year={2017},
  publisher={AIP Publishing}
}

@article{gales2018extreme,
  title={The extreme light infrastructure—nuclear physics (ELI-NP) facility: new horizons in physics with 10 PW ultra-intense lasers and 20 MeV brilliant gamma beams},
  author={Gales, S and Tanaka, KA and Balabanski, DL and Negoita, F and Stutman, D and Tesileanu, O and Ur, CA and Ursescu, D and Andrei, I and Ataman, S and others},
  journal={Reports on Progress in Physics},
  volume={81},
  number={9},
  pages={094301},
  year={2018},
  publisher={IOP Publishing}
}

@article{papadopoulos2016apollon,
  title={The Apollon 10 PW laser: experimental and theoretical investigation of the temporal characteristics},
  author={Papadopoulos, DN and Zou, JP and Le Blanc, C and Ch{\'e}riaux, G and Georges, Patrick and Druon, F and Mennerat, G and Ramirez, P and Martin, L and Fr{\'e}neaux, A and others},
  journal={High Power Laser Science and Engineering},
  volume={4},
  pages={e34},
  year={2016},
  publisher={Cambridge University Press}
}

@article{yoon2021realization,
  title={Realization of laser intensity over 1023 W/cm2},
  author={Yoon, Jin Woo and Kim, Yeong Gyu and Choi, Il Woo and Sung, Jae Hee and Lee, Hwang Woon and Lee, Seong Ku and Nam, Chang Hee},
  journal={Optica},
  volume={8},
  number={5},
  pages={630--635},
  year={2021},
  publisher={Optical Society of America}
}

@article{PhysRev.82.664,
  title = {On Gauge Invariance and Vacuum Polarization},
  author = {Schwinger, J.},
  journal = {Phys. Rev.},
  volume = {82},
  issue = {5},
  pages = {664--679},
  numpages = {0},
  year = {1951},
  month = {6},
  publisher = {American Physical Society},
  doi = {10.1103/PhysRev.82.664},
  url = {https://link.aps.org/doi/10.1103/PhysRev.82.664}
}

@article{Ruffini:2009hg,
    author = "Ruffini, R. and Vereshchagin, G. and Xue, S.",
    title = "{Electron-positron pairs in physics and astrophysics: from heavy nuclei to black holes}",
    doi = "10.1016/j.physrep.2009.10.004",
    journal = "Phys. Rept.",
    volume = "487",
    pages = "1",
    year = "2010"
}

@misc{nikishov62,
	author = "A. I. Nikishov",
	title = "{Absorption of high-energy photons in the universe}",
	howpublished ="{Sov. Phys. JETP 14, 393 (1962)}"
}

@article{Kouveliotou:1998ze,
      author         = "Kouveliotou, C. and others",
      title          = "{An X-ray pulsar with a superstrong magnetic field in the soft gamma-ray repeater SGR 1806-20.}",
      journal        = "Nature",
      volume         = "393",
      year           = "1998",
      pages          = "235",
      SLACcitation   = "%%CITATION = NATUA,393,235;%%"
}

@article{Harding_2006,
	doi = {10.1088/0034-4885/69/9/r03},
	url = {https://doi.org/10.1088/0034-4885/69/9/r03},
	year = 2006,
	volume = {69},
	number = {9},
	pages = {2631--2708},
	author = {A. K. Harding and D. Lai},
	title = {Physics of strongly magnetized neutron stars},
	journal = {Reports on Progress in Physics},
}

@article{Turolla:2015mwa,
    author = "Turolla, R. and Zane, S. and Watts, A.",
    title = "{Magnetars: the physics behind observations. A review}",
    doi = "10.1088/0034-4885/78/11/116901",
    journal = "Rept. Prog. Phys.",
    volume = "78",
    number = "11",
    pages = "116901",
    year = "2015"
}

@article{yakimenko2019prospect,
  title="{Prospect of Studying Nonperturbative {QED} with Beam-Beam Collisions}",
  author={Yakimenko, V and Meuren, S and Del Gaudio, F and Baumann, C and Fedotov, A and Fiuza, F and Grismayer, T and Hogan, M J and Pukhov, A and Silva, L O and others},
  journal={Phys. Rev. Lett.},
  volume={122},
  pages={190404},
  year={2019},
  publisher={APS},
  doi={https://doi.org/10.1103/PhysRevLett.122.190404}
}

@misc{Bucksbaum:2020,
  author = "Bucksbaum, P. H. and others",
  title = "{Probing QED Cascades and Pair Plasmas in Laboratory Experiments. LoI to Cosmic Frontier}",
  year = "2020",
  url = "https://www.snowmass21.org/docs/files/summaries/CF/SNOWMASS21-CF1-001.pdf",
  urldate = "2020-08-10"
}

@article{Akhmadaliev:2001ik,
    author = "Akhmadaliev, S. Z. and others",
    title = "{Experimental investigation of high-energy photon splitting in atomic fields}",
    eprint = "hep-ex/0111084",
    archivePrefix = "arXiv",
    reportNumber = "BUDKER-INP-2001-80",
    doi = "10.1103/PhysRevLett.89.061802",
    journal = "Phys. Rev. Lett.",
    volume = "89",
    pages = "061802",
    year = "2002"
}

@article{ivanov,
	author = {M. Y. Ivanov  and  M.   Spanner  and  O.   Smirnova},
	title = {Anatomy of strong field ionization},
	journal = {Journal of Modern Optics},
	volume = {52},
	number = {2-3},
	pages = {165},
	year  = {2005},
	publisher = {Taylor & Francis},
	doi = {10.1080/0950034042000275360},
	URL = {https://doi.org/10.1080/0950034042000275360}
}

@article{Blackburn:2021cuq,
    author = "Blackburn, T. G. and King, B.",
    title = "{Higher fidelity simulations of nonlinear Breit\textendash{}Wheeler pair creation in intense laser pulses}",
    eprint = "2108.10883",
    archivePrefix = "arXiv",
    primaryClass = "hep-ph",
    doi = "10.1140/epjc/s10052-021-09955-3",
    journal = "Eur. Phys. J. C",
    volume = "82",
    number = "1",
    pages = "44",
    year = "2022"
}

@article{Blackburn:2023mlo,
    author = "Blackburn, T. G. and King, B. and Tang, S.",
    title = "{Simulations of laser-driven strong-field QED with Ptarmigan: Resolving wavelength-scale interference and \ensuremath{\gamma}-ray polarization}",
    eprint = "2305.13061",
    archivePrefix = "arXiv",
    primaryClass = "hep-ph",
    doi = "10.1063/5.0159963",
    journal = "Phys. Plasmas",
    volume = "30",
    number = "9",
    pages = "093903",
    year = "2023"
}

@inproceedings{Chen:22,
booktitle = {Optica High-brightness Sources and Light-driven Interactions Congress 2022},
doi = {10.1364/HILAS.2022.HF4B.6},
journal = {Optica High-brightness Sources and Light-driven Interactions Congress 2022},
pages = {HF4B.6},
publisher = {Optica Publishing Group},
title = {Preparation of Strong-field QED Experiments at FACET-II},
url = {https://opg.optica.org/abstract.cfm?URI=HILAS-2022-HF4B.6},
year = {2022}
}

@article{ALLISON2016186,
	author = {J. Allison and K. Amako and J. Apostolakis and P. Arce and M. Asai and T. Aso and E. Bagli and A. Bagulya and S. Banerjee and G. Barrand and B.R. Beck and A.G. Bogdanov and D. Brandt and J.M.C. Brown and H. Burkhardt and Ph. Canal and D. Cano-Ott and S. Chauvie and K. Cho and G.A.P. Cirrone and G. Cooperman and M.A. Cort{\'e}s-Giraldo and G. Cosmo and G. Cuttone and G. Depaola and L. Desorgher and X. Dong and A. Dotti and V.D. Elvira and G. Folger and Z. Francis and A. Galoyan and L. Garnier and M. Gayer and K.L. Genser and V.M. Grichine and S. Guatelli and P. Gu{\`e}ye and P. Gumplinger and A.S. Howard and I. H{\v r}ivn{\'a}{\v c}ov{\'a} and S. Hwang and S. Incerti and A. Ivanchenko and V.N. Ivanchenko and F.W. Jones and S.Y. Jun and P. Kaitaniemi and N. Karakatsanis and M. Karamitros and M. Kelsey and A. Kimura and T. Koi and H. Kurashige and A. Lechner and S.B. Lee and F. Longo and M. Maire and D. Mancusi and A. Mantero and E. Mendoza and B. Morgan and K. Murakami and T. Nikitina and L. Pandola and P. Paprocki and J. Perl and I. Petrovi{\'c} and M.G. Pia and W. Pokorski and J.M. Quesada and M. Raine and M.A. Reis and A. Ribon and A. {Risti{\'c} Fira} and F. Romano and G. Russo and G. Santin and T. Sasaki and D. Sawkey and J.I. Shin and I.I. Strakovsky and A. Taborda and S. Tanaka and B. Tom{\'e} and T. Toshito and H.N. Tran and P.R. Truscott and L. Urban and V. Uzhinsky and J.M. Verbeke and M. Verderi and B.L. Wendt and H. Wenzel and D.H. Wright and D.M. Wright and T. Yamashita and J. Yarba and H. Yoshida},
	journal = {Nuclear Instruments and Methods in Physics Research Section A: Accelerators, Spectrometers, Detectors and Associated Equipment},
	pages = {186-225},
	title = {Recent developments in Geant4},
	volume = {835},
	year = {2016}
}

@ARTICLE{1610988,
  author={Allison, J. and Amako, K. and Apostolakis, J. and Araujo, H. and Arce Dubois, P. and Asai, M. and Barrand, G. and Capra, R. and Chauvie, S. and Chytracek, R. and Cirrone, G.A.P. and Cooperman, G. and Cosmo, G. and Cuttone, G. and Daquino, G.G. and Donszelmann, M. and Dressel, M. and Folger, G. and Foppiano, F. and Generowicz, J. and Grichine, V. and Guatelli, S. and Gumplinger, P. and Heikkinen, A. and Hrivnacova, I. and Howard, A. and Incerti, S. and Ivanchenko, V. and Johnson, T. and Jones, F. and Koi, T. and Kokoulin, R. and Kossov, M. and Kurashige, H. and Lara, V. and Larsson, S. and Lei, F. and Link, O. and Longo, F. and Maire, M. and Mantero, A. and Mascialino, B. and McLaren, I. and Mendez Lorenzo, P. and Minamimoto, K. and Murakami, K. and Nieminen, P. and Pandola, L. and Parlati, S. and Peralta, L. and Perl, J. and Pfeiffer, A. and Pia, M.G. and Ribon, A. and Rodrigues, P. and Russo, G. and Sadilov, S. and Santin, G. and Sasaki, T. and Smith, D. and Starkov, N. and Tanaka, S. and Tcherniaev, E. and Tome, B. and Trindade, A. and Truscott, P. and Urban, L. and Verderi, M. and Walkden, A. and Wellisch, J.P. and Williams, D.C. and Wright, D. and Yoshida, H.},
  journal={IEEE Transactions on Nuclear Science}, 
  title={Geant4 developments and applications}, 
  year={2006},
  volume={53},
  number={1},
  pages={270-278},
  doi={10.1109/TNS.2006.869826}
}

@article{AGOSTINELLI2003250,
	author = {S. Agostinelli and J. Allison and K. Amako and J. Apostolakis and H. Araujo and P. Arce and M. Asai and D. Axen and S. Banerjee and G. Barrand and F. Behner and L. Bellagamba and J. Boudreau and L. Broglia and A. Brunengo and H. Burkhardt and S. Chauvie and J. Chuma and R. Chytracek and G. Cooperman and G. Cosmo and P. Degtyarenko and A. Dell'Acqua and G. Depaola and D. Dietrich and R. Enami and A. Feliciello and C. Ferguson and H. Fesefeldt and G. Folger and F. Foppiano and A. Forti and S. Garelli and S. Giani and R. Giannitrapani and D. Gibin and J.J. {G{\'o}mez Cadenas} and I. Gonz{\'a}lez and G. {Gracia Abril} and G. Greeniaus and W. Greiner and V. Grichine and A. Grossheim and S. Guatelli and P. Gumplinger and R. Hamatsu and K. Hashimoto and H. Hasui and A. Heikkinen and A. Howard and V. Ivanchenko and A. Johnson and F.W. Jones and J. Kallenbach and N. Kanaya and M. Kawabata and Y. Kawabata and M. Kawaguti and S. Kelner and P. Kent and A. Kimura and T. Kodama and R. Kokoulin and M. Kossov and H. Kurashige and E. Lamanna and T. Lamp{\'e}n and V. Lara and V. Lefebure and F. Lei and M. Liendl and W. Lockman and F. Longo and S. Magni and M. Maire and E. Medernach and K. Minamimoto and P. {Mora de Freitas} and Y. Morita and K. Murakami and M. Nagamatu and R. Nartallo and P. Nieminen and T. Nishimura and K. Ohtsubo and M. Okamura and S. O'Neale and Y. Oohata and K. Paech and J. Perl and A. Pfeiffer and M.G. Pia and F. Ranjard and A. Rybin and S. Sadilov and E. {Di Salvo} and G. Santin and T. Sasaki and N. Savvas and Y. Sawada and S. Scherer and S. Sei and V. Sirotenko and D. Smith and N. Starkov and H. Stoecker and J. Sulkimo and M. Takahata and S. Tanaka and E. Tcherniaev and E. {Safai Tehrani} and M. Tropeano and P. Truscott and H. Uno and L. Urban and P. Urban and M. Verderi and A. Walkden and W. Wander and H. Weber and J.P. Wellisch and T. Wenaus and D.C. Williams and D. Wright and T. Yamada and H. Yoshida and D. Zschiesche},
	journal = {Nuclear Instruments and Methods in Physics Research Section A: Accelerators, Spectrometers, Detectors and Associated Equipment},
	number = {3},
	pages = {250-303},
	title = {Geant4---a simulation toolkit},
	volume = {506},
	year = {2003}
}

@article{PhysRevAccelBeams.22.101301,
  title = {FACET-II facility for advanced accelerator experimental tests},
  author = {Yakimenko, V. and Alsberg, L. and Bong, E. and Bouchard, G. and Clarke, C. and Emma, C. and Green, S. and Hast, C. and Hogan, M. J. and Seabury, J. and Lipkowitz, N. and O'Shea, B. and Storey, D. and White, G. and Yocky, G.},
  journal = {Phys. Rev. Accel. Beams},
  volume = {22},
  issue = {10},
  pages = {101301},
  numpages = {11},
  year = {2019},
  month = {10},
  doi = {10.1103/PhysRevAccelBeams.22.101301},
}

@InProceedings{Yakimenko:IPAC2016-TUOBB02,
  author       = {V. Yakimenko and others},
  title        = {{FACET-II} {A}ccelerator {R}esearch with {B}eams of {E}xtreme {I}ntensities},
  booktitle    = {Proc. of International Particle Accelerator Conference (IPAC'16),
                  Busan, Korea, May 8-13, 2016},
  pages        = {1067--1070},
  paper        = {TUOBB02},
  language     = {english},
  venue        = {Busan, Korea},
  series       = {International Particle Accelerator Conference},
  number       = {7},
  publisher    = {JACoW},
  address      = {Geneva, Switzerland},
  month        = {6},
  year         = {2016},
  isbn         = {978-3-95450-147-2},
  doi          = {doi:10.18429/JACoW-IPAC2016-TUOBB02},
  url          = {http://jacow.org/ipac2016/papers/tuobb02.pdf},
  note         = {doi:10.18429/JACoW-IPAC2016-TUOBB02},
}

@article{Bamber:1999zt,
    author = "Bamber, C. and others",
    title = "{Studies of nonlinear QED in collisions of 46.6-GeV electrons with intense laser pulses}",
    reportNumber = "SLAC-PUB-8063, PRINCETON-HEP-99-1, UR-1565, UTHEP-99-02",
    doi = "10.1103/PhysRevD.60.092004",
    journal = "Phys. Rev. D",
    volume = "60",
    pages = "092004",
    year = "1999"
}

@article{kuchiev_production_2007,
  title = {Production of High-Energy Particles in Laser and Coulomb Fields and the ${e}^{+}{e}^{\ensuremath{-}}$ Antenna},
  author = {Kuchiev, M. Yu.},
  journal = {Phys. Rev. Lett.},
  volume = {99},
  issue = {13},
  pages = {130404},
  year = {2007},
  month = {9},
  doi = {10.1103/PhysRevLett.99.130404},
}

@article{meuren_high-energy_2015,
  title = {High-Energy Recollision Processes of Laser-Generated Electron-Positron Pairs},
  author = {Meuren, Sebastian and Hatsagortsyan, Karen Z. and Keitel, Christoph H. and Di Piazza, Antonino},
  journal = {Phys. Rev. Lett.},
  volume = {114},
  issue = {14},
  pages = {143201},
  year = {2015},
  month = {4},
  doi = {10.1103/PhysRevLett.114.143201},
}

@article{pouyez_kinetic_2025,
  title = {Kinetic Structure of Strong-Field QED Showers in Crossed Electromagnetic Fields},
  author = {Pouyez, M. and Grismayer, T. and Grech, M. and Riconda, C.},
  journal = {Phys. Rev. Lett.},
  volume = {134},
  issue = {13},
  pages = {135001},
  year = {2025},
  month = {4},
  doi = {10.1103/PhysRevLett.134.135001},
}

@article{qu_signature_2021,
  title = {Signature of Collective Plasma Effects in Beam-Driven QED Cascades},
  author = {Qu, Kenan and Meuren, Sebastian and Fisch, Nathaniel J.},
  journal = {Phys. Rev. Lett.},
  volume = {127},
  issue = {9},
  pages = {095001},
  year = {2021},
  month = {8},
  doi = {10.1103/PhysRevLett.127.095001},
}

@article{bell_possibility_2008,
  title = {Possibility of Prolific Pair Production with High-Power Lasers},
  author = {Bell, A. R. and Kirk, John G.},
  journal = {Phys. Rev. Lett.},
  volume = {101},
  issue = {20},
  pages = {200403},
  year = {2008},
  month = {11},
  doi = {10.1103/PhysRevLett.101.200403},
}

@article{danson_petawatt_2019, 
title={Petawatt and exawatt class lasers worldwide}, 
volume={7}, 
DOI={10.1017/hpl.2019.36}, 
journal={High Power Laser Science and Engineering}, 
author={Danson, Colin N. and others}, 
year={2019}, 
pages={e54},
}

@article{karnako_current_2015,
      author        = {Karnakov, B. M. and Mur, V. D. and Popruzhenko, S. V. and Popov, V. S.},
      title         = {Current progress in developing the nonlinear ionization theory of atoms and ions},
      journal       = {Physics-Uspekhi},
      volume        = {58},
      year          = {2015},
      pages         = {3},
    doi             = {10.3367/UFNe.0185.201501b.0003},
}

@book{landau_quantum_1981,
	edition	= {2},
	title	= {{Q}uantum {E}lectrodynamics},
	publisher	= {Butterworth-Heinemann},
	author	= {Berestetskii, V. B. and Lifshitz, E. M. and Pitaevskii, L. P.},
	year	= {1982},
}

@article{keldysh_ionization_1965,
      author         = {Keldysh, L. V.},
      title          = {Ionization in the field of a strong electromagnetic wave},
      journal        = {Sov. Phys. JETP},
      volume         = {20},
      year           = {1965},
      pages          = {1307},
        url         = {http://www.jetp.ras.ru/cgi-bin/dn/e_020_05_1307.pdf},
}

@article{baier_processes_1968,
      author         = {Baier, V. N. and Katkov, V. M.},
      title          = {Processes involved in the motion of high energy particles in a magnetic field},
      journal        = {Sov. Phys. JETP},
      volume         = {26},
      year           = {1968},
      pages          = {854},
        url         = {http://jetp.ras.ru/cgi-bin/dn/e_026_04_0854.pdf},
}

@article{pellegrini_physics_2016,
  title = {The physics of x-ray free-electron lasers},
  author = {Pellegrini, C. and Marinelli, A. and Reiche, S.},
  journal = {Rev. Mod. Phys.},
  volume = {88},
  issue = {1},
  pages = {015006},
  numpages = {55},
  year = {2016},
  month = {3},
  doi = {10.1103/RevModPhys.88.015006},
}

@article{huang_review_2007,
  title = {Review of x-ray free-electron laser theory},
  author = {Huang, Zhirong and Kim, Kwang-Je},
  journal = {Phys. Rev. ST Accel. Beams},
  volume = {10},
  issue = {3},
  pages = {034801},
  numpages = {26},
  year = {2007},
  month = {3},
  doi = {10.1103/PhysRevSTAB.10.034801},
}

@article{aleksandrov_lcfa_2019,
  title = {Locally-constant field approximation in studies of electron-positron pair production in strong external fields},
  author = {Aleksandrov, I. A. and Plunien, G. and Shabaev, V. M.},
  journal = {Phys. Rev. D},
  volume = {99},
  issue = {1},
  pages = {016020},
  year = {2019},
  month = {1},
  doi = {10.1103/PhysRevD.99.016020},
}

@book{landau_fields_1975,
	edition	= {4},
	title	= {The classical theory of fields},
	publisher	= {Butterworth-Heinemann},
	author	= {Landau, L. D. and Lifshitz, E. M.},
	year	= {1975},
}

@phdthesis{meuren_2015,
  title={Nonlinear quantum electrodynamic and electroweak processes in strong laser fields},
  author={Meuren, Sebastian},
  year={2015},
  school={Heidelberg Univsersity},
 doi = {10.11588/heidok.00018971},
}

@article{baier_concept_2005,
title = {Concept of formation length in radiation theory},
journal = {Physics Reports},
volume = {409},
number = {5},
pages = {261-359},
year = {2005},
doi = {https://doi.org/10.1016/j.physrep.2004.11.003},
author = {V.N. Baier and V.M. Katkov},
}

@article{di_piazza_first-order_2017,
  title = {First-order strong-field QED processes in a tightly focused laser beam},
  author = {Di Piazza, A.},
  journal = {Phys. Rev. A},
  volume = {95},
  issue = {3},
  pages = {032121},
  numpages = {11},
  year = {2017},
  doi = {10.1103/PhysRevA.95.032121},
}

@article{brown_interaction_1964,
  title = {Interaction of Intense Laser Beams with Electrons},
  author = {Brown, Lowell S. and Kibble, T. W. B.},
  journal = {Phys. Rev.},
  volume = {133},
  issue = {3A},
  pages = {A705--A719},
  year = {1964},
  month = {2},
  doi = {10.1103/PhysRev.133.A705},
}

@article{PhysRevA.98.012134,
  title = {Implementing nonlinear Compton scattering beyond the local-constant-field approximation},
  author = {Di Piazza, A. and Tamburini, M. and Meuren, S. and Keitel, C. H.},
  journal = {Phys. Rev. A},
  volume = {98},
  issue = {1},
  pages = {012134},
  numpages = {10},
  year = {2018},
  month = {7},
  doi = {10.1103/PhysRevA.98.012134},
}

@article{baier_quantum_1989,
title = {Quantum radiation theory in inhomogeneous external fields},
journal = {Nuclear Physics B},
volume = {328},
number = {2},
pages = {387-405},
year = {1989},
doi = {https://doi.org/10.1016/0550-3213(89)90334-9},
author = {V.N. Baier and V.M. Katkov and V.M. Strakhovenko},
}

@article{Abelevetal:2014dna,
      author         = "ALICE Collaboration",
      collaboration  = "ALICE Collaboration",
      title          = "{Technical Design Report for the Upgrade of the ALICE
                        Inner Tracking System}",
      collaboration  = "ALICE",
      journal        = "J. Phys.",
      volume         = "G41",
      year           = "2014",
      pages          = "087002",
      doi            = "10.1088/0954-3899/41/8/087002",
      reportNumber   = "CERN-LHCC-2013-024, ALICE-TDR-017",
      SLACcitation   = "%%CITATION = JPAGA,G41,087002;%%"
}

@article{AGLIERIRINELLA2017583,
	title = {The ALPIDE pixel sensor chip for the upgrade of the ALICE Inner Tracking System},
	journal = {NIM. A.},
	volume = {845},
	pages = {583-587},
	year = {2017},
	note = {Proceedings of the Vienna Conference on Instrumentation 2016},
	issn = {0168-9002},
	doi = {https://doi.org/10.1016/j.nima.2016.05.016},
	url = {https://www.sciencedirect.com/science/article/pii/S0168900216303825},
	author = {Gianluca {Aglieri Rinella}}
}

@article{Spannagel:2018usc,
    author = {Spannagel, Simon and Wolters, Koen and Hynds, Daniel and Alipour Tehrani, Niloufar and Benoit, Mathieu and Dannheim, Dominik and Gauvin, Neal and N\"urnberg, Andreas and Sch\"utze, Paul and Vicente Barreto Pinto, Mateus},
    title = "{Allpix$^2$: A Modular Simulation Framework for Silicon Detectors}",
    eprint = "1806.05813",
    archivePrefix = "arXiv",
    primaryClass = "physics.ins-det",
    reportNumber = "CLICdp-Pub-2018-002, CLICDP-PUB-2018-002",
    doi = "10.1016/j.nima.2018.06.020",
    journal = "Nucl. Instrum. Meth. A",
    volume = "901",
    pages = "164--172",
    year = "2018"
}

@article{Dannheim:2020yna,
    author = {Dannheim, Dominik and Dort, Katharina and Hynds, Daniel and Munker, Magdalena and N\"urnberg, Andreas and Snoeys, Walter and Spannagel, Simon},
    title = "{Combining TCAD and Monte Carlo Methods to Simulate CMOS Pixel Sensors with a Small Collection Electrode using the Allpix$^2$ Squared Framework}",
    eprint = "2002.12602",
    archivePrefix = "arXiv",
    primaryClass = "physics.ins-det",
    reportNumber = "CLICdp-Pub-2019-008",
    doi = "10.1016/j.nima.2020.163784",
    journal = "Nucl. Instrum. Meth. A",
    volume = "964",
    pages = "163784",
    year = "2020"
}

@article{Spannagel:2018xhg,
    author = "Spannagel, Simon",
    editor = "Batignani, Giovanni and Grassi, Marco and Paoletti, Riccardo and Retico, Alessandra and Signorelli, Giovanni and Spagnolo, Paolo",
    collaboration = "CLICdp",
    title = "{Technologies for Future Vertex and Tracking Detectors at CLIC}",
    eprint = "1812.02625",
    archivePrefix = "arXiv",
    primaryClass = "physics.ins-det",
    reportNumber = "CLICdp-Conf-2018-003",
    doi = "10.1016/j.nima.2018.08.103",
    journal = "Nucl. Instrum. Meth. A",
    volume = "936",
    pages = "612--615",
    year = "2019"
}

@inproceedings{moore1959shortest,
  author    = {Moore, Edward F.},
  title     = {The shortest path through a maze},
  booktitle = {Proceedings of the International Symposium on the Theory of Switching},
  pages     = {285--292},
  year      = {1959},
  publisher = {Harvard University Press},
  address   = {Cambridge, MA},
}

@book{cormen2009introduction,
  title     = {Introduction to Algorithms},
  author    = {Cormen, Thomas H. and Leiserson, Charles E. and Rivest, Ronald L. and Stein, Clifford},
  edition   = {3rd},
  year      = {2009},
  publisher = {MIT Press},
  address   = {Cambridge, MA},
}

@article{Santra:2022sjw,
    author = "Santra, Arka and Tal Hod, Noam",
    title = "{A derivation of the electric field inside MAPS detectors from beam-test data and limited TCAD simulations}",
    eprint = "2209.03457",
    archivePrefix = "arXiv",
    primaryClass = "physics.ins-det",
    doi = "10.1088/1748-0221/18/05/P05007",
    journal = "JINST",
    volume = "18",
    number = "05",
    pages = "P05007",
    year = "2023"
}

@article{TowerJazz,
    Author = {TowerJazz},
    journal = {\url{https://towersemi.com/}},
    url = {https://towersemi.com/},
    doi = {},
    year = {}
}

@article{Collaboration_2008,
	doi = {10.1088/1748-0221/3/08/s08002},
	url = {https://doi.org/10.1088/1748-0221/3/08/s08002},
	year = 2008,
	month = {8},
	publisher = {{IOP} Publishing},
	volume = {3},
	number = {08},
	pages = {S08002--S08002},
	author = "ALICE Collaboration",
	collaboration  = "ALICE Collaboration",
	title = {The {ALICE} experiment at the {CERN} {LHC}},
	journal = {Journal of Instrumentation},
}

@article{MAGER2016434,
      title = "ALPIDE, the Monolithic Active Pixel Sensor for the ALICE ITS upgrade",
      journal = "Nuclear Instruments and Methods in Physics Research Section A: Accelerators, Spectrometers, Detectors and Associated Equipment",
      volume = "824",
      pages = "434 - 438",
      year = "2016",
      note = "Frontier Detectors for Frontier Physics: Proceedings of the 13th Pisa Meeting on Advanced Detectors",
      issn = "0168-9002",
      doi = "https://doi.org/10.1016/j.nima.2015.09.057",
      url = "http://www.sciencedirect.com/science/article/pii/S0168900215011122",
      author = "M. Mager",
}

@article{Ravasenga:2023yqd,
    author = "Ravasenga, Ivan",
    collaboration = "ALICE",
    title = "{Commissioning and Performance of the New ALICE Inner Tracking System in the First Phase of LHC Run 3}",
    eprint = "2302.00432",
    archivePrefix = "arXiv",
    primaryClass = "physics.ins-det",
    doi = "10.7566/JPSCP.42.011002",
    journal = "JPS Conf. Proc.",
    volume = "42",
    pages = "011002",
    year = "2024"
}

@article{ALICE:2023udb,
    author = "Acharya, Shreyasi and others",
    collaboration = "ALICE",
    title = "{ALICE upgrades during the LHC Long Shutdown 2}",
    eprint = "2302.01238",
    archivePrefix = "arXiv",
    primaryClass = "physics.ins-det",
    reportNumber = "CERN-EP-2023-009",
    doi = "10.1088/1748-0221/19/05/P05062",
    journal = "JINST",
    volume = "19",
    number = "05",
    pages = "P05062",
    year = "2024"
}

@article{YANG201561,
      title = "Low-power priority Address-Encoder and Reset-Decoder data-driven readout for Monolithic Active Pixel Sensors for tracker system",
      journal = "Nuclear Instruments and Methods in Physics Research Section A: Accelerators, Spectrometers, Detectors and Associated Equipment",
      volume = "785",
      pages = "61 - 69",
      year = "2015",
      issn = "0168-9002",
      doi = "https://doi.org/10.1016/j.nima.2015.02.063",
      url = "http://www.sciencedirect.com/science/article/pii/S0168900215002818",
      author = "P. Yang and others",
}

@article{SENYUKOV2013115,
      title = "Charged particle detection performances of CMOS pixel sensors produced in a 0.18~$\mu$m process with a high resistivity epitaxial layer",
      journal = "Nuclear Instruments and Methods in Physics Research Section A: Accelerators, Spectrometers, Detectors and Associated Equipment",
      volume = "730",
      pages = "115 - 118",
      year = "2013",
      note = "Proceedings of the 9th International Conference on Radiation Effects on Semiconductor Materials Detectors and Devices",
      issn = "0168-9002",
      doi = "https://doi.org/10.1016/j.nima.2013.03.017",
      url = "http://www.sciencedirect.com/science/article/pii/S0168900213002945",
      author = "S. Senyukov and others",
}

@article{AGLIERIRINELLA2021164859,
title = {Charge collection properties of TowerJazz 180nm CMOS Pixel Sensors in dependence of pixel geometries and bias parameters, studied using a dedicated test-vehicle: The Investigator chip},
journal = {Nuclear Instruments and Methods in Physics Research Section A: Accelerators, Spectrometers, Detectors and Associated Equipment},
volume = {988},
pages = {164859},
year = {2021},
issn = {0168-9002},
doi = {https://doi.org/10.1016/j.nima.2020.164859},
url = {https://www.sciencedirect.com/science/article/pii/S0168900220312560},
author = {G. {Aglieri Rinella} and G. Chaosong and A. {di Mauro} and J. Eum and H. Hillemanns and A. Junique and M. Keil and D. Kim and H. Kim and T. Kugathasan and S. Lee and M. Mager and V. Manzari and C.A. {Marin Tobon} and P. Martinengo and H. Mugnier and L. Musa and F. Reidt and J. Rousset and K. Sielewicz and W. Snoeys and M. Šuljić and J.W. {van Hoorne} and Q.W. Malik and P. Yang and I.-K. Yoo}
}

@article{DANNHEIM2019187,
title = {Comparison of small collection electrode CMOS pixel sensors with partial and full lateral depletion of the high-resistivity epitaxial layer},
journal = {Nuclear Instruments and Methods in Physics Research Section A: Accelerators, Spectrometers, Detectors and Associated Equipment},
volume = {927},
pages = {187-193},
year = {2019},
issn = {0168-9002},
doi = {https://doi.org/10.1016/j.nima.2019.02.049},
url = {https://www.sciencedirect.com/science/article/pii/S0168900219302372},
author = {Dominik Dannheim and Adrian Fiergolski and Jacobus {van Hoorne} and Daniel Hynds and Wolfgang Klempt and Thanushan Kugathasan and Magdalena Munker and Andreas Nürnberg and Krzysztof Sielewicz and Walter Snoeys and Simon Spannagel}
}

@article{ALICE-PUBLIC-2018-013,
      title         = "{Expression of Interest for an ALICE ITS Upgrade in LS3}",
      collaboration = "ALICE Collaboration",
      author        = "ALICE Collaboration",
      month         = "10",
      year          = "2018",
      reportNumber  = "ALICE-PUBLIC-2018-013",
      url           = "http://cds.cern.ch/record/2644611",
}

@article{LUXETDR,
	author = {LUXE Collaboration},
	month = "10",
	doi = {10.1140/epjs/s11734-024-01164-9},
	id = {Abramowicz2024},
	journal = {The European Physical Journal Special Topics},
	title = {Technical Design Report for the LUXE experiment},
	url = {https://doi.org/10.1140/epjs/s11734-024-01164-9},
	year = {2024}
}

@article{Hartin:2018sha,
    author = "Hartin, Anthony and Ringwald, Andreas and Tapia, Natalia",
    title = "{Measuring the Boiling Point of the Vacuum of Quantum Electrodynamics}",
    eprint = "1807.10670",
    archivePrefix = "arXiv",
    primaryClass = "hep-ph",
    reportNumber = "DESY-18-128, DESY 18-128",
    doi = "10.1103/PhysRevD.99.036008",
    journal = "Phys. Rev. D",
    volume = "99",
    number = "3",
    pages = "036008",
    year = "2019"
}

@article{Fedotov:2022ely,
    author = "Fedotov, A. and Ilderton, A. and Karbstein, F. and King, B. and Seipt, D. and Taya, H. and Torgrimsson, G.",
    title = "{Advances in QED with intense background fields}",
    doi = "10.1016/j.physrep.2023.01.003",
    journal = "Phys. Rept.",
    volume = "1010",
    pages = "1--138",
    year = "2023"
}

@article{refId0,
	author = {{Robertis, G. De} and {Fanizzi, G.} and {Loddo, F.} and {Manzari, V.} and {Rizzi, M.}},
	title = {A MOdular System for Acquisition, Interface and Control (MOSAIC) of detectors and their related electronics for high energy physics experiment},
	DOI= "10.1051/epjconf/201817407002",
	url= "https://doi.org/10.1051/epjconf/201817407002",
	journal = {EPJ Web Conf.},
	year = 2018,
	volume = 174,
	pages = "07002",
}

@article{ALPIDEManual,
      title = "ALPIDE Operations Manual",
      url = "http://sunba2.ba.infn.it/MOSAIC/ALICE-ITS/Documents/ALPIDE-operations-manual-version-0_3.pdf",
}

@article{Abovyan:2022qaz,
    author = "Abovyan, S. and others",
    title = "{Test of low-dropout voltage regulators with neutrons and protons}",
    doi = "10.1088/1748-0221/17/05/C05006",
    journal = "JINST",
    volume = "17",
    number = "05",
    pages = "C05006",
    year = "2022"
}

@article{Liu_2019,
	doi = {10.1088/1748-0221/14/10/p10033},
	url = {https://doi.org/10.1088/1748-0221/14/10/p10033},
	year = 2019,
	month = {10},
	publisher = {{IOP} Publishing},
	volume = {14},
	number = {10},
	pages = {P10033--P10033},
	author = {Y. Liu and M.S. Amjad and P. Baesso and D. Cussans and J. Dreyling-Eschweiler and R. Ete and I. Gregor and L. Huth and A. Irles and H. Jansen and K. Krueger and J. Kvasnicka and R. Peschke and E. Rossi and A. Rummler and F. Sefkow and M. Stanitzki and M. Wing and M. Wu},
	title = {{EUDAQ}2{\textemdash}A flexible data acquisition software framework for common test beams},
	journal = {Journal of Instrumentation}
}

@article{EUDAQ2020,
	doi = {10.1088/1748-0221/15/01/p01038},
	url = {https://doi.org/10.1088/1748-0221/15/01/p01038},
	year = 2020,
	month = Jan,
	publisher = {{IOP} Publishing},
	volume = {15},
	number = {01},
	pages = {P01038--P01038},
	author = {P. Ahlburg and S. Arfaoui and J.-H. Arling and H. Augustin and D. Barney and M. Benoit and T. Bisanz and E. Corrin and D. Cussans and D. Dannheim and J. Dreyling-Eschweiler and T. Eichhorn and A. Fiergolski and I.-M. Gregor and J. Grosse-Knetter and D. Haas and L. Huth and A. Irles and H. Jansen and J. Janssen and M. Keil and J.S. Keller and M. Kiehn and H.J. Kim and J. Kroll and K. Krüger and S. Kulis and J. Kvasnicka and J. Lange and Y. Liu and F. Lütticke and C. Marinas and P. Martinengo and A. Nurnberg and B. Paschen and H. Perrey and R. Peschke and D. Pitzl and D.-L. Pohl and A. Quadt and T. Quast and F. Reidt and E. Rossi and I. Rubinsky and A. Rummler and H. Schreeck and P. Schütze and B. Schwenker and S. Spannagel and M. Stanitzki and U. Stolzenberg and T. Suehara and M. Suljic and G. Troska and M. Varga-Kofarago and J. Weingarten and P. Wieduwilt},
	title = {{EUDAQ}{\textemdash}a data acquisition software framework for common beam telescopes}
}

@article{ai2022common,
  title={A common tracking software project},
  author={Ai, Xiaocong and Allaire, Corentin and Calace, Noemi and Czirkos, Ang{\'e}la and Elsing, Markus and Ene, Irina and Farkas, Ralf and Gagnon, Louis-Guillaume and Garg, Rocky and Gessinger, Paul and others},
  journal={Computing and Software for Big Science},
  volume={6},
  number={1},
  pages={8},
  year={2022},
  publisher={Springer},
doi = {10.1007/s41781-021-00078-8},
}

@article{abramson1982bandwidth,
  title={On bandwidth variation in kernel estimates-a square root law},
  author={Abramson, Ian S},
  journal={The annals of Statistics},
  pages={1217--1223},
  year={1982},
  publisher={JSTOR}
}

@book{silverman2018density,
  title={Density estimation for statistics and data analysis},
  author={Silverman, Bernard W},
  year={2018},
  publisher={Routledge}
}

@article{sheather1991reliable,
  title={A reliable data-based bandwidth selection method for kernel density estimation},
  author={Sheather, Simon J and Jones, Michael C},
  journal={Journal of the Royal Statistical Society: Series B (Methodological)},
  volume={53},
  number={3},
  pages={683--690},
  year={1991},
  publisher={Wiley Online Library}
}

@article{LMA2,
    author = "Chen, P. and Horton-Smith, G. and Ohgaki, T. and Weidemann, A. W. and Yokoya, K.",
    title = "{CAIN: Conglomerat d'ABEL et d'interactions nonlineaires}",
    reportNumber = "SLAC-PUB-6583",
    doi = "10.1016/0168-9002(94)01186-9",
    journal = "Nucl. Instrum. Meth. A",
    volume = "355",
    pages = "107--110",
    year = "1995"
}

@article{LMA3,
    author = "Hartin, Anthony",
    title = "{Strong field QED in lepton colliders and electron/laser interactions}",
    eprint = "1804.02934",
    archivePrefix = "arXiv",
    primaryClass = "hep-ph",
    doi = "10.1142/S0217751X18300119",
    journal = "Int. J. Mod. Phys. A",
    volume = "33",
    number = "13",
    pages = "1830011",
    year = "2018"
}

@article{LMA4,
    author = "Heinzl, T. and King, B. and Macleod, A. J.",
    title = "{The locally monochromatic approximation to QED in intense laser fields}",
    eprint = "2004.13035",
    archivePrefix = "arXiv",
    primaryClass = "hep-ph",
    doi = "10.1103/PhysRevA.102.063110",
    journal = "Phys. Rev. A",
    volume = "102",
    pages = "063110",
    year = "2020"
}

@article{LMA5,
    author = "Torgrimsson, Greger",
    title = "{Loops and polarization in strong-field QED}",
    eprint = "2012.12701",
    archivePrefix = "arXiv",
    primaryClass = "hep-ph",
    doi = "10.1088/1367-2630/abf274",
    journal = "New J. Phys.",
    volume = "23",
    number = "6",
    pages = "065001",
    year = "2021"
}

@article{billoir1992fast,
  title={Fast vertex fitting with a local parametrization of tracks},
  author={Billoir, Pierre and Qian, Sijin},
  journal={Nuclear Instruments and Methods in Physics Research Section A: Accelerators, Spectrometers, Detectors and Associated Equipment},
  volume={311},
  number={1-2},
  pages={139--150},
  year={1992},
  publisher={Elsevier}
}

@techreport{fruhwirth2004adaptive,
  title={Adaptive Multi-vertex fitting},
  author={Fr{\"u}hwirth, R and Waltenberger, W},
  year={2004},
  institution={CERN}
}

@phdthesis{schlag2022advanced,
  title={Advanced algorithms and software for primary vertex reconstruction and search for flavor-violating supersymmetry with the ATLAS experiment},
  author={Schlag, Bastian},
  year={2022},
  school={Johannes Gutenberg-Universit{\"a}t Mainz}
}

@article{aniculaesei2023acceleration,
    author = {Aniculaesei, Constantin and Ha, Thanh and Yoffe, Samuel and Labun, Lance and Milton, Stephen and McCary, Edward and Spinks, Michael M. and Quevedo, Hernan J. and Labun, Ou Z. and Sain, Ritwik and Hannasch, Andrea and Zgadzaj, Rafal and Pagano, Isabella and Franco-Altamirano, Jose A. and Ringuette, Martin L. and Gaul, Erhart and Luedtke, Scott V. and Tiwari, Ganesh and Ersfeld, Bernhard and Brunetti, Enrico and Ruhl, Hartmut and Ditmire, Todd and Bruce, Sandra and Donovan, Michael E. and Downer, Michael C. and Jaroszynski, Dino A. and Hegelich, Bjorn Manuel},
    title = {The acceleration of a high-charge electron bunch to 10 GeV in a 10-cm nanoparticle-assisted wakefield accelerator},
    journal = {Matter and Radiation at Extremes},
    volume = {9},
    number = {1},
    pages = {014001},
    year = {2023},
    month = {11},
    issn = {2468-2047},
    doi = {10.1063/5.0161687},
    url = {https://doi.org/10.1063/5.0161687},
    eprint = {https://pubs.aip.org/aip/mre/article-pdf/doi/10.1063/5.0161687/18286148/014001\_1\_5.0161687.pdf},
}

@article{tanaka2020eli,
    author = {Tanaka, K. A. and Spohr, K. M. and Balabanski, D. L. and Balascuta, S. and Capponi, L. and Cernaianu, M. O. and Cuciuc, M. and Cucoanes, A. and Dancus, I. and Dhal, A. and Diaconescu, B. and Doria, D. and Ghenuche, P. and Ghita, D. G. and Kisyov, S. and Nastasa, V. and Ong, J. F. and Rotaru, F. and Sangwan, D. and Söderström, P.-A. and Stutman, D. and Suliman, G. and Tesileanu, O. and Tudor, L. and Tsoneva, N. and Ur, C. A. and Ursescu, D. and Zamfir, N. V.},
    title = {Current status and highlights of the ELI-NP research program},
    journal = {Matter and Radiation at Extremes},
    volume = {5},
    number = {2},
    pages = {024402},
    year = {2020},
    month = {03},
    issn = {2468-2047},
    doi = {10.1063/1.5093535},
    url = {https://doi.org/10.1063/1.5093535},
    eprint = {https://pubs.aip.org/aip/mre/article-pdf/doi/10.1063/1.5093535/15750185/024402\_1\_online.pdf},
}

@article{kim2013zsa,
    author = "Kim, Hyung Taek and Pae, Ki Hong and Cha, Hyuk Jin and Kim, I Jong and Yu, Tae Jun and Sung, Jae Hee and Lee, Seong Ku and Jeong, Tae Moon and Lee, Jongmin",
    title = {Enhancement of electron energy to multi-GeV regime by a dual-stage laser-wakefield accelerator pumped by petawatt laser pulses},
    eprint = "1307.4159",
    archivePrefix = "arXiv",
    primaryClass = "physics.plasm-ph",
    doi = "10.1103/PhysRevLett.111.165002",
    journal = "Phys. Rev. Lett.",
    volume = "111",
    number = "16",
    pages = "165002",
    year = "2013"
}

@article{rezaei2023laser,
  title = {Laser Wakefield Electron Acceleration with Polarization-Dependent Ionization Injection},
  author = {Rezaei-Pandari, Mohammad and Mirzaie, Mohammad and Hojbota, Calin Ioan and Pak, Tae Gyu and Kim, Sang Beom and Lee, Geon Woo and Massudi, Reza and Niknam, Ali Reza and Lee, Seong Ku and Kim, Ki-Yong and Nam, Chang Hee},
  journal = {Phys. Rev. Appl.},
  volume = {20},
  issue = {3},
  pages = {034026},
  numpages = {9},
  year = {2023},
  month = {9},
  publisher = {American Physical Society},
  doi = {10.1103/PhysRevApplied.20.034026},
  url = {https://link.aps.org/doi/10.1103/PhysRevApplied.20.034026}
}

@article{kim2021multigev,
  article-number = {5831},
  author = {Kim, Hyung Taek and Pathak, Vishwa Bandhu and Hojbota, Calin Ioan and Mirzaie, Mohammad and Pae, Ki Hong and Kim, Chul Min and Yoon, Jin Woo and Sung, Jae Hee and Lee, Seong Ku},
  doi = {10.3390/app11135831},
  issn = {2076-3417},
  journal = {Applied Sciences},
  number = {13},
  title = {Multi-GeV Laser Wakefield Electron Acceleration with PW Lasers},
  url = {https://www.mdpi.com/2076-3417/11/13/5831},
  volume = {11},
  year = {2021}
}

@article{hojbota2019accurate,
    author = {Hojbota, C. I. and Kim, Hyung Taek and Shin, Jung Hun and Aniculaesei, C. and Rao, B. S. and Nam, Chang Hee},
    title = {Accurate single-shot measurement technique for the spectral distribution of GeV electron beams from a laser wakefield accelerator},
    journal = {AIP Advances},
    volume = {9},
    number = {8},
    pages = {085229},
    year = {2019},
    month = {08},
    issn = {2158-3226},
    doi = {10.1063/1.5117311},
    url = {https://doi.org/10.1063/1.5117311},
    eprint = {https://pubs.aip.org/aip/adv/article-pdf/doi/10.1063/1.5117311/12911799/085229\_1\_online.pdf},
}

@article{miao2022multigev,
  title = {Multi-GeV Electron Bunches from an All-Optical Laser Wakefield Accelerator},
  author = {Miao, B. and Shrock, J. E. and Feder, L. and Hollinger, R. C. and Morrison, J. and Nedbailo, R. and Picksley, A. and Song, H. and Wang, S. and Rocca, J. J. and Milchberg, H. M.},
  journal = {Phys. Rev. X},
  volume = {12},
  issue = {3},
  pages = {031038},
  numpages = {17},
  year = {2022},
  month = {9},
  publisher = {American Physical Society},
  doi = {10.1103/PhysRevX.12.031038},
  url = {https://link.aps.org/doi/10.1103/PhysRevX.12.031038}
}

@article{winkler2025active,
   author = {Winkler, P. and Trunk, M. and H{\"u}bner, L. and Martinez de la Ossa, A. and Jalas, S. and Kirchen, M. and Agapov, I. and Antipov, S. A. and Brinkmann, R. and Eichner, T. and Ferran Pousa, A. and H{\"u}lsenbusch, T. and Palmer, G. and Schnepp, M. and Schubert, K. and Th{\'e}venet, M. and Walker, P. A. and Werle, C. and Leemans, W. P. and Maier, A. R.},
   year = {2025},
   month = {04},
   doi = {10.1038/s41586-025-08772-y},
   id = {Winkler2025},
   journal = {Nature},
   number = {8060},
   pages = {907--910},
   title = {Active energy compression of a laser-plasma electron beam},
   url = {https://doi.org/10.1038/s41586-025-08772-y},
   volume = {640},
   year = {2025}
}

@article{PhysRevLett.133.255001,
  title = {Matched Guiding and Controlled Injection in Dark-Current-Free, 10-GeV-Class, Channel-Guided Laser-Plasma Accelerators},
  author = {Picksley, A. and Stackhouse, J. and Benedetti, C. and Nakamura, K. and Tsai, H. E. and Li, R. and Miao, B. and Shrock, J. E. and Rockafellow, E. and Milchberg, H. M. and Schroeder, C. B. and van Tilborg, J. and Esarey, E. and Geddes, C. G. R. and Gonsalves, A. J.},
  journal = {Phys. Rev. Lett.},
  volume = {133},
  issue = {25},
  pages = {255001},
  numpages = {8},
  year = {2024},
  month = {12},
  publisher = {American Physical Society},
  doi = {10.1103/PhysRevLett.133.255001},
  url = {https://link.aps.org/doi/10.1103/PhysRevLett.133.255001}
}

@article{PhysRevX.8.031004,
  title = {Experimental Signatures of the Quantum Nature of Radiation Reaction in the Field of an Ultraintense Laser},
  author = {Poder, K. and Tamburini, M. and Sarri, G. and Di Piazza, A. and Kuschel, S. and Baird, C. D. and Behm, K. and Bohlen, S. and Cole, J. M. and Corvan, D. J. and Duff, M. and Gerstmayr, E. and Keitel, C. H. and Krushelnick, K. and Mangles, S. P. D. and McKenna, P. and Murphy, C. D. and Najmudin, Z. and Ridgers, C. P. and Samarin, G. M. and Symes, D. R. and Thomas, A. G. R. and Warwick, J. and Zepf, M.},
  journal = {Phys. Rev. X},
  volume = {8},
  issue = {3},
  pages = {031004},
  numpages = {11},
  year = {2018},
  month = {7},
  publisher = {American Physical Society},
  doi = {10.1103/PhysRevX.8.031004},
  url = {https://link.aps.org/doi/10.1103/PhysRevX.8.031004}
}

@article{PhysRevX.8.011020,
  title = {Experimental Evidence of Radiation Reaction in the Collision of a High-Intensity Laser Pulse with a Laser-Wakefield Accelerated Electron Beam},
  author = {Cole, J. M. and Behm, K. T. and Gerstmayr, E. and Blackburn, T. G. and Wood, J. C. and Baird, C. D. and Duff, M. J. and Harvey, C. and Ilderton, A. and Joglekar, A. S. and Krushelnick, K. and Kuschel, S. and Marklund, M. and McKenna, P. and Murphy, C. D. and Poder, K. and Ridgers, C. P. and Samarin, G. M. and Sarri, G. and Symes, D. R. and Thomas, A. G. R. and Warwick, J. and Zepf, M. and Najmudin, Z. and Mangles, S. P. D.},
  journal = {Phys. Rev. X},
  volume = {8},
  issue = {1},
  pages = {011020},
  numpages = {11},
  year = {2018},
  month = {2},
  publisher = {American Physical Society},
  doi = {10.1103/PhysRevX.8.011020},
  url = {https://link.aps.org/doi/10.1103/PhysRevX.8.011020}
}

@article{los2024observation,
  title={Observation of quantum effects on radiation reaction in strong fields},
  author={Los, EE and Gerstmayr, Elias and Arran, Christopher and Streeter, Matthew JV and Colgan, Cary and Cobo, Claudia C and Kettle, Brendan and Blackburn, Thomas G and Bourgeois, Nicolas and Calvin, Luke and others},
  journal={arXiv preprint arXiv:2407.12071},
  year={2024}
}

@article{PhysRevLett.122.084801,
  title = {Petawatt Laser Guiding and Electron Beam Acceleration to 8 GeV in a Laser-Heated Capillary Discharge Waveguide},
  author = {Gonsalves, A. J. and Nakamura, K. and Daniels, J. and Benedetti, C. and Pieronek, C. and de Raadt, T. C. H. and Steinke, S. and Bin, J. H. and Bulanov, S. S. and van Tilborg, J. and Geddes, C. G. R. and Schroeder, C. B. and T\'oth, Cs. and Esarey, E. and Swanson, K. and Fan-Chiang, L. and Bagdasarov, G. and Bobrova, N. and Gasilov, V. and Korn, G. and Sasorov, P. and Leemans, W. P.},
  journal = {Phys. Rev. Lett.},
  volume = {122},
  issue = {8},
  pages = {084801},
  numpages = {6},
  year = {2019},
  month = {2},
  publisher = {American Physical Society},
  doi = {10.1103/PhysRevLett.122.084801},
  url = {https://link.aps.org/doi/10.1103/PhysRevLett.122.084801}
}

@article{PhysRevLett.132.195001,
  title = {Multi-GeV Electron Acceleration in Wakefields Strongly Driven by Oversized Laser Spots},
  author = {P\~oder, K. and Wood, J. C. and Lopes, N. C. and Cole, J. M. and Alatabi, S. and Backhouse, M. P. and Foster, P. S. and Hughes, A. J. and Kamperidis, C. and Kononenko, O. and Mangles, S. P. D. and Palmer, C. A. J. and Rusby, D. and Sahai, A. and Sarri, G. and Symes, D. R. and Warwick, J. R. and Najmudin, Z.},
  journal = {Phys. Rev. Lett.},
  volume = {132},
  issue = {19},
  pages = {195001},
  numpages = {7},
  year = {2024},
  month = {5},
  publisher = {American Physical Society},
  doi = {10.1103/PhysRevLett.132.195001},
  url = {https://link.aps.org/doi/10.1103/PhysRevLett.132.195001}
}

\clearpage
\newpage

\begin{appendices}

\section{The different regimes of SF-QED}
\label{app:sfqed}
Here, we give an overview of the most important parameters that characterize different regimes of strong-field QED.
Since photon emission by an electron/positron and photon-induced electron-positron pair production are closely related, we cover both processes.
In the following we focus on laser-electron collisions, which are relevant for E320.
The arguments presented are more intuitive than detailed; for a comprehensive review, we refer the reader to the literature, in particular Refs.~\cite{Fedotov:2022ely,RevModPhys.94.045001,di_piazza_extremely_2012,Ritus}.

From classical electrodynamics, we know that electromagnetic waves and electrically charged particles interact.
Once we accept that the photon field should be quantized, it follows that the simplest quantum process is linear Compton scattering: it describes the scattering of a single photon off an electron~\cite{landau_quantum_1981}.
In general, the electron’s energy and direction both change as a result of the scattering.
While the Thomson-scattering limit of vanishing recoil can be obtained from classical electrodynamics, the more general case requires the assumption of quantized photons.
Linear Compton scattering with a significant recoil is therefore a genuine quantum effect.

\subsection{Nonlinear interactions with the photon field}
When investigating collisions of an electron and a photon beam, one question naturally arises: when is the photon density so high, that nonlinear effects become important?
To estimate the relevant scale, we note that after absorbing a laser photon, an electron (or a positron) is no longer a real particle, as its four-momentum $p^\mu = (\epsilon, \mathbf{p})$ is no longer on shell ($p^2 = m_e^2$).
This is obvious from the four-momentum conservation law $p'^\mu = p^\mu + \kappa^\mu$: after squaring, we obtain $m_e^2 = m_e^2 + 2p\cdot \kappa$, where $p\cdot \kappa > 0$.
Thus, the electron needs to change its four-momentum, for example by radiating a high-energy NCS photon.

We now estimate the so‐called ``formation volume'', i.e., the space-time volume in which the virtual electron state, created by absorbing a single laser photon, can exist.
From a mathematical point of view the scattering probability is obtained from a matrix element that contains a space-time integral over oscillating functions~\cite{landau_quantum_1981}.
The formation volume represents the space-time region over which this integral obtains a significant contribution.
In what follows, we will approximate the relevant (1D) phase‐space volume by $\pi$, since this corresponds to the phase shift required to go from constructive to destructive interference (recalling that $\hbar=c=1$)~\cite{PhysRevA.98.012134}.
However, this choice is merely illustrative, as we do not expect to determine order‐one factors precisely.
Note that the derivation of the formation region is conceptually closely related to the derivation of the uncertainty principle in non-relativistic quantum mechanics.
More details are given in Ref.~\cite{meuren_2015,baier_concept_2005}.

In the following we focus on an ultra-relativistic electron with energy $\epsilon$, propagating along the $z$-direction ($\epsilon \approx p_z$, $p_x = p_y = 0$) that absorbs a counter-propagating photon of energy $\omega$.
As a result, the energy of the electron increases by $\omega$ while its longitudinal momentum decreases by the same amount.
Consequently, the electron’s new four-momentum is off-shell.
To leading order, the on-shell condition is violated by $\Delta\epsilon \approx 2\omega$.
To be explicit, if $\epsilon'=\epsilon+\omega$ and $p_z'=p_z-\omega$, then we can define $\Delta\epsilon = \epsilon'-\sqrt{m_e^2+p_z'^2}$.
The last square-root term represents the off-shell particle's energy, assuming that three-momentum conservation holds.
To first order, this square-root term becomes $\epsilon-\omega$ and hence we see that $\Delta\epsilon \approx 2\omega$.
Therefore, the electron's lifetime is $\Delta t \sim 1/\Delta\epsilon$.
Thus, since $c=1$, we obtain $\Delta x_{\rm{L}} = \Delta{}t\sim \pi/\Delta{}\epsilon = \pi/(2\omega)$ for the longitudinal formation length of the process.

To estimate the transverse dimension of the formation volume, we need to determine how much transverse momentum can be acquired during the scattering process itself.
To this end we point out that the transverse momentum of the electron changes continuously in the laser field, due to its classical motion.
Here, however, we estimate only the momentum transfer that happens during the formation of the core quantum transition itself (more details can be found in~\cite{PhysRevD.93.085028}). Therefore, we consider now the instantaneous momentum of the electron and assume, by definition, that the electron has no transverse momentum in its initial state (this re-defines the coordinate system).
Even if the momentum transfer from the background field can be neglected, the final state may have transverse momentum, provided it is shared between the emitted photon and the scattered electron.
Thus, we impose $k_x = p_x'$ and $k_y = p_y'$ in the following.
However, it is important to point out that introducing transverse momentum is not free: the more transverse momentum the final state has, the more energy the external field needs to contribute to realize the transition~\cite{meuren_2015}. 

To understand the scale at which transverse momentum becomes a significant factor and eventually suppresses the probability for the transition, we expand the relativistic dispersion relation for the final electron, $\epsilon'^2 = m_e^2 + \mathbf{p}'^2$, as $\epsilon' \approx p_z' + M'^2/(2p_z')$, where $M'^2 = m_e^2 + p_x'^2 + p_y'^2$.
From this expansion, we see that the effective mass $M'$ sets the scale at which the ``photon‐like'', initial dispersion relation $\epsilon' \approx p_z$ is violated.
The contribution $p_x'^2 + p_y'^2$ to $M'^2$ can be treated as small as long as $p_{x,y}'^2 \ll m_e^2$.
By defining the transverse momentum $p_{x,y}=p_{\rm{T}}$ and setting $\Delta p_{\rm{T}} \sim m_e$, we obtain, applying the same argumentation as above, that $\Delta x_{\rm{T}} \sim \pi/\Delta p_{\rm T} = \pi/m_e$ for the transverse scale of the formation volume.
We note that $1/m_e$ is the reduced Compton wavelength~\cite{di_piazza_first-order_2017,baier_quantum_1989,Ritus}.
Combining everything, the formation volume scales as $\Delta V = \Delta x_{\rm{L}} (\Delta x_{\rm{T}})^2 
\sim \pi^3/(2\omega m_e^2)$.

Using the time-averaged energy density $\mathbf{E}^2/2$ of a plane-wave laser field with electric field $\mathbf{E}$, we conclude that the average photon number density is $\mathbf{E}^2/(2\omega)$~\cite{landau_fields_1975}.
As the probability to interact with one photon scales as the fine structure constant $\alpha = e^2/(4\pi) \approx 1/137$, we see that the probability to absorb another laser photon, while the electron/positron is still virtual from the first interaction, is $\mathcal{P} \sim \alpha \Delta V \mathbf{E}^2/(2\omega)$.

Combining the three terms derived above, we find that $\mathcal{P} \sim \left[e|\mathbf{E}|/(\omega m_e)\right]^2$. 
Here, we have neglected the order‐one factor $\pi^2/16 \approx 0.6$.
Proper calculations substantiate this estimate and show that $a_0^2 \gtrsim 1$ is indeed the condition for multi-photon processes to become non-perturbative~\cite{brown_interaction_1964,Ritus}.
Here, we have introduced the classical nonlinearity parameter, $a_0=e|\mathbf{E}|/(m_e \omega)$, which does not depend on $\hbar$, even if $c$ and $\hbar$ are properly restored.
This shows that nonlinear processes can also be obtained in the realm of classical electrodynamics, e.g., nonlinear Thomson scattering, see, e.g., Refs.~\cite{yan_high_2017,PhysRevE.48.3003,mourou_optics_2006}.
If the induced recoil is large, we obtain the nonlinear generalization of Compton scattering, a genuine quantum process, which was first observed by E144~\cite{Bamber:1999zt}.

\subsection{Strong-field regime}
\label{app:strong_field_regime}
Next, we consider the regime $a_0^2 \gtrsim 1$, in which an electron interacts with many laser photons during the emission of a single non-laser photon.
In this regime, we expect that the collective, classical electromagnetic field of the laser is more important than the quantized nature of individual laser photons.
As a result, radiation and pair production can be treated in the quasi-classical approximation~\cite{baier_processes_1968}.

To understand how the longitudinal formation length changes in the regime $a_0^2 \gtrsim 1$, we consider the classical energy-momentum transfer induced by the Lorentz force.
Over a distance of a laser wavelength $\lambda$, an (approximately) plane-wave laser field can transfer a transverse momentum of the order of $a_0 
m_e$~\cite{landau_fields_1975}.
This implies that we need to revisit the above estimation of the longitudinal formation length, as the assumption $\Delta p_{\rm{T}} \ll m_e$ for the transverse momentum is now violated.
As a result, the longitudinal formation length for typically emitted photons will be shortened by a factor of $\sim 1/a_0$, i.e., $\Delta x_L \sim \lambda/a_0$, such that the transverse momentum transfer over the formation region remains small in comparison with $m_e$.
For a rigorous derivation/discussion of this result, see Refs.~\cite{di_piazza_first-order_2017,PhysRevD.93.085028}.
This is the reason why we can often apply the so-called local constant field approximation (LCFA) for $a_0 \gg 1$ (see, however,~\cite{PhysRevA.98.012134,baier_quantum_1989}).

In the case of negligible recoil, we can give a classical interpretation of this result: as we consider ultra-relativistic electrons, high-energy gamma photons are emitted into a $1/\gamma = m_e/\epsilon$ cone \cite{landau_fields_1975}.
As soon as the electron gains a transverse momentum that is large in comparison with $m_e$, its trajectory will have a transverse amplitude that extents beyond the opening of the $1/\gamma$ cone and hard gamma photons that are emitted at different parts of the trajectory can, in general, no longer interfere, as their respective $1/\gamma$ cones do no longer overlap. 

In the context of free electron laser (FELs), the two regimes are known as the undulator ($a_0 \lesssim 1$) and the wiggler regime ($a_0 \gg 1$), respectively, where in the context of a magnetic field the parameter equivalent to $a_0$ is typically called $K$~\cite{pellegrini_physics_2016,huang_review_2007}. 

\subsection{Quantum corrections to radiation}
One can argue that radiation is always a quantum effect, as an on-shell electron with 4-momentum $p^\mu$ ($p^2 = m_e^2$) cannot radiate a real photon with 4-momentum $k^\mu$ while remaining on-shell.
To see this, we square the 4-momentum conservation law $p^\mu = k^\mu + p'^\mu$ and find that $p^2 = p'^2 + 2k\cdot p'$ cannot be exactly fulfilled, as $k\cdot p' > 0$.
However, by reducing the frequency $\omega_k$ of the radiated photon, the violation can be made arbitrarily small. Thus, one can understand why radiation is indeed possible in the classical limit. 

From the classical treatment of radiation by an electron inside an oscillatory laser field with average photon 4-momentum $\kappa^\mu$ one finds that the radiated photon spectrum is characterized by the so-called ``critical'' angular frequency $\omega_c$, beyond which radiation is exponentially suppressed.
The calculation is completely analogous to synchrotron radiation inside a magnetic field~\cite{landau_quantum_1981,landau_fields_1975}.
Explicitly, one obtains the relation $\omega_c \sim \epsilon \chi_e/\hbar$, where $\chi_e = a_0 (\kappa\cdot p/m^2) = |\mathbf{E}^*|/E_{\text{cr}}$ is the quantum parameter of the radiating electron with energy $\epsilon$ (we have written $\sim$ here to allow for factors like $3/2$ in the definition of $\omega_c$). Here $|\mathbf{E}^*|$ denotes the electric field of the laser in the electron rest frame and $E_{\text{cr}}$ is the Schwinger QED critical field strength~\cite{PhysRev.82.664,Sauter}.
Note that $\chi_e$ depends explicitly on $\hbar$ (for the sake of clarity we have not used $\hbar=1$ here).

Because this result is derived from a fully classical calculation, classical electrodynamics already indicates that $\chi_e \gtrsim 0.1$ marks the onset of significant quantum effects~\cite{Fedotov:2022ely,RevModPhys.94.045001}.
This follows from the fact that an electron cannot emit photons with energy exceeding its own, assuming the external field contributes a negligible amount of energy. To arrive at this conclusion, we only need to know that the radiated photons are quantized with an energy $\hbar\omega_k$.

\subsection{Threshold for pair production}
\label{app:pair_production_threshold}
So far we have focused on the emission of gamma photons by an electron, as this process has a classical limit.
Given that the Feynman diagrams for photon emission by an electron and for electron–positron pair production by a photon inside a classical background field are related by crossing symmetry, it is straight forward to apply the above considerations also to pair production. 

In the following we focus on the strong-field regime $a_0^2 \gtrsim 1$, where the particles can absorb many laser photons from the field.
To estimate the scale at which pair production by a high-energy photon inside the laser field becomes relevant, we argue in the laboratory frame, where the pair will be born with a large longitudinal momentum.
Therefore, we can expand the energy-momentum relation using $|\mathbf{p}_\pm| \approx \epsilon_\pm + m_e^2/(2\epsilon_\pm)$.
From $k^\mu = p_+^\mu + p_-^\mu$ we obtain $\Delta\epsilon \sim m_e^2/\omega_k$, using that $p_+^\mu \approx p_-^\mu$ at threshold and assuming that three-momentum is conserved (see the similar derivation of $\Delta\epsilon$ above).

Using the same reasoning as above, we find that the lifetime of a photo-produced, virtual electron-positron pair is given by $\Delta t \sim \omega_k / m_e^2$.
To estimate when pair production becomes likely, we have to compare this value with the formation time in the strong-field regime, $\sim 1/(\omega a_0)$.

To clarify the distinction between lifetime and formation time, we note that the phase of the matrix element for pair production contains both a linear and a nonlinear term~\cite{di_piazza_first-order_2017,PhysRevD.93.085028}.
The nonlinear term arises from the acceleration of the pair by the laser field and determines the formation time of the process -- the integration is effectively cut off when the phase becomes nonlinear (see Sec.~\ref{app:strong_field_regime}).
In contrast, the linear term governs whether a significant probability amplitude is obtained.
If the lifetime, derived from the linear term, is much shorter than the formation time, the linear phase becomes highly oscillatory, leading to an exponentially small probability.

As soon as the lifetime of the virtual state matches the formation time, pair production is no longer exponentially suppressed by a tunnel exponent.
Therefore, the relevant parameter/regime is obtained from $ \omega_k / m_e^2 \gtrsim 1/(\omega a_0)$, which implies $\chi_\gamma = a_0 k\cdot \kappa/m_e^2 \gtrsim 1$.

\subsection{Multi-photon vs.\ tunneling regime}
\label{app:tunneling}
Electron-positron pair production and ionization of atoms are closely related phenomena.
For the latter, Keldysh developed a theory that treats both the multi-photon and the strong-field tunneling regime in a generalized framework~\cite{karnako_current_2015,keldysh_ionization_1965}. Central to his approach is the ``Keldysh parameter'' $\gamma_K = \sqrt{2I m_e} \omega/(e |\mathbf{E}|)$, where $I$ denotes the ionization potential of the atomic system.
For $\gamma \ll 1$ we find that ionization is a tunneling phenomena, whereas for $\gamma \gg 1$ it happens via multi-photon absorption.

As pair production can be viewed as an ionization process with an ionization potential of $I \sim m_ec^2$, we see that the parameter $1/a_0$ can be considered as a generalized Keldysh parameter for pair production~\cite{aleksandrov_lcfa_2019}.
It is important that this analogy can be substantiated further.
The Keldysh parameter is essentially the ratio of two time scales: the formation time of ionization or pair production, and the oscillation period of the laser field.
It is important to note that we use the formation time -- not the lifetime of the virtual electron-positron pair -- in the definition of the Keldysh parameter (see Sec.~\ref{app:pair_production_threshold} for a comparison of these two concepts)~\cite{aleksandrov_lcfa_2019}.
This choice is justified by the fact that the size of the formation region determines whether the process occurs in an oscillating or effectively static field.
For $a_0 \gg 1$ (which corresponds to $\gamma_K \ll 1$) we can, in general, apply the local constant field approximation (LCFA)~\cite{aleksandrov_lcfa_2019}, which implies that pair production happens inside a quasi-constant field, and we obtain a situation that is analogue to tunnel ionization in atomic systems.
We call the regime $a^2_0 \gg 1$, where corrections in $1/a_0$ can be neglected to leading order~\cite{PhysRevD.101.056017}, the deep tunneling regime.
In the opposite limit the formation region extends over multiple laser periods, which means that we are in the multi-photon regime.

\section{NBW positron kinematics at $a_0 = 5$ and $a_0 = 10$}
\label{app:a0510signal}
\begin{figure*}[pos=!ht]
\centering
\begin{overpic}[width=0.99\textwidth]{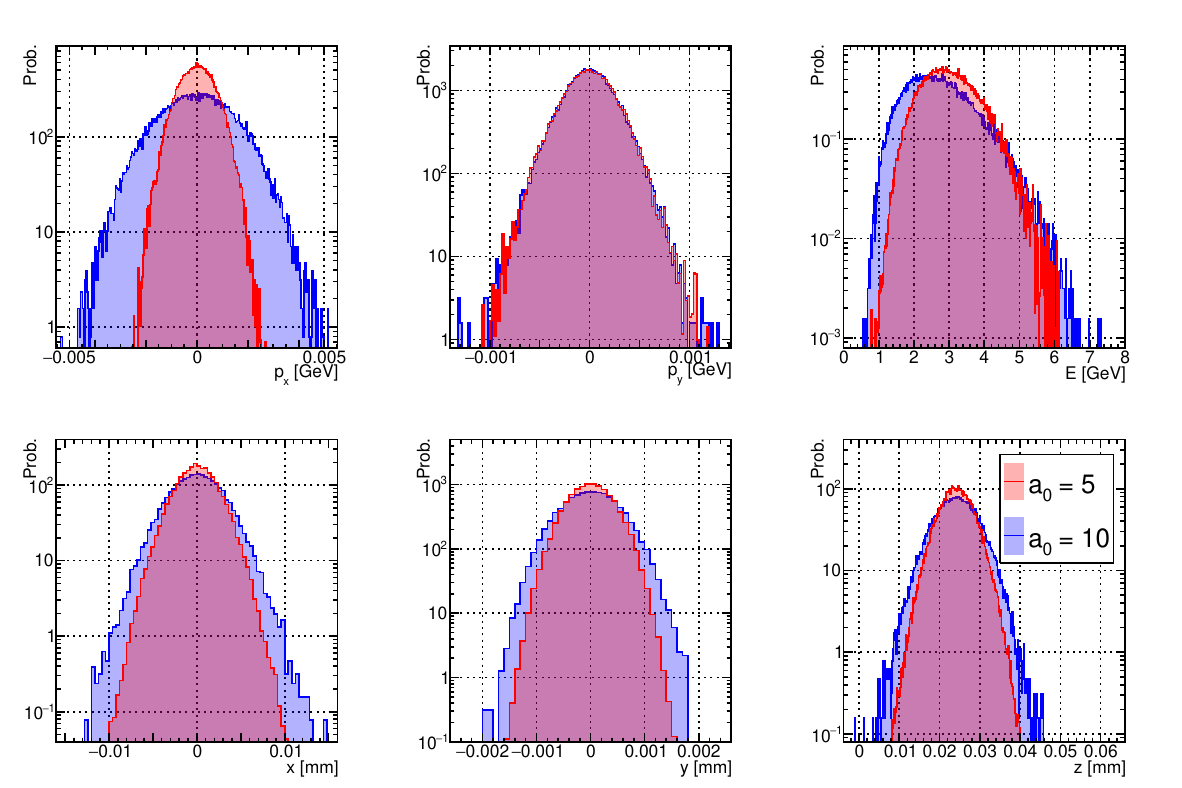}\end{overpic}
\caption{
Comparison of the IP kinematic and vertex distributions for NBW positrons produced at $a_0 = 5$ and $a_0=10$. Top: $x$ and $y$ components of the IP momentum and the energy spectrum. Bottom: $x$, $y$, and $z$ components of the vertex. $a_0 = 5$ distributions are shown in red. $a_0=10$ distributions are shown in blue.}
\label{fig:a0510_kinematics}
\end{figure*}

The analysis in this study is done with the assumption of $a_0=10$, while, as stated in Sec.~\ref{sec:fullsim}, the relevant value in E320 for the coming few years is smaller, at $a_0=4\textsf{--}5$.
While the positron yields are 3-4 order of magnitude apart, we argue that the kinematic behavior between the two scenarios is effectively the same for the purpose of this study.
A comparison of the relevant distributions for NBW positrons produced at $a_0 = 5$ and $a_0 = 10$ is shown in Fig.~\ref{fig:a0510_kinematics}.
For the $a_0 = 5$ case, the distributions are obtained from 1000 BXs simulated with \PTARMIGAN only, such that the sample includes $\sim 1.3{\rm M}$ positrons, for illustration.
To achieve this large number, \PTARMIGAN is configured to increase the pair creation rate (by a factor of $w=10^9$ here), while decreasing the weight of any created electrons and positrons, by the same factor.
This helps to resolve the positron spectrum when the total probability is much smaller than $1/N$, where $N$ is the number of primary particles.
For the $a_0 = 10$ case, the distributions are derived from the unweighted BXs used in the main analysis, i.e. \PTARMIGAN is configured such that each positron in the sample has a weight of 1.
As can be seen, the energy distributions, which have the most impact on the reconstruction pipeline differ only slightly.
The spectrum (which virtually coincides in shape with the dominant $z$-component of the momentum) of positrons produced at $a_0 = 5$ is only slightly harder than that for $a_0 = 10$.
The distributions of the $y$-component of the momentum coincide for both $a_0$ values.
However, the distribution of the $x$-component of the momentum is roughly twice as narrow at $a_0=5$ compared to $a_0=10$.
This does not impact the reconstruction in any significant way, however, as at $\sim 16$~m distance from the detector to the IP the reconstruction pipeline is completely insensitive to this difference in the distribution widths.
Finally, the vertex distribution components are only slightly narrower in the $a_0=5$ case.

\section{Detector material description}
\label{app:material_mapping}
For track reconstruction, the material properties of the detector are approximated using effective material bulk descriptions assigned to the surfaces.
This is achieved by simulating non-interacting particle propagation through the \GEANT setup and recording material characteristics at each propagation step.
The resulting ``material tracks'' provide material-properties data that is mapped onto the ACTS surfaces using material grids, imposed onto each sensitive surface.
For each material track, bins of the material grids intersected by it are identified.
The track's material recordings are then projected into the closest intersected bin along the track's direction.
The accumulated material properties within each bin are then averaged, resulting in effective material descriptions that account for uncertainties such as energy loss and multiple scattering during reconstruction.

\begin{figure*}[pos=!ht] 
\centering 
\begin{subfigure}{0.452\textwidth} 
\begin{overpic}[width=0.98\textwidth]{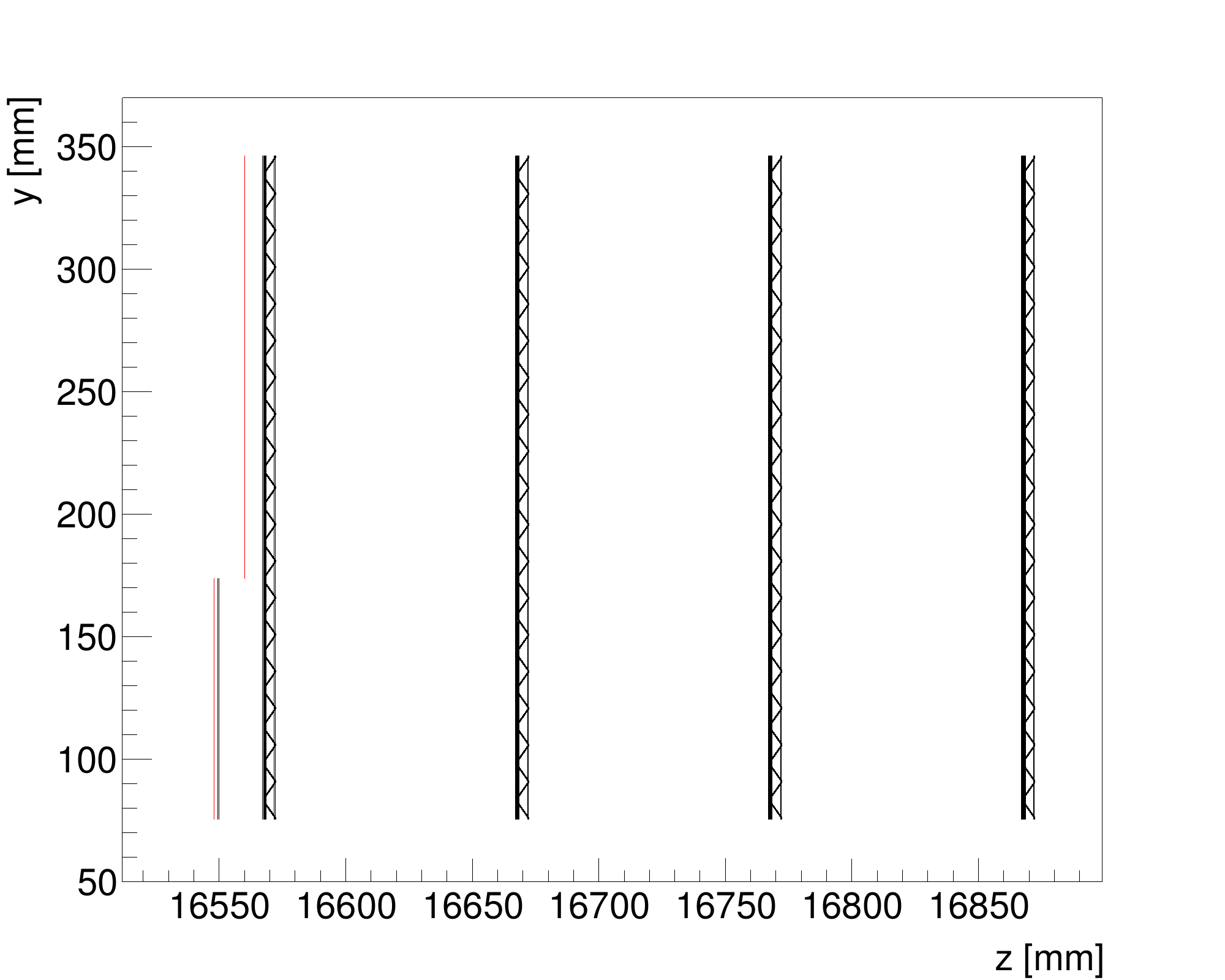}\end{overpic} 
\end{subfigure} 
\begin{subfigure}{0.348\textwidth} 
\begin{overpic}[width=0.98\textwidth]{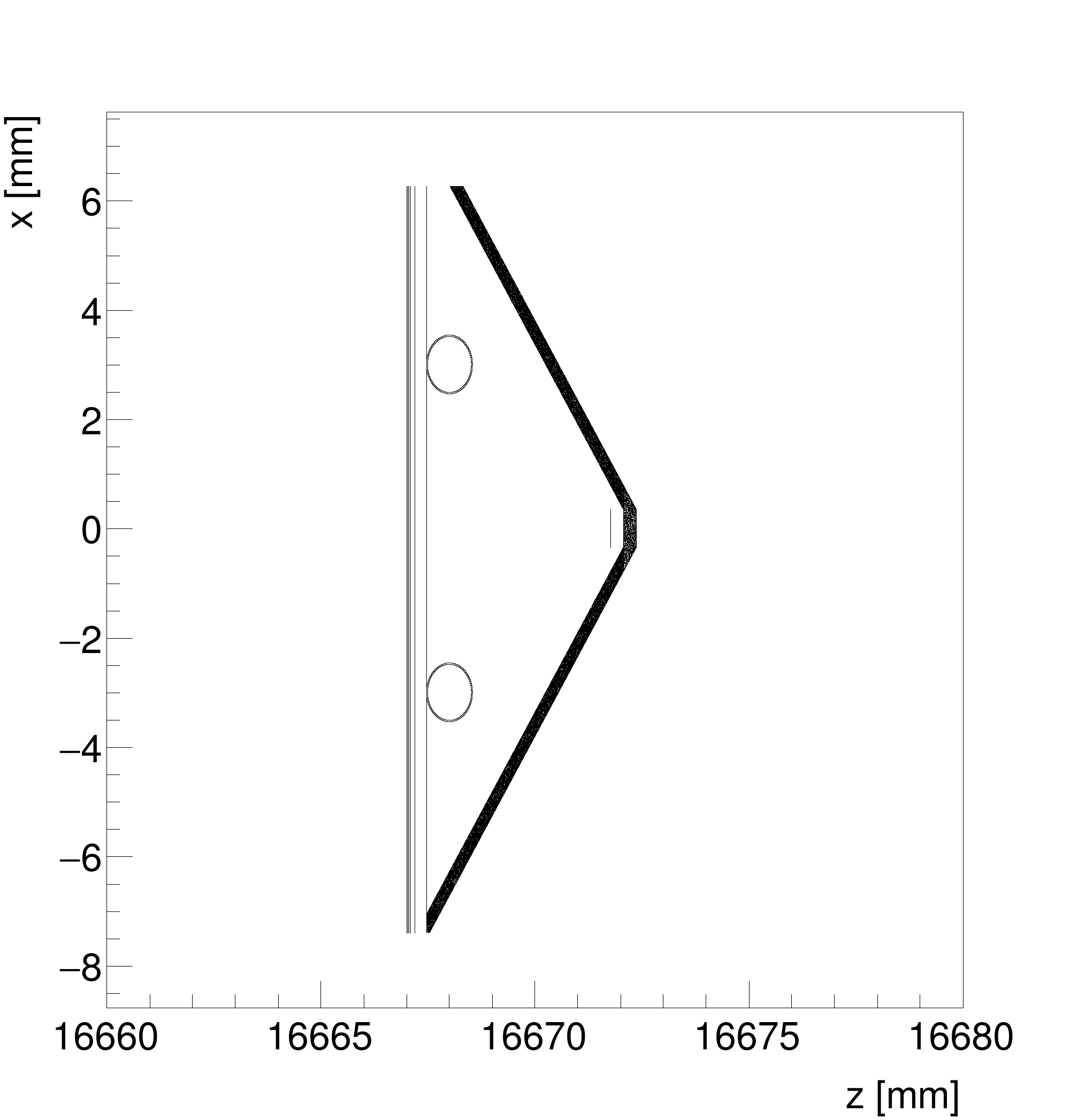}\end{overpic} 
\end{subfigure} 
\caption{
Tracking detector material map. 
Each point corresponds to a sample of the material properties of the respective \GEANT volume.
Vertices of the ``material tracks'' are shown in red.
Left: $y\textsf{--}z$ projection of the entire detector.
Right: $x\textsf{--}z$ projection of one of the staves.
The shape of the stave's cooling pipes and carbon fiber support structure is clearly visible.} \label{fig:geant4_material_map} 
\end{figure*}

A \GEANT simulation of $10^6$ material-recording particles is performed, with particle vertices uniformly distributed in a plane parallel to the tracking layers.
The distribution's bounds match the rectangle occupied by the sensitive surfaces of the ALPIDE chips in the $x\textsf{--}y$ plane, and the particles' momentum is directed along the $z$-axis.
Fig.~\ref{fig:geant4_material_map} shows the distribution of material sampling points.

Each sensitive surface in ACTS representing an ALPIDE sensor is assigned a $256 \times 128$ grid, with the larger bin count corresponding to the chip's longer side.
The material properties of the vacuum chamber window are directly extracted from the \GEANT description, as sampling is unnecessary for this homogeneous bulk.

\begin{figure*}[pos=!ht]
\centering
\begin{overpic}[width=0.8\textwidth]{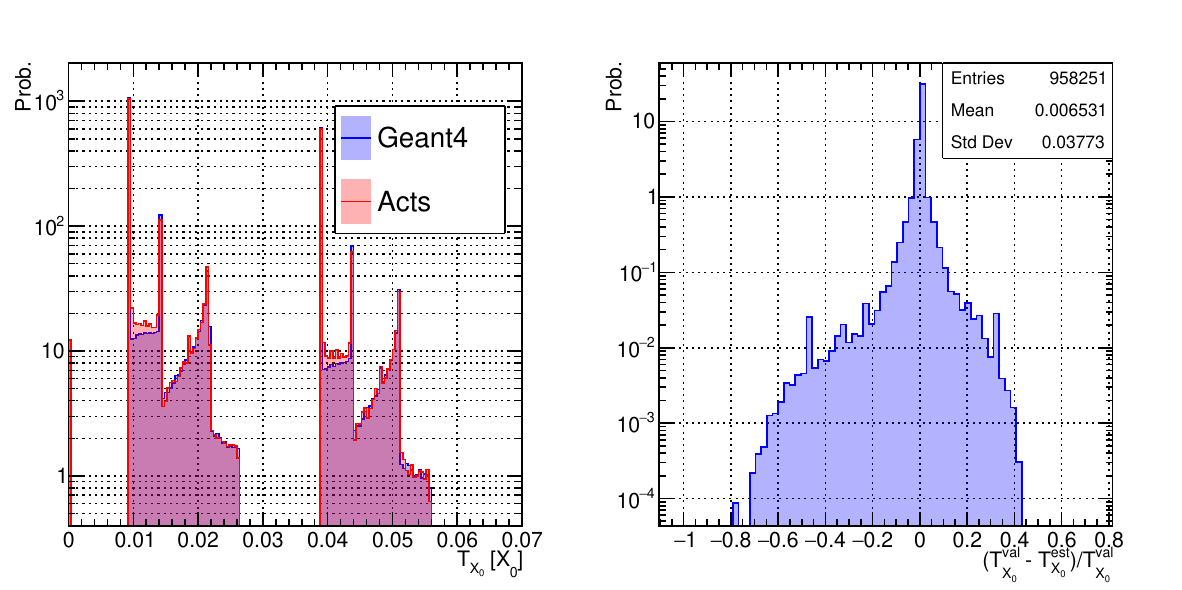}\end{overpic}
\caption{
Material thickness in units of radiation length encountered by material-recording particles in the detector implementations.
Left: Distribution of total material thickness encountered by the particles in the \GEANT and ACTS implementations.
The left distribution bulk corresponds to particles generated downstream of the vacuum exit window.
The right distribution bulk corresponds to particles generated upstream of the vacuum exit window, such that they traverse the window as well.
Right: Response distribution of the material thickness encountered by the material-recording particles in the \GEANT and ACTS implementations of the detector.}
\label{fig:material_diff}
\end{figure*}

A separate sample of material tracks is generated to validate the material grids.
The sampling follows the same procedure as described above, except for the simulated particles originating within a rectangle coinciding with the vacuum chamber window boundary in the $x\textsf{--}y$ plane.
For these particles, the $z$-coordinate of the vertices is shifted upstream of the window to ensure their interaction with the window's material.
The kinematic features of these generated particles are recorded and supplied as input to the ACTS simulation.
Using this input, the corresponding material-recording particles are propagated through the ACTS geometry implementation with the material grid imposed.
Fig.~\ref{fig:material_diff} compares the total material thickness in units of effective radiation length, $T_{X_0}$, encountered by particles in the \GEANT and ACTS descriptions.
As can be seen, the distribution exhibits two bulks of the same shape, corresponding to the tracks that interact with and those that do not interact with the vacuum chamber window material.
The relative error between the ACTS and \GEANT $T_{X_0}$ is centered around zero with a $\sim$4\% resolution, demonstrating good agreement between the material properties of the detector descriptions.

\section{FastSim cluster size sampling}
\label{app:cluster_size_sampling}
\begin{figure*}[pos=!ht]
\centering
\begin{overpic}[width=1.0\linewidth]{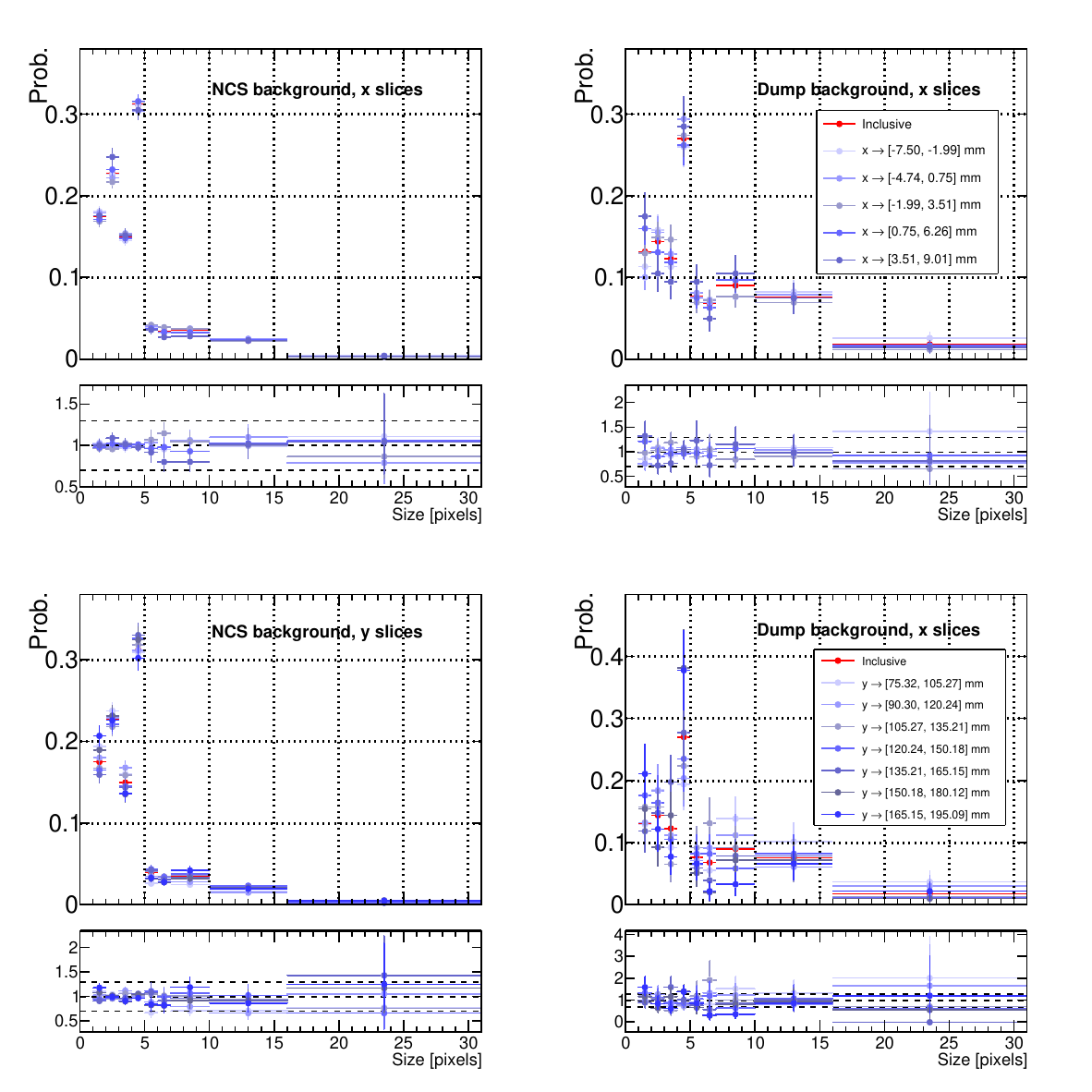}\end{overpic}
\caption{Comparison of the inclusive cluster size distribution with the cluster size distributions constructed in a set of slices along the $x$ and $y$ axes for NCS and dump background sources.
The global coordinate ranges of the slices are indicated in the legend.
All distributions are constructed based on the FullSim data.
For each plot, the bin-wise ratio of the sliced distributions to the inclusive distribution is also shown.
Top: $x$-sliced distributions of the NCS (left) and the dump (right) backgrounds.
Bottom: $y$-sliced distributions of the NCS (left) and dump (right) backgrounds.}
\label{fig:cl_size_corr}
\end{figure*}

The FastSim employs coordinate-independent cluster size distributions to estimate and assign errors to the simulated background measurements.
That is, the cluster sizes are sampled from the corresponding FullSim distributions of the two background types (NCS background and dump background), as discussed in Sec.~\ref{sec:fastsim}.
The use of coordinate-independent sampling is justified by the negligible correlation between cluster size and position of the particle hit, as shown in Fig.~\ref{fig:cl_size_corr}.
The comparison seen is between the inclusive cluster size distribution (in red) and a set of analogous distributions constructed in a few slices along the $x$ and $y$ axes (for both types of backgrounds).
At the bottom of each plot we show the bin-wise ratio of the sliced distributions to the inclusive one.
Due to lack of statistics when slicing the data, we apply the slicing with overlaps, so that the sample sizes of the sliced distributions are kept reasonably high, while the evolution, if any, may be apparent.
The slice width in $x$ is equal to 40\% of the chip length along the $x$ axis, with the slice centers spaced at 20\% of the chip length.
The slice widths in $y$ correspond to the full chip length along the $y$ axis, with the slice centers spaced at half the chip length.
The plot covers the full chip width in $x$ and the entire signal region in $y$.

As can be seen, the differences between the inclusive distribution and the sliced distributions are negligible.
The agreement is particularly good for the NCS background.
The dump background distributions exhibit a much broader spread due to the limited statistics, but the sliced distributions agree with the inclusive one within the statistical uncertainties.
Thus, we conclude that the inclusive cluster size distribution is a valid representation across the relevant ranges of cluster sizes and coordinates for this analysis.

\section{Initial state correlations}
\label{app:fast_sim_inital_corr}
\begin{figure*}[!ht]
\centering
\begin{overpic}[width=0.99\textwidth]{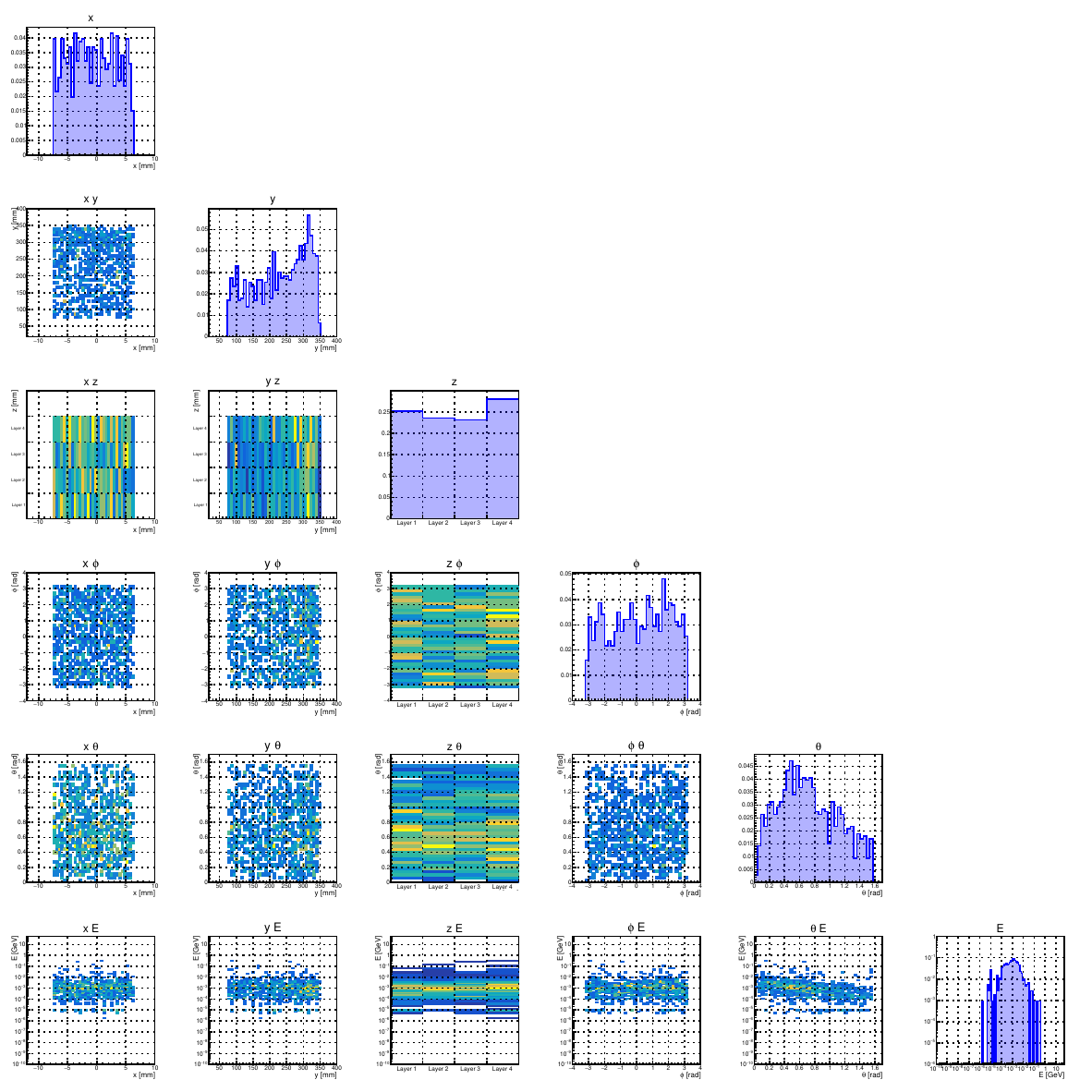}\end{overpic}
\caption{Pair-wise correlations of the particles' features at the initial state of their propagation for the FullSim dump background. 0.32 BX are shown.}
\label{fig:corr_dump_FullSim}
\end{figure*}

\begin{figure*}[!ht]
\centering
\begin{overpic}[width=0.99\textwidth]{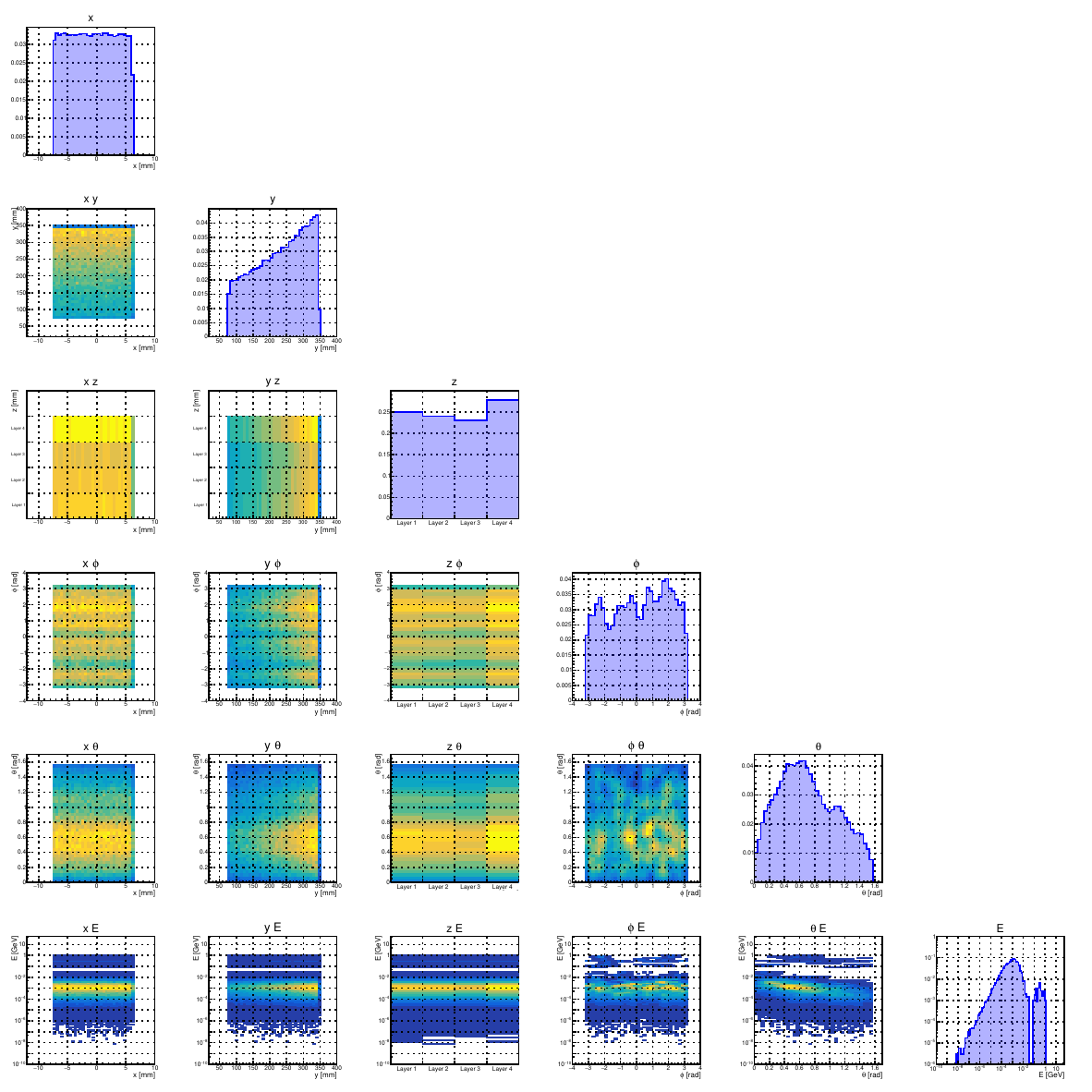}\end{overpic}
\caption{Pair-wise correlations of the particles' features at the initial state of their propagation for the FastSim dump background. 100 BXs are shown.}
\label{fig:corr_dump_FastSim}
\end{figure*}

\begin{figure*}[!ht]
\centering
\begin{overpic}[width=0.99\textwidth]{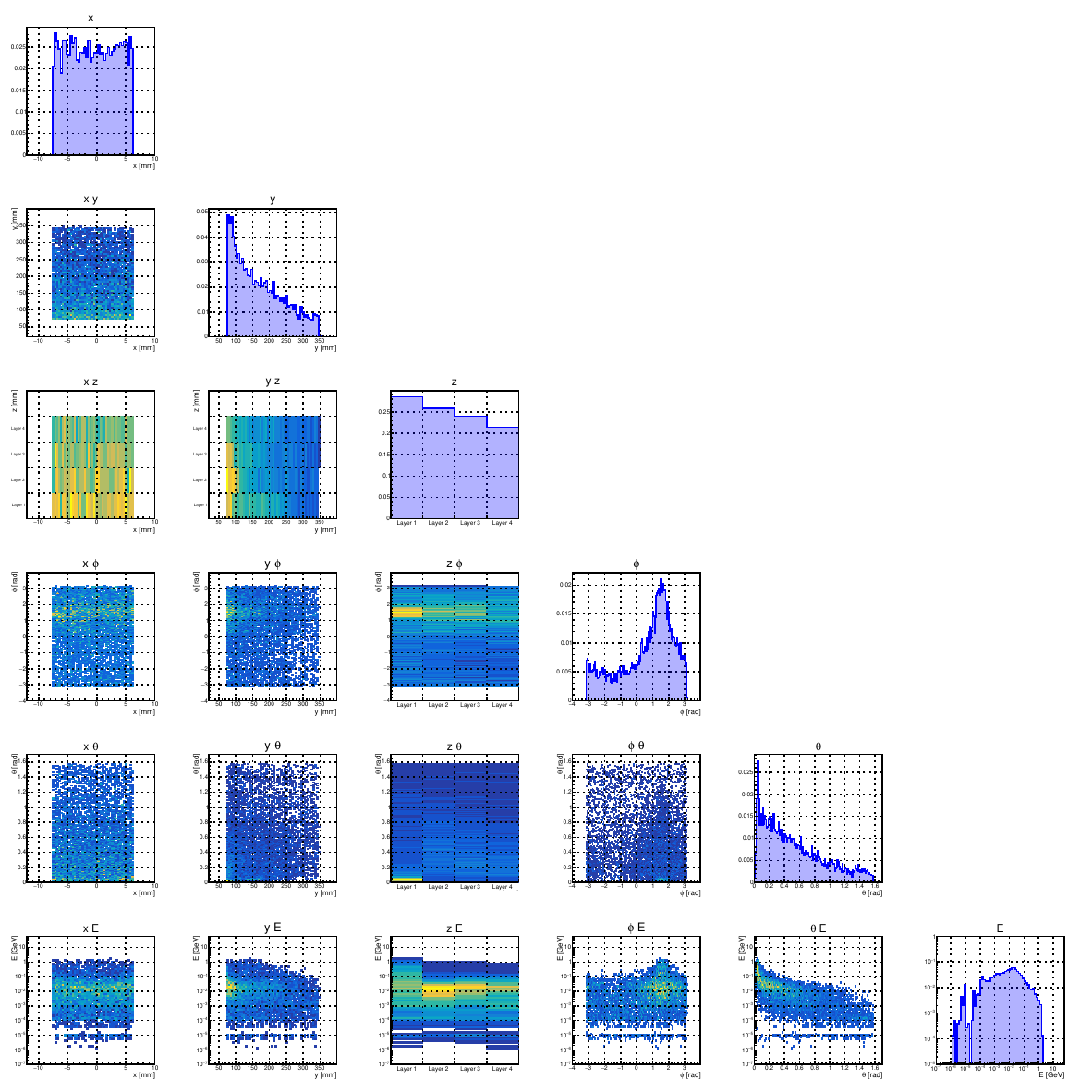}\end{overpic}
\caption{Pair-wise correlations of the particles' features at the initial state of their propagation for the FullSim NCS background. 12 BXs are shown.}
\label{fig:corr_NCS_FullSim}
\end{figure*}

\begin{figure*}[!ht]
\centering
\begin{overpic}[width=0.99\textwidth]{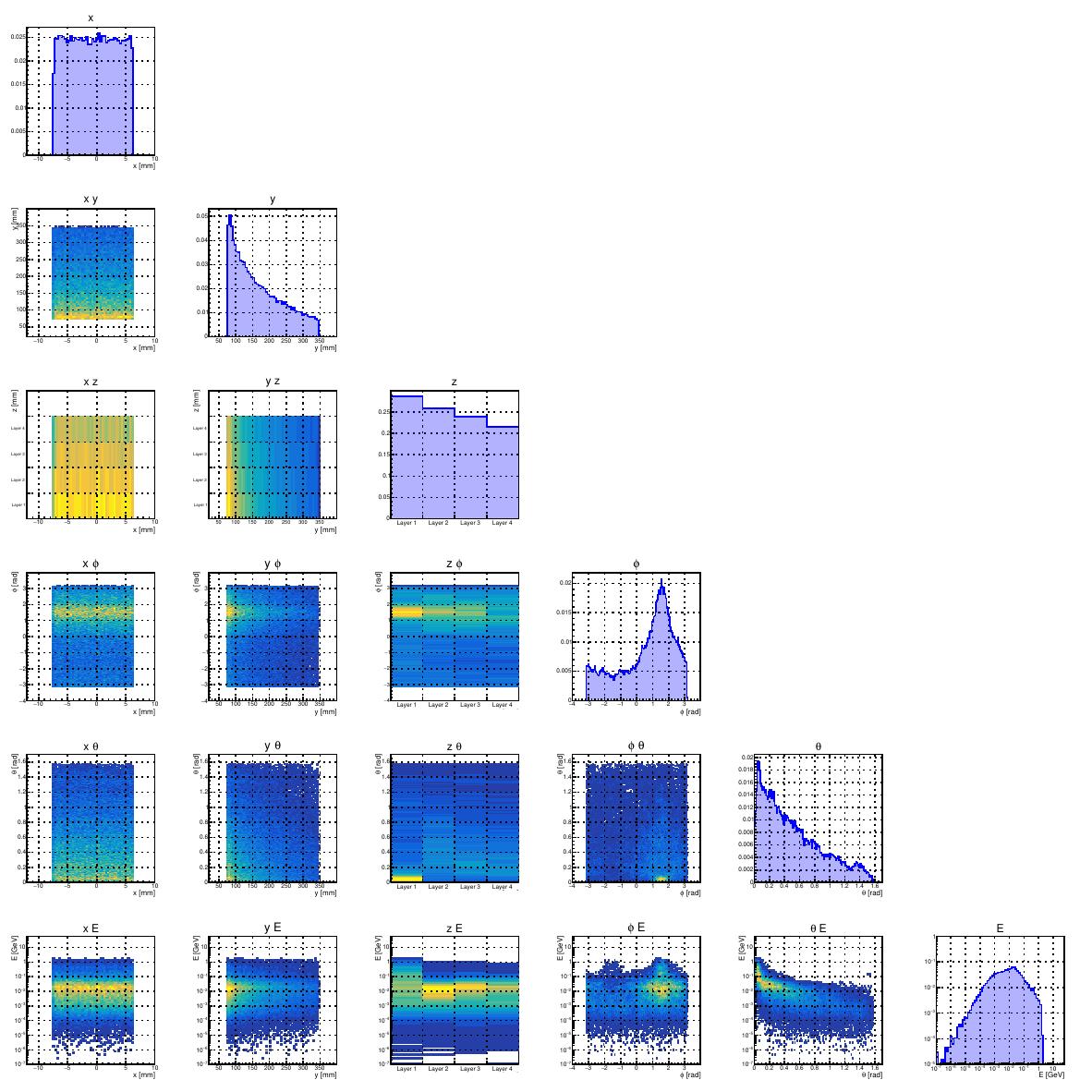}\end{overpic}
\caption{Pair-wise correlations of the particles' features at the initial state of their propagation for the FastSim NCS background. 100 BXs are shown.}
\label{fig:corr_NCS_FastSim}
\end{figure*}

The pair-wise correlations of the particles' features at the initial state of their propagation for the dump background in FullSim and FastSim are shown in Fig.~\ref{fig:corr_dump_FullSim} and~\ref{fig:corr_dump_FastSim}, respectively.
Similarly, the pair-wise correlations of the particles' features at the initial state of their propagation for the NCS background in FullSim and FastSim are shown in Fig.~\ref{fig:corr_NCS_FullSim} and~\ref{fig:corr_NCS_FastSim}, respectively.

While the comparison between FullSim and FastSim in that respect is qualitative, it is still possible to see that the FastSim process reproduces the correlations seen in the FullSim within the statistical error margins.

\section{Simulation pipeline overview}
\label{app:simflow}
\begin{figure*}[!ht]
\centering
\begin{overpic}[width=0.99\textwidth]{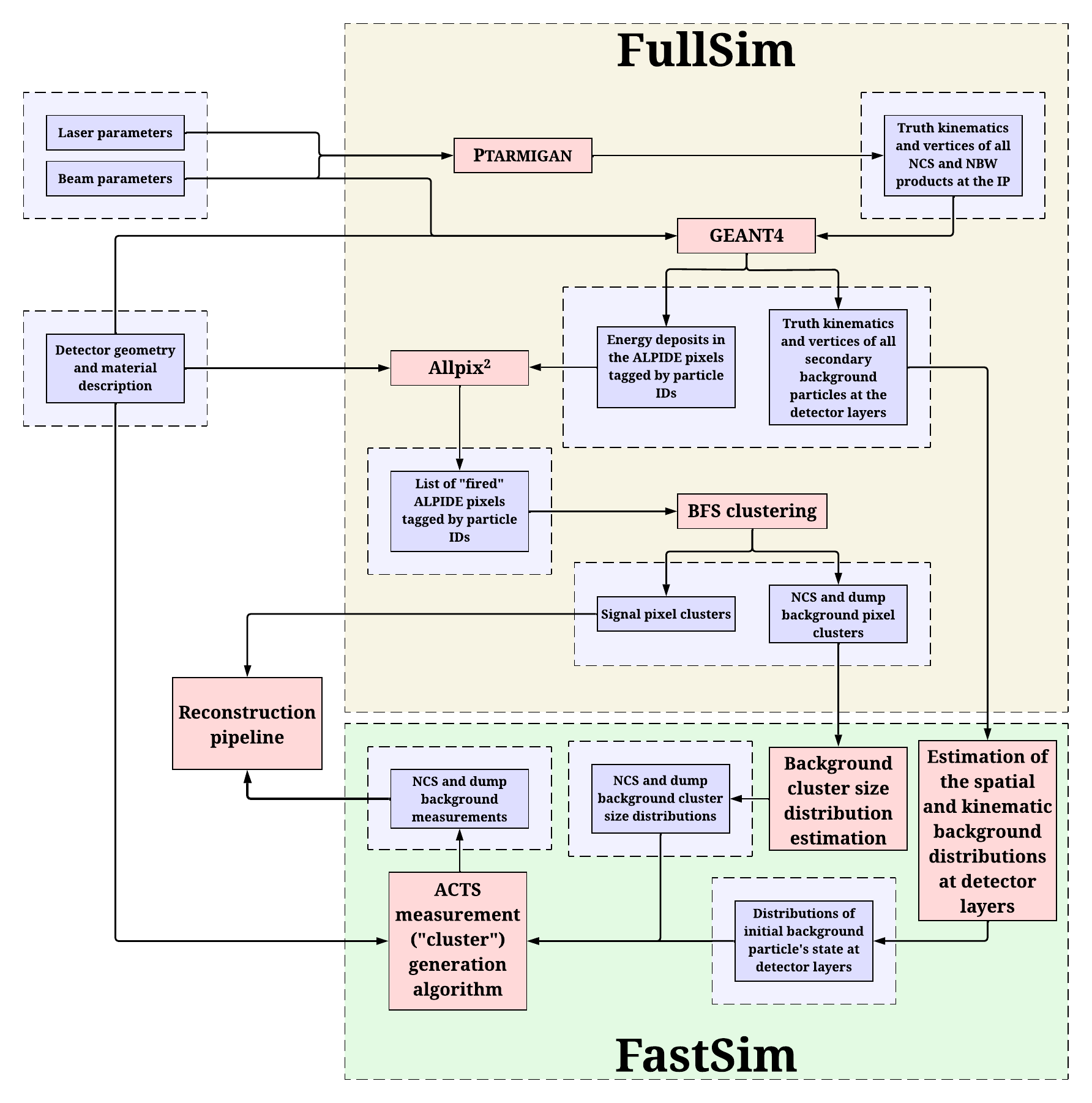}\end{overpic}
\caption{Schematic representation of the simulation pipeline, including both FullSim and FastSim steps.
The simulation steps are shown in red, while the inputs and outputs of individual steps are shown in blue. 
The output of both FullSim and FastSim serves as input for the reconstruction pipeline.}
\label{fig:simulation_pipeline}
\end{figure*}

Fig.~\ref{fig:simulation_pipeline} shows a schematic representation of the simulation pipeline, including both FullSim and FastSim steps, as described in Sec.~\ref{sec:simulation} and~\ref{sec:background_fastsim}. The inputs to this pipeline are shown in blue (laser and beam parameter as well as the geometry) outside of the main two shaded rectangles (FullSim and FastSim). The output of this simulation pipeline is fed to the reconstruction pipeline which is discussed in Sec.~\ref{sec:reconstruction}.

\section{Kinematic map construction}
\label{app:kinematics_guess}
\begin{figure*}[!ht]
\centering
\begin{overpic}[width=0.99\textwidth]{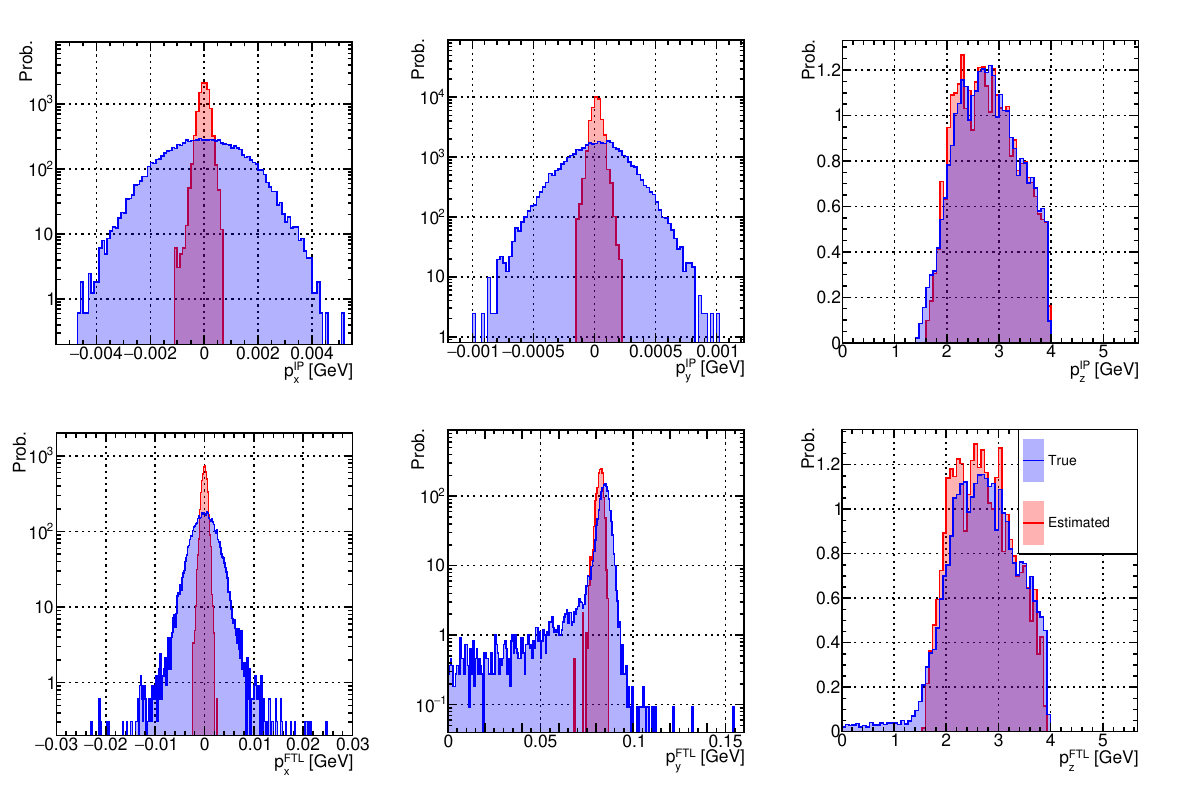}\end{overpic}
\caption{
Comparison of the true and estimated momentum components for the NBW positrons' clusters. 
Top: $x$-, $y$-, and $z$-momentum components at the IP. 
Bottom: $x$-, $y$-, and $z$-momentum components at the FTL. 
The true distributions are shown in blue. 
The estimated distributions are shown in red.}
\label{fig:kinematics_diff}
\end{figure*}

\begin{figure*}[!ht]
\centering
\begin{overpic}[width=0.99\textwidth]{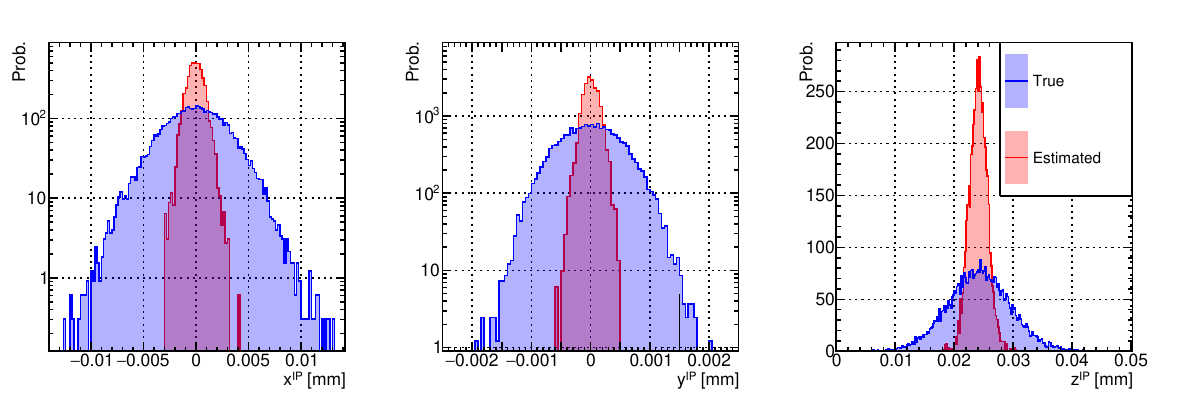}\end{overpic}
\caption{
Comparison of the true and estimated vertex components for the NBW positrons' clusters. 
Top: $x$-, $y$-, and $z$-vertex components at the IP. 
Bottom: $x$-, $y$-, and $z$-vertex components at the FTL. 
The true distributions are shown in blue. 
The estimated distributions are shown in red.}
\label{fig:vertex_diff}
\end{figure*}

The quality of the constructed map $\mathcal{M}$ is assessed by inferring the track parameters of signal NBW positrons from their associated pivot clusters and comparing these estimates to the true vertex positions and momenta. 
Fig.~\ref{fig:kinematics_diff} presents the comparison of true and inferred momenta at both the IP and the FTL, while Fig.~\ref{fig:vertex_diff} shows the corresponding vertex comparison.

The $z$-component of the momentum, which dominates and thus determines the positron energy to leading order, is inferred with good accuracy. 
The mean of the $y$-component of the FTL momentum is also well estimated, although the shape of the true distribution is not reproduced.

The $y$-component of the IP momentum, as well as the $x$-components of both the IP and FTL momenta, are centered at zero with widths significantly narrower than those of the true distributions. 
This behavior arises because positrons with both positive and negative $x$- and $y$-momenta fall into the same $y$-bins of the one-dimensional $\mathcal{M}$ map.
The non-zero widths of the inferred distributions reflect the finite size of the sample rather than the underlying momentum spread. 
The same effect explains the behavior observed in the vertex distributions.

\section{Reconstructed kinematics}
\label{app:reco_kinematics}
\begin{figure*}[!ht]
\centering
\includegraphics[width=0.8\textwidth]{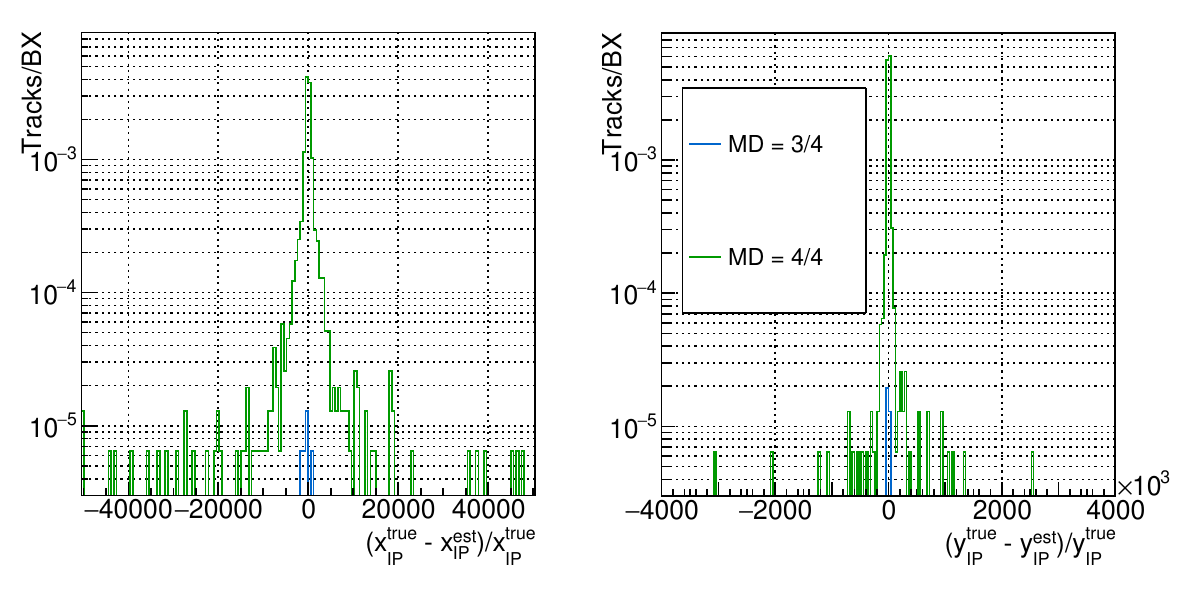}
\caption{
Vertex position response distributions for reconstructed tracks of different matching degrees. 
Left: $x$-vertex response distribution. 
Right: $y$-vertex response distribution.}
\label{fig:sig_bkg_vertex}
\end{figure*}

Fig.~\ref{fig:sig_bkg_vertex} presents the response distributions for the estimated  $x$- and $y$-vertex positions of reconstructed tracks with different matching degrees, following the application of selection criteria described in Sec.~\ref{sec:acts_track_analysis}. 
The $z$-position is not included in this estimation, as $z=0$ defines the plane where back-propagation terminates and is therefore fixed.

As can be seen, vertex estimations exhibit significant uncertainties, limiting their reliability for precise vertex reconstruction. 
This behavior is expected given the experimental setup: the tracking detector is positioned 16~m from the IP, while the vertex position scale is on the order of $\sim100~\mu$m. 
Errors introduced at the detector level are further amplified by the presence of quadrupole magnets, which are sensitive to the transverse position of the particle.

Further studies are required to enhance the pipeline's vertex reconstruction capabilities. 
This challenge is well known in the field of track reconstruction, and dedicated vertex-finding algorithms have been developed that could be adapted to E320~\cite{billoir1992fast,fruhwirth2004adaptive,schlag2022advanced}.

\begin{figure*}[!ht]
\centering
\includegraphics[width=0.99\textwidth]{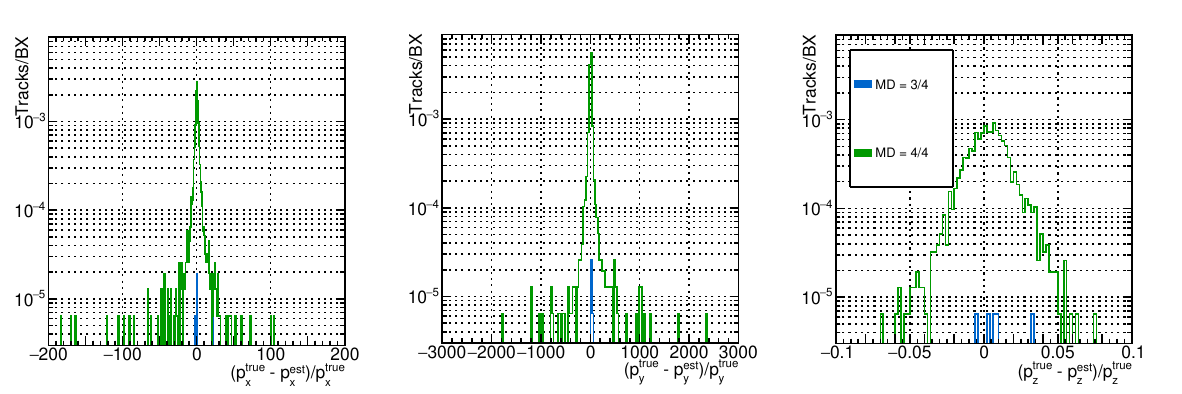}
\caption{
Response distributions for the reconstructed track momentum components and energy at the IP, for different matching degrees.}
\label{fig:sig_bkg_momentum}
\end{figure*}

Fig.~\ref{fig:sig_bkg_momentum} presents the response distributions for the estimated $x$-, $y$-, and $z$-components of the IP momentum for reconstructed tracks of different matching degrees after applying the selection criteria.

As can be seen, the deviations from the truth for the $x$- and $y$-momentum components are relatively high. 
This is again attributed to the influence of quadrupole magnets and the long back-propagation distance of 16~m. 
These errors are expected to be reduced significantly with implementation of a more rigorous statistical model of the IP kinematics. 
In contrast, the dominant $z$-component of the momentum is in good agreement with the truth. 
Furthermore, the resolution of the particles' energy remains small even for tracks with matching degrees below one. 
This indicates that although such tracks are present in the final reconstructed sample, they should not be classified as background. 
Instead, they may represent part of the signal and warrant further investigation.

\end{appendices}

\end{document}